\documentclass[acmsmall,screen]{acmart}
\AtBeginDocument{%
  }

\usepackage{algorithm}
\usepackage[noend]{algpseudocode}
\usepackage[normalem]{ulem}
\usepackage{amsfonts}
\usepackage{bm}
\usepackage{stmaryrd}
\usepackage{algorithm}
\usepackage{xspace}
\usepackage[capitalise]{cleveref}
\usepackage{tikz}
\usetikzlibrary{quantikz}
\usepackage{subcaption}
\usepackage{pgfplotstable}
\usepackage{wrapfig}
\usepackage{prooftree}
\usepackage{numprint}
\usepackage{csvsimple}
\usepackage{placeins}
\npthousandsep{,}

\renewcommand{\vec}[1]{\bm{\mathrm{#1}}}

\newcommand{\circuit}{C}

\newcommand{\param}{\rho}
\newcommand{\vparam}{{\vec{\rho}}}

\newcommand{\unitc}{\mathbb{S}}
\newcommand{\ours}{\textsc{queso}\xspace}
\newcommand{\ourspp}{\textsc{queso-pp}\xspace}
\newcommand{\oursrz}{\textsc{queso-rz-obj}\xspace}
\newcommand{\cnot}{\textrm{CX}}
\newcommand{\x}{X}
\newcommand{\h}{H}
\newcommand{\rz}{R_z}
\newcommand{\rx}{R_x\xspace}
\newcommand{\ry}{R_y}
\newcommand{\rxx}{R_{xx}}
\newcommand{\cz}{\textrm{CZ}}
\newcommand{\uone}{\textrm{U1}}
\newcommand{\utwo}{\textrm{U2}}
\newcommand{\uthree}{\textrm{U3}}
\newcommand{\poly}{p}
\renewcommand{\paragraph}[1]{\vspace{.03in}\noindent\textbf{{#1}.~~~}}
\newcommand{\gapply}[3]{#1 ~ #2 ~ #3}
\newcommand{\sem}[1]{\llbracket #1 \rrbracket}
\newcommand{\symb}{S}
\newcommand{\cvar}{v}
\newcommand{\cvars}{\vec{\cvar}}
\newcommand{\Z}{\mathbb{Z}_2}
\newcommand{\Zn}{\Z^n}
\newcommand{\sspace}{R}
\newcommand{\pit}{\textsc{pit}\xspace}
\newcommand{\pif}{\textsc{pif}\xspace}
\newcommand{\ibm}{\textsc{ibm}\xspace}
\newcommand{\voqc}{\textsc{voqc}\xspace}
\newcommand{\tket}{\textsc{tket}\xspace}

\newcommand{\cost}{\textsc{cost}}
\newcommand{\applymax}{\textsc{apply-max}}
\newcommand{\matchsymb}{\textsc{match-symb}\xspace}
\newcommand{\maxbeam}{\textsc{max-beam}\xspace}
\newcommand{\qaoa}{\textsc{qaoa}\xspace}
\renewcommand{\leq}{\leqslant}
\renewcommand{\geq}{\geqslant}
\Crefname{figure}{Fig.}{Figs.}
\Crefname{tabular}{Tab.}{Tabs.}
\Crefname{section}{\S}{\S}
\Crefname{theorem}{Thm.}{Thms.}
\Crefname{lemma}{Lem.}{Lems.}
\Crefname{corollary}{Cor.}{Cors.}

\Crefname{algorithm}{Alg.}{Algs.}

\setcopyright{rightsretained}
\acmPrice{}
\acmDOI{10.1145/3591254}
\acmYear{2023}
\copyrightyear{2023}
\acmSubmissionID{pldi23main-p190-p}
\acmJournal{PACMPL}
\acmVolume{7}
\acmNumber{PLDI}
\acmArticle{140}
\acmMonth{6}
\received{2022-11-10}
\received[accepted]{2023-03-31}

\begin{document}

%%
%% The "title" command has an optional parameter,
%% allowing the author to define a "short title" to be used in page headers.
\title{Synthesizing Quantum-Circuit Optimizers}

\bibliographystyle{ACM-Reference-Format}
%% Citation style
%% Note: author/year citations are required for papers published as an
%% issue of PACMPL.
\citestyle{acmauthoryear}   %% For author/year citations

%%
%% The "author" command and its associated commands are used to define
%% the authors and their affiliations.
%% Of note is the shared affiliation of the first two authors, and the
%% "authornote" and "authornotemark" commands
%% used to denote shared contribution to the research.
\author{Amanda Xu}
\affiliation{%
  \institution{University of Wisconsin-Madison}%
  \city{Madison, WI}
  \country{USA}%
}
\email{axu44@wisc.edu}

\author{Abtin Molavi}
\affiliation{%
  \institution{University of Wisconsin-Madison}%
  \city{Madison, WI}
  \country{USA}%
}
\email{amolavi@wisc.edu}

\author{Lauren Pick}
\affiliation{%
  \institution{University of Wisconsin-Madison}%
  \city{Madison, WI}
  \country{USA}%
}
\email{lpick2@wisc.edu}

\author{Swamit Tannu}
\affiliation{%
  \institution{University of Wisconsin-Madison}%
  \city{Madison, WI}
  \country{USA}%
}
\email{stannu@wisc.edu}

\author{Aws Albarghouthi}
\affiliation{%
  \institution{University of Wisconsin-Madison}%
  \city{Madison, WI}
  \country{USA}%
}
\email{aws@cs.wisc.edu}

%%
%% By default, the full list of authors will be used in the page
%% headers. Often, this list is too long, and will overlap
%% other information printed in the page headers. This command allows
%% the author to define a more concise list
%% of authors' names for this purpose.

%%
%% The abstract is a short summary of the work to be presented in the
%% article.
\begin{abstract}
  Near-term quantum computers are expected to work in an environment where each operation is noisy, with no error correction.
  Therefore, quantum-circuit optimizers are applied to minimize the number of noisy operations.
  Today, physicists are constantly experimenting with novel devices and architectures.
  For every new physical substrate and for every modification of a quantum computer, we need to modify or rewrite major pieces of the optimizer to run successful experiments.
  In this paper, we present \ours, an efficient approach for automatically synthesizing a quantum-circuit optimizer for a given quantum device.
  For instance, in 1.2 minutes, \ours can synthesize an optimizer with high-probability correctness guarantees for \ibm computers that significantly outperforms leading compilers, such as \ibm's Qiskit and \tket, on the majority (85\%) of the circuits in a diverse benchmark suite.

  A number of theoretical and algorithmic insights underlie \ours : 
  (1) An algebraic approach for representing rewrite rules and their semantics.
This facilitates reasoning about complex \emph{symbolic} rewrite rules that are beyond the scope of existing techniques.
(2) A fast approach for probabilistically verifying equivalence of quantum circuits by reducing the problem to a special form of \emph{polynomial identity testing}.
(3) A novel probabilistic data structure, called a \emph{polynomial identity filter} (\pif), for efficiently synthesizing rewrite rules.
(4) A beam-search-based algorithm that efficiently applies the synthesized symbolic rewrite rules to optimize quantum circuits.
  
\end{abstract}

  


\maketitle

\section{Introduction}
The dream of quantum computing has been around for decades, but it is only recently that we have begun to witness promising physical realizations of quantum computers.
Quantum computers enable efficient simulation of quantum mechanical phenomena,  potentially opening the door to advances in quantum physics, chemistry, material design, and beyond.
Near-term quantum computers with several dozens of qubits are expected to operate in a noisy environment without error correction, in a model of computation called {\em Noisy Intermediate Scale Quantum} (\textsc{nisq}) computing~\cite{preskill2018quantum}.  

In \textsc{nisq} computers, each operation is noisy.
Therefore,
powerful quantum-circuit optimizers are absolutely crucial: we want to produce smaller circuits that are more tolerant to noise.
Without careful optimization, one can easily end up with a circuit whose results are indistinguishable from random noise.
However, the state of quantum hardware is in flux. There are so many physical realizations of quantum computers, and physicists are constantly experimenting with new devices and architectures---\emph{neutral atoms}, \emph{superconducting circuits}, \emph{semiconductor devices}~\cite{saffman2019next,wilen2021correlated,watson2018programmable}.
\textbf{For every new physical substrate and for every modification of a quantum computer, we need to modify or rewrite major pieces of the optimizer to run experiments.}
This is a bottleneck in our progress towards a quantum computing future:
writing optimizers is a tedious, iterative, heuristic process,
and one that is error-prone~\cite{paltenghi2022bugs}. 

Our goal in this paper is to answer the following question: 
\begin{center}
  \emph{Given a specification of a quantum architecture, can we automatically synthesize an efficient and correct quantum-circuit optimizer?}
\end{center}
Recent developments only partially address this question:
The  quantum-circuit optimizer, \voqc~\cite{hietala2021verified}, is manually written with machine-checked correctness proofs, and therefore is not automatically extensible to new quantum architectures.
The superoptimizer, Quartz~\cite{xu2022quartz}, automatically synthesizes semantics-preserving circuit rewrite rules;
however, it can only synthesize simple rewrite rules and, as a superoptimizer, is heavily dependent on hand-crafted, device-specific optimization passes without which the synthesized rules have little impact.

We present \ours, a new technique that rapidly synthesizes sophisticated, correct rewrite rules.
\ours then efficiently applies the synthesized rules to optimize quantum circuits.
\ours builds upon four critical ideas:
(1) An algebraic approach for representing rewrite rules and their semantics.
This allows us to reason about and synthesize complex optimizations  beyond the scope of existing techniques.
(2) A fast probabilistic verification approach for checking rewrite-rule correctness by reducing the problem to a special form of \emph{polynomial identity testing}
and demonstrating that the standard fast randomized algorithm applies. 
(3) A probabilistic data structure for efficiently synthesizing equivalent pairs of circuits without incurring a quadratic explosion.
(4) A beam-search-based algorithm that efficiently applies synthesized rewrite rules to optimize  circuits.

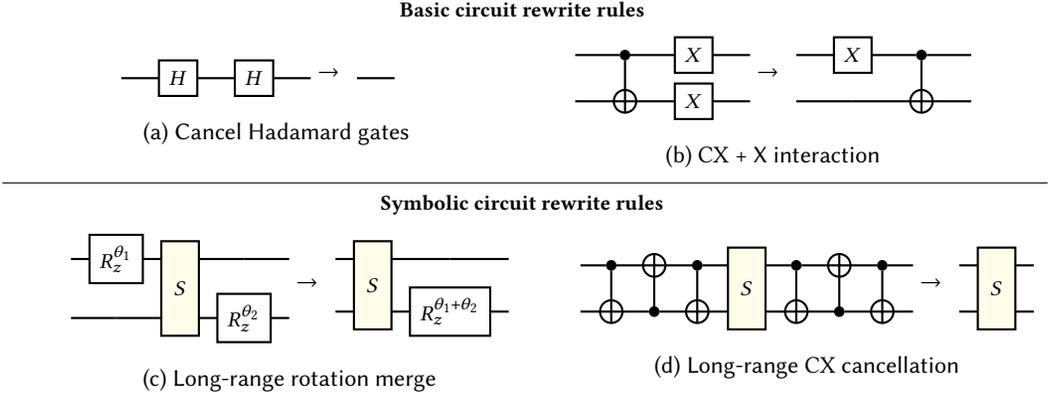
\begin{figure}[t!]\footnotesize
  \textbf{Basic circuit rewrite rules}  \vspace{.5em}

  \begin{subfigure}{.47\textwidth}
    \centering
      \begin{quantikz}[row sep = 4pt]
      & \gate{H} & \gate{H} & \qw
      \end{quantikz}
      $\to$
      \begin{quantikz}[row sep = 4pt]
      & \qw &
      \end{quantikz}
      \caption{Cancel Hadamard gates}\label{fig:gatec}
  \end{subfigure}
  \begin{subfigure}{.47\textwidth}
    \centering
  \begin{quantikz}[row sep = 4pt]
  & \ctrl{1} & \gate{X} & \qw \\
  & \targ{} & \gate{X} & \qw
  \end{quantikz}
  $\to$
  \begin{quantikz}[row sep = 4pt]
  & \gate{X} & \ctrl{1} & \qw \\
  & \ghost{X}\qw & \targ{} & \qw
  \end{quantikz}
  \caption{$\cnot$ + X interaction}\label{fig:cnotx}
  \end{subfigure}

  \vspace{.5em}
  \hrule
  \vspace{.5em}

  \textbf{Symbolic circuit rewrite rules}  \vspace{.5em}

  \begin{subfigure}{.54\textwidth}
    \centering
      \begin{quantikz}[row sep = 4pt, column sep=7pt]
      & \gate{R_z^{\theta_1}} &  \gate[wires=2,style={fill=yellow!10}][.5cm]{S} & \qw & \qw\\
      &  \qw &  & \gate{R_z^{\theta_2}} & \qw
      \end{quantikz}
      $\to$
      \begin{quantikz}[row sep = 4, column sep=7pt]
      &  \gate[wires=2,style={fill=yellow!10}][.5cm]{S} & \qw & \qw\\
      &  & \gate{R_z^{\theta_1 + \theta_2}} & \qw
      \end{quantikz}
      \caption{Long-range rotation merge}\label{fig:roti}
  \end{subfigure}%
  \begin{subfigure}{.45\textwidth}
    \centering
    \begin{quantikz}[row sep = 4, column sep=7pt]
    & \ctrl{1} & \targ{} & \ctrl{1} & \gate[wires=2,style={fill=yellow!10}][.5cm]{S} & \ctrl{1} & \targ{} & \ctrl{1} & \qw \\
    & \targ{} & \ctrl{-1} & \targ{} &  & \targ{} & \ctrl{-1} & \targ{} & \qw
    \end{quantikz}
    $\to$
    \begin{quantikz}[row sep = 4, column sep=7pt]
    & \gate[wires=2,style={fill=yellow!10}][.5cm]{S} & \qw \\
    &  & \qw
    \end{quantikz}
    \caption{Long-range $\cnot$ cancellation}\label{fig:cnot-long}
    \end{subfigure}
  
  \caption{Some optimizations \ours can synthesize/verify ($S$ is a symbolic gate satisfying some constraints)}
  \label{fig:rules}
  \end{figure}

\paragraph{Symbolic rules}
Typically, quantum-circuit optimizers apply a schedule of rewrite rules to shrink a given circuit.
In its simplest form, a rewrite rule 
matches a specific subcircuit
and rewrites it into a smaller, equivalent subcircuit.
For instance, \cref{fig:gatec} shows two equivalent circuits:
if we see two Hadamard gates ($H$) applied to the same qubit,
we can eliminate them because they cancel each other out.
\cref{fig:cnotx} shows a  rewrite rule over subcircuits with two qubits.

Automatically synthesizing such rules is relatively simple:
enumerate pairs of circuits and verify their equivalence.
This approach that has been applied in other domains, e.g., machine learning~\cite{jia2019taso}, traditional compilers~\cite{sasnauskas2017souper},
and recently quantum-circuit superoptimization~\cite{xu2022quartz}.
However, there are complex and critical rules that cannot be discovered
this way:  subcircuits can have parameters, e.g., angles of rotations,
or \emph{completely unknown subcircuits}. Thus, we need a \emph{symbolic approach} for reasoning about such rules.
For instance, \cref{fig:roti} shows a rewrite rule in which two rotations about the $z$-axis on different qubits can be merged into a single rotation, even if they are separated by arbitrarily many operations, denoted $S$, so long as $S$ satisfies certain conditions. We think of $S$ as an unknown, \emph{symbolic} gate.
Similarly, \cref{fig:cnot-long} shows a rule in which two distant sequences of $\cnot$ gates can be cancelled.

To reason about symbolic circuits and rules,
we utilize \emph{path-sum}-based semantics. 
First introduced by Feynman,
path sums compactly capture the semantics of a quantum system as an expression.
Intuitively, one can think of a path sum as a transition relation specifying how a quantum system's state evolves.
Indeed, path sums have been used for quantum-circuit verification \cite{amy2018towards, chareton2021qbricks}. 
In this paper, we exploit the algebraic nature of path sums to reason about circuits with unknown parameters and unknown subcircuits.
By reasoning using path sums, we show that we are able to synthesize \emph{long-range} rules, like \cref{fig:roti,fig:cnot-long}, that cancel out far-apart quantum gates.

\paragraph{Probabilistic verification and synthesis}
\ours synthesizes rewrite rules following the standard synthesize-and-verify story:
we enumerate circuits with symbolic components and verify equivalence of pairs of circuits.
If two circuits $C_1$ and $C_2$ are proven equivalent, then we can soundly rewrite  $C_1$ to $C_2$ or vice versa.
Na\"ively following this recipe, of course, does not scale due to the large space of pairs of circuits.
Our first, and perhaps most critical, observation is that the equivalence-checking 
problem for two circuits can be reduced to a \emph{constrained} form of \emph{polynomial identity testing} (\pit)---%
the problem of checking equivalence of two polynomials---where the constraints are on the domain of the variables.
We then demonstrate that this problem can be directly solved by the foundational 
Schwartz--Zippel randomized algorithm for \pit~\cite[Ch. 7]{motwani1995randomized}, 
which is very fast, because it relies on a single random instantiation of the 
variables of a polynomial. 

With this insight, we present a probabilistic data structure---the \emph{polynomial identity filter} (\pif)---for constructing equivalence classes of circuits.
The \pif builds upon the high-probability guarantees of Schwartz--Zippel to eliminate the quadratic explosion of checking equivalence of pairs of circuits. 
The \pif therefore enables fast construction of rewrite rules from equivalence classes.

\begin{figure}[t!]
  \includegraphics[scale=1.18]{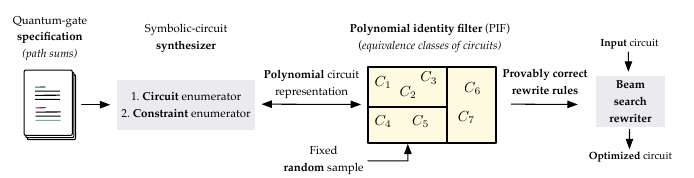}
  \caption{Overview of the \ours approach}\label{fig:overview}
\end{figure}

\begin{sloppypar}
\paragraph{Applying rewrite rules}
  Quantum-circuit optimizers, like \voqc~\cite{hietala2021verified} and \tket~\cite{tket}, use a fixed schedule for applying optimizations that is chosen by the compiler developer. 
In our setting, however, we synthesize tens of thousands of rules, and we simply cannot ask a developer to experiment with different schedules.
We demonstrate that a simple, beam-search-based algorithm can quickly optimize quantum circuits
by applying sequences of rewrite rules.
The most algorithmically challenging piece is matching symbolic rewrite rules, which can match arbitrarily large subcircuits.
\end{sloppypar}

\paragraph{Evaluation}
We implemented \ours and used it to synthesize optimizers for four different quantum architectures with different operations (or gate sets)---including \ibm, Rigetti, and \emph{ion trap} computers.
\ours can synthesize all rewrite rules in about 2 minutes.
Our results demonstrate that \ours is able to outperform or match handwritten optimizers, like \voqc~\cite{hietala2021verified}, \tket~\cite{tket}, and \ibm Qiskit~\cite{Qiskit}.
For instance, in 1.2 minutes, \ours can synthesize an optimizer for \ibm computers that significantly outperforms leading compilers, such as \ibm's Qiskit and \tket, on the majority (85\%) of the circuits in a diverse benchmark suite, and can outperform or match \voqc on 72\% of the benchmarks, outperforming it in 51\% of the benchmarks.
In comparison to the superoptimizer, Quartz~\cite{xu2022quartz}, we demonstrate  (1) that \ours is radically faster at rule synthesis and (2) the critical importance of symbolic rewrite rules.
For instance, on \ibm, in a head-to-head comparison of synthesized rules (i.e., excluding any preprocessing),
\ours synthesizes rules an order of magnitude faster than Quartz and
 outperforms Quartz on 97\% of the benchmarks. 

% contributions
\paragraph{Contributions}
In summary, we make the following contributions:
\begin{itemize}
  \item A path-sum-based circuit semantics that can  compactly capture circuits with unknown, symbolic subcircuits.
  This enables us to synthesize sophisticated, long-range rewrite rules that are critical for quantum-circuit optimization. (\cref{sec:ps})

  \item  A fast quantum-circuit equivalence verifier that reduces the problem to a constrained form of polynomial identity testing,
  which can be solved using a fast randomized algorithm. (\cref{sec:verifier})
  
  \item A fast rule synthesizer that uses a novel probabilistic data structure, called \emph{polynomial identity filter} (\pif), to avoid the quadratic explosion of rule enumeration. (\cref{sec:synth})
  
  \item A beam-search-like algorithm that applies symbolic rewrite rules to optimize a  circuit. (\cref{sec:opt})
  
  \item A thorough evaluation of \ours on four quantum architectures. Our results demonstrate that \ours can outperform or match state-of-the-art optimizers. (\cref{sec:imp})
\end{itemize}

\section{Background and Overview}

\subsection{Quantum Circuits Background}\label{sec:background}
\paragraph{Quantum state}
A quantum bit (qubit) can be in state 0 or 1, the \emph{computational basis states}, which are represented by the 2-dimensional vectors 
$\ket{0} = \left[\begin{smallmatrix}1\\0\end{smallmatrix}\right]$ and 
$\ket{1} = \left[\begin{smallmatrix}0\\1\end{smallmatrix}\right]$, respectively.
A qubit can also be in a linear combination (\emph{superposition}) of the basis states,
$
\alpha
\ket{0}
+
\beta
\ket{1}
= \left[\begin{smallmatrix}\alpha\\\beta\end{smallmatrix}\right],
$
where $\alpha, \beta$ are complex numbers, called the \emph{amplitudes},
such that $|\alpha|^2 + |\beta|^2 =1$.
% If one were to \emph{measure} the state of a qubit, they will read 
% 0 with probability $|\alpha|^2$ and 1 with probability $|\beta|^2$.
%
The state of two qubits is a vector of four complex numbers, where each number is
the amplitude of one of the basis states, $\ket{00}, \ket{01}, \ket{10},$ and $\ket{11}$.
The state of $n$ qubits is a vector of  $2^n$ complex numbers. 
% Key to the computational power of a quantum computer is that a linear increase
% in the number of qubits results in an exponential increase in the size of the state.
% In quantum programs, we apply quantum operations to a set of $N$ qubits.

\paragraph{Quantum gates}
Quantum operations (or \emph{gates}) transform the state of the qubits of a system.
Unlike in Boolean circuits, there are infinitely many possible
quantum gate combinations that can be used to produce a \emph{universal} quantum computer---%
one that can approximate arbitrary \emph{unitary} transformations
(the class of state transformations the rules of quantum mechanics admit) to arbitrary precision.
Because of a variety of engineering challenges, different quantum computers provide different gate sets. We give examples of some standard gates below.

A classical gate like NOT (denoted $X$) can be applied to a single qubit.
If the qubit state is $\ket{0}$, it becomes $\ket{1}$, and vice versa, just like on a classical circuit.
However, if the state of the qubit is a superposition $\alpha\ket{0}+\beta\ket{1}$,
applying $X$ results in the state $\beta\ket{0} + \alpha\ket{1}$,
i.e., swaps the amplitudes.
The \emph{Hadamard} gate, denoted $H$, 
takes a qubit from a basis state and puts it in superposition;
for example, given the basis state $\ket{0}$, applying $H$ results in 
$\frac{1}{\sqrt{2}}\ket{0} + \frac{1}{\sqrt{2}} \ket{1}$.

\paragraph{Path sums}
Since quantum operations are \emph{linear} transformations, 
they are represented uniquely by how they transform
the basis states. We use the traditional \emph{path-sum} notation~\cite{amy2018towards},
which can be seen as a compact representation of a state-transition relation.
For example, the path-sum representation of the $X$, $H$, and $R_z$ gates are defined as follows:
$$X: \ket{x} \to \ket{\neg x} 
\hspace{1.5cm} 
H: \ket{x} \to \frac{1}{\sqrt{2}}\sum_{y \in \{0,1\}}e^{i\pi xy}\ket{y}
\hspace{1.5cm}
R_z^\theta : \ket{x} \to e^{i(2x-1)\theta}\ket{x}
$$
These are read as follows:
Applying gate $X$ to basis state $\ket{x}$ results in the state $\ket{\neg x}$;
applying $H$ to $\ket{x}$ results in the state $\frac{1}{\sqrt{2}}\ket{0} + \frac{1}{\sqrt{2}}e^{i\pi x}\ket{1}$.
The $R_z$ gate is parameterized by an angle $\theta$, and only changes the amplitude of a given basis state.
%
% To build a quantum computer, we  need gates that operate on pairs of qubits.
The \emph{controlled X} (\cnot) gate can \emph{entangle} two qubits,
a critical operation in quantum computing:
$\ket{x_1x_2} \to \ket{x_1(x_1\oplus x_2)}$.
Given a basis state $\ket{x_1x_2}$, $\cnot$ produces
the basis state $\ket{x_1(x_1\oplus x_2)}$,
where $\oplus$ is XOR.

\begin{wrapfigure}{r}{.42\textwidth}\footnotesize\centering
	\begin{subfigure}[c]{.18\textwidth}
		\vspace{-0.15in}
		\begin{quantikz}[column sep = 7pt, row sep = 3pt]
			x_1~ & \ctrl{1} & \gate{X} & \qw \\
			x_2~ & \targ{} & \gate{R_z^{\frac{\pi}{2}}} & \qw
		\end{quantikz}
	\end{subfigure}
	\begin{subfigure}[c]{.18\textwidth}
		\vspace{-0.20in}
		\begin{align*}
			&\cnot ~ x_1 ~ x_2;\\  &X ~x_1;\\  &R_z^{\pi/2} ~x_2
		\end{align*}
	\end{subfigure}
	\caption{Circuit and its linear representation}\label{fig:ecircuit}
\end{wrapfigure}
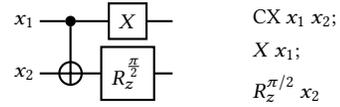

\paragraph{Quantum circuits}
Quantum circuits are combinations of quantum gates.
Consider the circuit in \cref{fig:ecircuit}~(left) over  two qubits, $x_1$ and $x_2$, represented  by the two horizontal \emph{wires}.
The circuit is read from left to right.
The first gate is a $\cnot$ gate applied to the two qubits.
Then, an $X$ gate and an $R_z^{\pi/{2}}$ gate are applied to $x_1$ and $x_2$ in parallel.

We will use a linear representation of the circuit as a sequence of gates---\cref{fig:ecircuit}~(right).
Since  $X$ and $R_z$  are applied in parallel to two different qubits, 
we can safely swap them in the linear representation.
The semantics of the circuit can be represented in path-sum notation
by \emph{composing} the path-sum representation of the constituent gates.

%, as follows:
% \begin{align*}
%   \ket{x_1,x_2} \to & \ket{x_1, x_1 \oplus x_2} && \text{apply \cnot}\\
%    \to & \ket{\neg x_1, q1 \oplus x_2} && \text{apply $X$}\\
%    \to & e^{\frac{(1-2x_2) \pi}{4}} \ket{\neg x_1, q1 \oplus x_2} && \text{apply $R_z$}
% \end{align*}
%%%%%%%%%%%%%%%%%%%%%%%%%%%%%%%%%%%%%%%%%%%%%%%%%%%%%%%%%%%%%%%%%%%%%%%%%%
%
%%%%%%%%%%%%%%%%%%%%%%%%%%%%%%%%%%%%%%%%%%%%%%%%%%%%%%%%%%%%%%%%%%%%%%%%%
\paragraph{Executing circuits on hardware}
On quantum hardware, operations are imperfect and prone to errors. Typically, quantum computers provide single- and two-qubit gates, which are optimized to minimize errors. Despite these optimizations, the average single-qubit gate error rate is about 0.1\%, \textit{while the two-qubit gate error is about 10-100x higher} on most industrial quantum computers~\cite{googledatasheet,toronto}. 
Therefore, to minimize circuit error rate, we need to eliminate as many 
two-qubit gates as possible, which not only corrupt the qubits involved in the operation, but also impose \emph{crosstalk} errors on neighboring qubit devices, significantly degrading circuit reliability. See \cref{app:hardware} for more details on hardware and errors.

\subsection{Overview of \ours}
\paragraph{High-level view}
\cref{fig:overview} provides a high-level illustration of \ours. 
First, the user provides a  specification of the gate set of some quantum architecture, e.g., \ibm's gate set, in the form of path sums.
Then, a synthesis algorithm starts enumerating (symbolic) circuits along with constraints on the symbolic components.
All circuits are inserted into a probabilistic data structure---the polynomial identity filter---which groups circuits into correct equivalence classes with high-probability guarantees.
Finally, we construct rewrite rules from equivalent pairs of circuits and apply them following a beam-search-based algorithm to optimize a given circuit.
We illustrate these pieces with examples.

\paragraph{Example A}
Let us consider the rewrite rule  in \cref{fig:roti}.
The two circuits on either side of the rule have three unknowns:
the rotation angles, $\theta_1$ and $\theta_2$, and the highlighted gate $S$.
Our synthesis algorithm enumerates such circuits in order to discover equivalent pairs of circuits.
However, in our example, the gate $S$ is completely unconstrained---we know nothing about it.
We call $S$ a \emph{symbolic} gate.
Therefore, \ours needs to answer the following \emph{abduction} question: 
\begin{center}
\emph{Under what conditions on $S$ are the two circuits equivalent?}
\end{center}
For this specific example, \ours abduces the following constraint on $S$:
$$S: \ket{x_1 x_2\ldots} \to \phi(x_1x_2\ldots) \ket{x_2x_1\ldots}$$ 
In other words, all we need to know about $S$ for these two circuits to be equivalent is that $S$ swaps the values of $x_1$ and $x_2$.
Note that $S$ is allowed to change the amplitudes (denoted by an unconstrained function $\phi$, called the \emph{amplitude transformer}) and may even apply gates to other qubits (denoted by the $\ldots$).

\paragraph{Example B}
\cref{fig:cnot-long} shows another rewrite rule that \ours can synthesize.
This rewrite rule cancels two distant sequences of $\cnot$ gates, separated by a symbolic gate $S$.
\ours abduces the following constraint on $S$.
$$S: \ket{x_1 x_2\ldots} \to \phi(x_1x_2\ldots) \ket{x_1x_2\ldots}$$ 
Informally, $S$ may change the amplitudes but should not change the first two bits of the state, $x_1x_2$.

\paragraph{Proving equivalence}
To prove equivalence of two circuits, we observe that we can reduce the problem to a constrained form of \emph{polynomial identity testing} (\pit), the problem of checking equivalence of two polynomials.
Specifically, the amplitudes of every basis state will be represented as a  polynomial over the complex field.
In Example A, the polynomials describing the amplitudes will be over the variables $\theta_1, \theta_2,$  and the function $\phi$ (from the  constraint on $S$).

We show that our constrained \pit problem can be solved with the standard randomized algorithm---following the Schwartz--Zippel lemma~\cite{motwani1995randomized}.
Simply, randomly sample values for the variables
and check if the two polynomials evaluate to the same value.
If they do not, then we have a counterexample; if they do, then we have a high-probability guarantee that the polynomials are equivalent.

To give some intuition,
for Example A (\cref{fig:roti}), the amplitudes of the basis state $\ket{11}$
for the left and right circuits are as follows (a full derivation is shown in \cref{ex:roti1}):
$$e^{i\theta_1} \cdot \phi(11) \cdot e^{i\theta_2} = \phi(11) \cdot e^{i(\theta_1 + \theta_2)} $$
where  $\phi$ is the unknown amplitude transformer of $S$.
While these two expressions are clearly equivalent, it is not always immediate,
and one generally requires algebraic manipulation to prove equivalence.
Further, these two expressions are not polynomials.
Luckily, as we show in \cref{sec:verifier}, we demonstrate that we can treat the above expressions as a special form of polynomials over the complex field and use Schwartz--Zippel to show their equivalence. 
Specifically, $\phi(11)$ can be viewed as a complex variable and terms of the form $e^{i\cdot}$ as complex variables \emph{constrained} to the unit circle.

\paragraph{Efficient synthesis}
To synthesize rewrite rules, we can simply enumerate pairs of circuits, abduce constraints, and verify their equivalence.
But this blows up quadratically---say there are 1 million circuits, then we will need to consider $10^{12}$ pairs.

To avoid the quadratic explosion we present a simple and efficient probabilistic data structure, the \emph{polynomial identity filter} (\pif).
The \pif  takes a set of polynomials and returns a set of equivalence classes.
The key idea underlying the \pif is that we can use a single random  valuation of variables to evaluate each circuit's polynomial representation, and store circuits with equal valuations in the same equivalence class.
Using the guarantees of Schwartz--Zippel, the \pif data structure
ensures that all of its equivalence classes are correct with a high probability (\cref{thm:pif}).
From the generated equivalence classes, we construct a set of rewrite rules like those shown in \cref{fig:rules}.

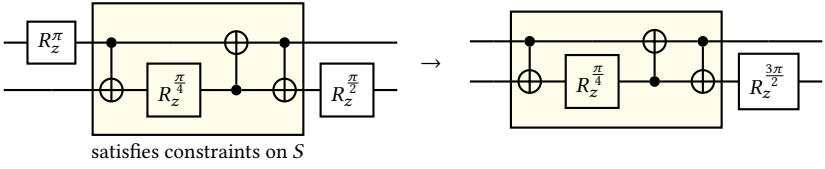
\begin{figure}
  \footnotesize
  \begin{quantikz}[row sep = 0pt, column sep=9pt]
    & \gate{R_z^{\pi}} &  \ctrl{1}\gategroup[2,steps=4,style={fill=yellow!10, inner xsep=0pt},
    background,label style={label position=below,anchor=
    north,yshift=-0.2cm}]{{satisfies constraints on} $S$} & \qw & \targ{} & \ctrl{1} & \qw & \qw\\
    &  \qw & \targ{} &  \gate[style={fill=yellow!10}]{R_z^{\frac{\pi}{4}}} & \ctrl{-1} & \targ{} & \gate{R_z^{\frac{\pi}{2}}} & \qw
  \end{quantikz}
  \ \ \ $\to$\ \ \ 
  \begin{quantikz}[row sep = 0pt, column sep=9pt]
    & \qw &  \ctrl{1}\gategroup[2,steps=4,style={fill=yellow!10, inner xsep=0pt},
    background,label style={label position=below,anchor=
    north,yshift=-0.2cm}]{} & \qw & \targ{} & \ctrl{1} & \qw & \qw\\
    &  \qw & \targ{} &  \gate[style={fill=yellow!10}]{R_z^{\frac{\pi}{4}}} & \ctrl{-1} & \targ{} & \gate{R_z^{\frac{3\pi}{2}}} & \qw
  \end{quantikz}
  \caption{Application of the long-range rotation merge optimization from \cref{fig:roti}.}\label{fig:apply}
\end{figure}

\paragraph{Applying rules}
Given a set of rewrite rules, \ours uses a beam search approach to traverse the space of rewrite sequences
and optimize the circuit by minimizing the number of gates.
The main challenge that \ours addresses is a generic algorithm for applying symbolic rewrite rules.
Specifically, it needs to discover a subcircuit that satisfies the constraints on the symbolic gate $S$ in our examples.
\cref{fig:apply} demonstrates an application of the long-range rotation merge rule (\cref{fig:roti})
to a circuit. The highlighted part of the circuit
satisfies the constraints on the symbolic gate $S$,
namely, the path-sum representation of the highlighted subcircuit is of the form
$\ket{x_1 x_2\ldots} \to \phi(x_1x_2) \ket{x_2x_1\ldots}$.

\section{Path-sum-based Circuit Semantics}\label{sec:ps}
We now formally define  (symbolic) quantum circuits and their semantics using \emph{path sums}.
Our semantics is a direct adaptation of that of \citet{amy2018towards}.
The novelty in this section is defining circuits with unknown, symbolic gates  and what it means for such circuits to be equivalent (\cref{sec:symbolic}).

\subsection{States, Gates, and Path Sums}
\paragraph{Quantum states}
The state of a qubit is a linear combination of the computational basis states, $\ket{0}$ and $\ket{1}$, written $\alpha \ket{0} + \beta\ket{1}$,
where $\alpha,\beta \in \mathbb{C}$.
The state of $n$ qubits is a term of the form
$\sum_{\vec{x} \in \Zn} \alpha_{\vec{x}} \ket{\vec{x}}$,
where $\Zn$ is the set of $n$-bit vectors and $\alpha_{\vec{x}} \in \mathbb{C}$.
For bit vector $\vec{x}$, we use $x_i$ to denote the $i$th bit of $\vec{x}$.

\paragraph{Gates and path sums}
We  consider two kinds of quantum gates, single- and multi-qubit gates.
We will use $G^\param$ to denote a gate that takes a parameter $\param \in \mathbb{C}$ (e.g., angle of rotation),
or simply $G$ if the gate does not take parameters.
For simplicity, we assume that gates have at most 1 parameter.

The semantics of a gate are defined in \emph{path-sum} notation~\cite{amy2018towards} which shows how a gate transforms a basis state.
From a verification perspective, a path sum is an expression defining the \emph{transition relation}.
Specifically, a single-qubit gate $ G^\param $ is defined in the following fashion:
\begin{equation}\label{eq:g1}
\ket{x} \to \sum_{y \in \mathbb{Z}_2} \phi(x,y,\param) \ket{f(x,y)}
\end{equation}
This path-sum specification says that for any basis state $\ket{x}$, applying $G^\param$ to 
$\ket{x}$ transforms the state into $\sum_{y \in \mathbb{Z}_2} \phi(x,y,\param) \ket{f(x,y)}$, where 
\begin{itemize}
\item $\phi \in \Z \times \Z \times \mathbb{C} \to \mathbb{C}$ is the \emph{amplitude transformer}, and
\item $f\in \Z \times \Z \to \Z$ is the \emph{state transformer}.
\end{itemize}

\begin{example}\label{ex:hadamard}
  For Hadamard $H: \ket{x} \to \sum_{y \in \{0,1\}}\frac{1}{\sqrt{2}}e^{i\pi xy}\ket{y}$,
   $\phi(x,y) = \frac{1}{\sqrt{2}}e^{i\pi x y}$ and $f(x,y) = y$.
\end{example}

\paragraph{$n$-qubit gates}
$n$-qubit gates are analogously defined; a gate $G^\param$ has a path sum of the form
$$ \ket{\vec{x}} \to \sum_{\vec{y} \in \Zn} \phi(\vec{x},\vec{y},\param) \ket{f(\vec{x},\vec{y})}$$
where  
$\phi \in \Zn \times \Zn \times \mathbb{C} \to \mathbb{C}$ and
$f \in \Zn\times\Zn \to \Zn$.
We  write $f(\vec{x},\vec{y})_i$ for the $i$th bit of  $f(\vec{x},\vec{y})$.

\paragraph{Monomial gates}
Some gates do not transform a basis state into \emph{superposition} (non-trivial linear combination of basis states), and therefore their path-sums can be simplified as follows:
$$\ket{\vec{x}} \to  \phi(\vec{x},\param) \ket{f(\vec{x})}$$
We call such gates \emph{monomial} gates because their matrix representation is a monomial matrix (or a generalized permutation matrix).
E.g., $\cnot$ is a monomial gate with path sum
$\ket{x_1x_2} \to \ket{x_1(x_1\oplus x_2)}$.

\paragraph{Path sums are expressions}
Observe how the right-hand side of $\to$ in a  path sum  is an expression of a  quantum state parameterized
by two variables, $\vec{x}$ and $\param$.
Henceforth, for a gate $G^\param$, we shall use $\sem{G^\param}$ to denote the 
expression on the right-hand side of its path sum.

\begin{example}
    $\sem{H}$ is 
    $\sum_{y \in \{0,1\}}\frac{1}{\sqrt{2}}e^{i\pi xy}\ket{y}$.
\end{example}

We will use the notation $\sem{G_1^\param} \equiv \sem{G_2^\param}$ 
to denote that 
$\forall \vec{x}, \param \ldotp \sem{G_1^\param} = \sem{G_2^\param}$.
In other words, the two path sums define the same quantum state for every valuation of the variables $\vec{x}$ and $\param$. 

\begin{figure}[t!]\footnotesize
  \begin{prooftree}
    \sem{G^\param} \equiv \sum_{y \in \mathbb{Z}_2} \phi(x,y,\param) \ket{f(x,y)}
    \justifies
    \sem{G^\param~i} \equiv
    \sum_{y \in \mathbb{Z}_2} \phi(x_i,y,\param) \ket{x_1\ldots x_{i-1} f(\vec{x},y)x_{i+1}\ldots x_n}
    \using \textsc{extend}
    \end{prooftree}
    
    \vspace{1em}

    \begin{prooftree}
      \sem{C_1^\vparam} \equiv \sum_{\vec{y_1} \in \Zn} \phi_1(\vec{x},\vec{y_1},\vparam) \ket{f_1(\vec{x},\vec{y_1})}
\quad \quad
\sem{C_2^\vparam} \equiv \sum_{\vec{y_2} \in \Zn} \phi_2(\vec{x},\vec{y_2},\vparam) \ket{f_2(\vec{x},\vec{y_2})}
 \justifies 
 \sem{C^{\param}_1 ; C^{\param}_2}
\equiv
  \sum_{\vec{y_1} \in \Zn} \phi_1(\vec{x},\vec{y_1},\vparam) (\sem{C_2^{\vparam}}[\vec{x} \gets f_1(\vec{x},\vec{y_1})])
  \using \textsc{seq}
\end{prooftree}   
\caption{Path sum circuit semantics. $\sem{C_2^{\vparam}}[\vec{x} \gets f_1(\vec{x},\vec{y})]$ is $\sem{C_2^{\vparam}}$ with every instance of $\vec{x}$ replaced by  $f_1(\vec{x},\vec{y})$.
} \label{fig:sem}
\end{figure}

\subsection{Circuit Semantics}

\paragraph{Circuits}
A quantum circuit over $n$ qubits is a sequence of gates.
Without loss of generality, we restrict ourselves to one- and two-qubit gates.
We will write $\gapply{G^\param}{i}{}$ to denote the single-qubit gate $G^\param$
applied to the $i$th qubit, and $\gapply{G^\param}{i}{j}$ to denote the two-qubit gate $G^\param$ applied to the $i$th and $j$th qubits.
A circuit $C$ is defined by the following grammar:
\begin{align}\label{eq:grammar}
  C ~ \coloneq ~ G^\param~i ~\mid~  \gapply{G^\param}{i}{j} ~\mid~ C_1; C_2
\end{align}
where $i,j \in [1,n]$.
We will use $C^{\vparam}$ to denote that the circuit has a number of parameters, a vector $\vparam$ containing the parameters of the circuit's constituent gates.

\paragraph{Semantics of circuits}
The path-sum representation of a circuit $\sem{C^{\vparam}}$
 follows the grammar recursively:
Either it is the path sum of a single gate, $\sem{ G^\param~i}$ or $\sem{\gapply{G^\param}{i}{j}}$, or the composition of the path sums of two circuits, $\sem{C_1; C_2}$.
See \cref{fig:sem} for the definitions.

Consider the rule \textsc{extend} in \cref{fig:sem}.
Given a single-qubit $G^\param$ gate,
\textsc{extend} defines the path sum to denote that $G^\param$ is applied to the $i$th qubit of an $n$-qubit system.
Intuitively, the state transformer of $G^\param~i$  modifies the $i$th qubit, leaving the rest intact. 
\textsc{extend} for two-qubit gates is analogous.
\begin{example}
Recall the Hadamard gate, where $\sem{H}$ is $\sum_{y \in \{0,1\}}\frac{1}{\sqrt{2}}e^{i\pi xy}\ket{y}$.
Following the \textsc{extend} rule, $\sem{H~i}$ is $\sum_{y \in \{0,1\}}\frac{1}{\sqrt{2}}e^{i\pi x_iy}\ket{x_1\ldots x_{i-1}yx_{i+1}\ldots x_n}$.
\end{example}

The rule \textsc{seq} defines the semantics of composition.
Informally, composing two path sums stitches together the ``final basis states'' of the first with the ``initial basis states'' of the second.

\begin{example}\label{ex:seq}
Consider $\sem{H}$ and $\sem{R_z^\theta}$, which are $\sum_{y \in \{0,1\}}\frac{1}{\sqrt{2}}e^{i\pi xy}\ket{y}$ and $e^{i(2x-1)\theta}\ket{x}$, respectively.
For the single-qubit circuit $H;R_z^\theta$, $\sem{H;R_z^\theta}$ is the following:
$\sum_{y \in \{0,1\}}\frac{1}{\sqrt{2}}e^{i\pi xy}\underbrace{e^{i(2y-1)\theta}\ket{y}}_{\sem{R_z^\theta}[x \gets y]}$.

\end{example}

\paragraph{Circuit equivalence}
Two cicuits are equivalent if they have equivalent path sums.
\begin{definition}[Circuit equivalence]\label{def:eq}
Consider two circuits $C_1^{\vparam}$ and $C_2^{\vparam}$ over the same set of parameters, $\vparam$.
We say that the two circuits are equivalent iff
  $\sem{C_1^{\vparam}} \equiv \sem{C_2^{\vparam}}$.
\end{definition}

\begin{example}
  Consider the two equivalent single-qubit circuits, 
  $R_z^{\theta_1}; R_z^{\theta_2}
  \text{~~~and~~~} R_z^{\theta_1 + \theta_2}$,
  We have 
  \begin{align*}
  \sem{R_z^{\theta_1}; R_z^{\theta_2}}  \equiv 
  e^{i(2x-1)\theta_1}e^{i(2x-1)\theta_2}\ket{x} \quad \quad \quad
  \sem{R_z^{\theta_1 + \theta_2}}  \equiv 
  e^{i(2x-1)(\theta_1+\theta_2)}\ket{x}
  \end{align*}
  It is easy to see that for any values of $x$ and the parameters $\theta_1$ and $\theta_2$, we have 
  $\sem{R_z^{\theta_1}; R_z^{\theta_2}} = 
  \sem{R_z^{\theta_1 + \theta_2}}$.
\end{example}

\subsection{Symbolic Circuits}\label{sec:symbolic}

\paragraph{Symbolic gates}
We will often use unknown gates in a circuit.
We will treat those \emph{symbolic gates} as path sums where the amplitude and state transformers are undefined (or \emph{uninterpreted}). 
We will use $\symb$ to refer to a symbolic gate,
where $\sem{S}$ is of the form:
\begin{equation}\label{eq:symb}
  \phi^u(\vec{x}) \ket{f^u(\vec{x})}
\end{equation}
where $\phi^u$ and $f^u$ are uninterpreted.
Observe that $S$ is a monomial gate;
this practical assumption helps us restrict the space of interpretations of the state transformers.

\paragraph{Symbolic circuits}
When a circuit uses a symbolic gate, we will call it a \emph{symbolic circuit}.
Given two symbolic circuits, ideally, we would like to discover constraints on the uninterpreted amplitude and state transformers under which the two circuits are equivalent.
We simplify this problem and only consider constraints on state transformers, $f^u$, treating the amplitude transformers, $\phi^u$, as parameters of the circuit.
The simplification is due to the fact that there are finitely many possible constraints on the Boolean function, $f^u$, while it is challenging to discover constraints on the complex-valued transformer, $\phi^u$.

\paragraph{Interpretations of state transformers}
We will use $I$ to denote an interpretation of the uninterpreted state transformers in a symbolic circuit $C$.
We will use $C(I)$ to denote $C$ with all uninterpreted state transformers of symbolic gates replaced with their interpretation in $I$.

\begin{example}
  Consider the symbolic gate $S$ where $\sem{S} \equiv   \phi^u(\vec{x}) \ket{f^u(\vec{x})}$.
  Let the interpretation $I$ set $f^u(\vec{x})$ to the identity function.
  Then, $\sem{S(I)} \equiv   \phi^u(\vec{x}) \ket{\vec{x}}$.
\end{example}

\begin{definition}[Unifying interpretations]
  Consider two symbolic circuits $C_1^{\vparam}$ and $C_2^{\vparam}$ 
  that use the same symbolic gates with
  amplitude transformers $\phi^u_1,\ldots,\phi^u_k$.
  We say that $I$ is a \emph{unifying interpretation}
  of the two circuits if
  \begin{align}
    \forall \phi^u_1,\ldots,\phi^u_k, \vec{x}, \vparam \ldotp 
    \sem{C_1^{\vparam}(I)} = \sem{C_2^{\vparam}(I)}.
  \end{align}

\end{definition}
Intuitively, a unifying interpretation $I$ creates two circuits that are equivalent, following \cref{def:eq}.
The idea is that we can treat the amplitude transformers as if they are circuit parameters.

\begin{example}\label{ex:roti1}
Recall the two circuits in \cref{fig:roti};
call them $C_1$ and $C_2$.
We have 
\begin{align*}
\sem{C_1}     \equiv
  e^{i(2x_1-1)\theta_1} \cdot
  \phi^u(\vec{x}) \cdot
    e^{i(2f^u(\vec{x})_2 - 1)\theta_2} \ket{\vec{x}}
    \hspace{3em}
\sem{C_2}  \equiv 
  \phi^u(\vec{x}) \cdot
    e^{i(2f^u(\vec{x})_2 - 1)(\theta_1+\theta_2)} \ket{\vec{x}}
\end{align*}
Consider the unifying interpretation 
$I$ where $f^u(x_1x_2) = x_2x_1$.
\begin{align*}
  \sem{C_1(I)}   \equiv 
    e^{i(2x_1 - 1)\theta_1} \cdot 
    \phi^u(\vec{x}) \cdot
    e^{i(2x_1 - 1)\theta_2} \ket{\vec{x}}
    \hspace{4em}
  \sem{C_2(I)}   \equiv  
    \phi^u(\vec{x}) \cdot
    e^{i(2x_1 - 1)(\theta_1+\theta_2)} \ket{\vec{x}}
\end{align*}
Observe that $\sem{C_1(I)} \equiv \sem{C_2(I)}$.
However, e.g., $f^u(x_1x_2)=x_1x_2$
is \emph{not} a unifying interpretation.
\end{example}

\section{Circuit Equivalence Verifier}\label{sec:verifier}

We now present a fast approach for probabilistically verifying equivalence of two circuits, which
 will be key for synthesizing rewrite rules.
We reduce the problem to a constrained form of  \emph{polynomial identity testing} (\pit),
and demonstrate that it can be solved using a standard randomized algorithm for checking equivalence of polynomials.
We begin with the foundations of  polynomial identity testing.

\subsection{Polynomial Identity Testing \& Schwartz--Zippel}
\label{sec:pit}
We are given two polynomials $\poly_1$ and $\poly_2$
over the same set of $n$ complex-valued variables.
We want to check if $\poly_1(\cvars) = \poly_2(\cvars)$
for all values of $\cvars \in \mathbb{C}^n$, concisely denoted as $\poly_1 = \poly_2$. We will use $d$ to denote the maximum degree of the two polynomials.

We can verify equivalence of $\poly_1$ and $\poly_2$ with high probability
as follows.
\begin{enumerate}
  \item Let $\sspace \subset \mathbb{C}$ be a finite subset of the complex numbers.
  \item Sample $n$ independent values, $\alpha_1,\ldots,\alpha_n$, 
  from the uniform distribution over $\sspace$.
  \item Return True if $\poly_1(\vec{\alpha})  = \poly_2(\vec{\alpha})$; otherwise, return False.
\end{enumerate}

The correctness of the above algorithm directly follows from the Schwartz--Zippel lemma~\cite[Ch. 7]{motwani1995randomized},
which we adapt to our purposes here:
\begin{theorem}\label{theorem:pit}
  If $\poly_1 = \poly_2$, the algorithm returns True.
  If $\poly_1 \neq \poly_2$, the algorithm returns True (false positive) with probability at most $d / |\sspace|$ (conversely, the algorithm returns False with probability at least $1-d/|R|$).
\end{theorem}
\begin{proof}
  First case ($\poly_1 = \poly_2)$:
  the algorithm returns True since for any $\vec{\alpha}$ we have $\poly_1(\vec{\alpha}) = \poly_2(\vec{\alpha})$.
  Second case ($\poly_1 \neq \poly_2$):
  The Schwartz--Zippel lemma says that if we sample $\alpha_1,\ldots,\alpha_n$ independently and uniformly from the finite set $R$, 
  the probability that $\poly_1(\vec{\alpha}) = \poly_2(\vec{\alpha})$ is at most $d / |R|$.
\end{proof}
Observe that the algorithm has a small probability of a false positive:
If the algorithm returns False, then we know that $\poly_1 \neq \poly_2$, since $\vec{\alpha}$ serves as a counterexample to equivalence.
However, given $\poly_1 \neq \poly_2$, the algorithm may return the wrong answer (True) with a small probability.
 The probability of failure $d / |\sspace|$ can be made arbitrarily small by sampling from a larger finite domain $\sspace$.

\begin{example}
Suppose $\poly_1$ and $\poly_2$
are inequivalent, degree 10 polynomials.
If we take $\sspace$ to be the set of 64-bit integers,
we will have a failure probability on the order of $10^{-19}$.
\end{example}

\paragraph{Constrained identity testing}
A nice property of \pit is that we can readily apply it to checking equivalence of two polynomials under the constraint that \emph{some} variables have a restricted domain in $\mathbb{C}$.
This is critical in our setting, since we will have variables constrained to the unit circle, denoted $\unitc = \{c \in \mathbb{C} \mid |c| = 1\}$. 
Specifically, say we want to prove the following: 
\begin{equation}\label{eq:const}
\poly_1(\vec{u},\cvars) = \poly_2(\vec{u}, \cvars)
\text{ for all } \vec{u} \in \mathbb{C}^n \text{ and } \cvars \in Z^m, \text{ where } Z \subset \mathbb{C}
\end{equation}
Then, we can simply apply \pit by using a finite sample space $\sspace \subseteq Z$.

\begin{corollary}[Constrained \pit]\label{thm:cpit}
    Consider \cref{eq:const}.
    Apply \pit to check $p_1 = p_2$ with $\sspace \subseteq Z$.
    If for all $\vec{u} \in \mathbb{C}^n \text{ and } \cvars \in Z^m$  we have 
$\poly_1(\vec{u},\cvars) = \poly_2(\vec{u}, \cvars)$, the algorithm returns True.
    Otherwise, if there exists $\vec{u} \in \mathbb{C}^n \text{ and } \cvars \in Z^m$  such that
    $\poly_1(\vec{u},\cvars) \neq \poly_2(\vec{u}, \cvars)$, the algorithm returns True with probability at most $d / |\sspace|$.
\end{corollary}

\begin{proof}
First case ($\poly_1 = \poly_2$): The algorithm will correctly return True because it samples each $\alpha_i$ from $\sspace \subseteq Z \subset \mathbb{C}$,
and we know  from \cref{eq:const} that 
$\poly_1(\vec{u},\cvars) = \poly_2(\vec{u}, \cvars)
\text{ for all } \vec{u} \in \mathbb{C}^n \text{ and } \cvars \in Z^m$.
Second case ($\poly_1 \neq \poly_2$):  following Schwartz--Zippel (\cref{theorem:pit}), the algorithm returns True with probability $\leq d / |\sspace|$.
\end{proof}
Observe how in the case the algorithm correctly returns True, 
you actually get a more general result than needed---a probabilistic guarantee that $\poly_1 = \poly_2$ with no constraints
on the $\cvars$ variables.

\subsection{Circuit-Equivalence Verification as Constrained Identity Testing}
To prove equivalence of two circuits, we will check the equivalence of
their amplitudes for every input basis state.
While there are exponentially many amplitudes in the number of qubits, for synthesizing rewrite rules, we only care about circuits with a relatively small number of qubits, and so we do not suffer an exponential explosion.
We will demonstrate that amplitude expressions are \emph{constrained} polynomials, and therefore reduce the equivalence problem to \pit.

The following approach works for symbolic and non-symbolic pairs of circuits, under the assumption that all state transformers are interpreted.
For simplicity, we assume that uninterpreted amplitude transformers are part of the circuit parameters.

\paragraph{Amplitude equivalence}
Consider two circuits  $C_1^{\vparam}$ and $C_2^{\vparam}$.
Fix a constant $\vec{a} \in \Zn$, which we will use as the initial basis state.
We can write $\sem{C_1^\vparam}[\vec{x}\gets\vec{a}]$ and $\sem{C_2^\vparam}[\vec{x}\gets\vec{a}]$ as follows:
$$ \sum_{\vec{y} \in \Zn} \psi_1^{\vec{a}}(\vec{y},\vparam) \ket{\vec{y}}
\hspace{2cm}
 \sum_{\vec{y} \in \Zn} \psi_2^{\vec{a}}(\vec{y},\vparam) \ket{\vec{y}}
$$
Intuitively, $\psi_1^{\vec{a}}(\vec{y},\vparam)$ is an expression
of the amplitude of basis state $\vec{y}$ if we apply $C_1^\vparam$ to state $\ket{\vec{a}}$.
\begin{example}\label{ex:psi}
From \cref{ex:seq}, $\sem{H; R_z^\theta}$ is
$\sum_{y \in \{0,1\}} 
    \underbrace{
\frac{1}{\sqrt{2}}
    e^{i\pi xy}e^{i(2y-1)\theta}}_{\psi^{x}(y,\theta)} \ket{y}$.
\end{example}

The following lemma reframes circuit equivalence as checking the equality of amplitudes:
\begin{lemma}\label{lemma:eq1}
  $C_1^\vparam$ and $C_2^\vparam$ are equivalent iff for all $\vec{x}$, $\vec{y}$, and $\vparam$,
  $\psi_1^{\vec{x}}(\vec{y},\vparam) = \psi_2^{\vec{x}}(\vec{y},\vparam)$.
\end{lemma}

We can eliminate the universal quantifier over $\vec{x}$ and $\vec{y}$ in  \cref{lemma:eq1} by turning it into a finite summation,
following the basic arithmetic fact:
\begin{center}
If $\alpha = \beta$ and $\alpha' = \beta'$, then $z \alpha + z'\alpha' = z\beta + z'\beta'$ for all $z,z'$.
\end{center}

\begin{theorem}\label{lemma:poly}
  For every $\vec{a},\vec{b} \in \Zn$,
  create a fresh complex-valued variable $\cvar_{\vec{a},\vec{b}}$.
  Then, $C_1^\vparam$ and $C_2^\vparam$ are equivalent iff
    for all values of the parameters $\vparam$ and the fresh variables,
  \begin{equation}\label{eq:sum}
    \sum_{\vec{a},\vec{b} \in \Zn} \cvar_{\vec{a},\vec{b}} \cdot \psi_1^{\vec{a}}(\vec{b},\vparam)
  =
      \sum_{\vec{a},\vec{b} \in \Zn} \cvar_{\vec{a},\vec{b}} \cdot \psi_2^{\vec{a}}(\vec{b},\vparam)
  \end{equation}
\end{theorem}

Observe that  \cref{eq:sum} is only over the freshly introduced ($\cvar$) variables and the parameters $\vparam$.

\begin{example}
    Continuing \cref{ex:psi}, if $H;R_z^\theta$ is circuit $C_1$ in \cref{lemma:poly}, then the left-hand side of \cref{eq:sum} will be
    $\sum_{a,b}\cvar_{a,b}
    \underbrace{
\frac{1}{\sqrt{2}}
    e^{i\pi ab}e^{i(2b-1)\theta}}_{\psi^{\vec{a}}(b,\theta)} \ket{b}$.
After expanding: 
    $$
    \left(\cvar_{0,0} \cdot
    \frac{1}{\sqrt{2}} \cdot
    e^{-i\theta} \right)
    + \left(\cvar_{0,1} \cdot
    \frac{1}{\sqrt{2}} \cdot
    e^{i\theta} \right)
    + \left(\cvar_{1,0} \cdot
    \frac{1}{\sqrt{2}} \cdot
    e^{-i\theta}\right)
    + \left( \cvar_{1,1} \cdot
    \frac{e^{i\pi}}{\sqrt{2}} \cdot
    e^{i\theta} \right)
$$
\end{example}

\paragraph{Amplitudes are polynomials}
We make the observation that for all quantum gates of interest,
the two sides of \cref{eq:sum}
can be reduced to constrained polynomials.
Therefore, we can reduce checking \cref{eq:sum} to constrained \pit.
Specifically, each side of \cref{eq:sum} can be written in the form
$$\sum_j c_j \prod_k t_k$$
where $c_j$ is a constant and $t_k$ is a term that can have three different forms:
\begin{enumerate} 
    \item variables $\cvar_{\vec{a},\vec{b}}$, introduced in \cref{lemma:poly},
    \item $(\phi^u({\vec{a},\vec{b}}))^n$, where $n \in \mathbb{N}$,
    \item or $(e^{i\theta})^n$, where $\theta$ is a parameter of the circuit and $n \in \mathbb{Z}$.
\end{enumerate}
Observe therefore that \cref{eq:sum} very much resembles a polynomial
over complex variables,
but only one of the three kinds of terms $t_k$ is a complex variable ($v_{\vec{a},\vec{b}}$).
We now show how to transform the rest of the terms into complex variables.
\begin{itemize}
\item
First, applications of amplitude transformers, 
        $\phi^u({\vec{a},\vec{b}})$,
of symbolic gates.
Since the domain of $\phi^u$ is finite,
we  replace each application of the form
$\phi^u({\vec{a},\vec{b}})$ with a complex
variable $\phi^u_{\vec{a},\vec{b}}$.
\item
Second, terms of the form $e^{i\theta}$,
which come from the gates in the circuit.
Note that $e^{i\theta}$ 
 is a  point on the complex 
unit circle, $\unitc$, parameterized by the angle $\theta$.
Therefore, for every unique term $e^{i\theta}$,
we  replace $e^{i\theta}$ with a complex-valued variable
$v_\theta$ and constrain it to $\unitc$.\footnote{We can handle terms with negative exponents like $e^{-i\theta}$ by multiplying both polynomials by $e^{i\theta}$. Terms with expressions such as $e^{i(\theta_1 + \theta_2)}$ can be expanded to $e^{i\theta_1}e^{i\theta_2}$. }
\end{itemize}

The above transformation 
reduces the problem in \cref{lemma:poly}
to constrained \pit,
which can be solved using Schwartz--Zippel (\cref{thm:cpit}).

\begin{theorem}[Reduction to constrained \pit]\label{thm:trans}
    In the context of \cref{lemma:poly},
     assume that \cref{eq:sum}  has
    $k$ unique terms of the form $e^{i\theta_1},\ldots e^{i\theta_k}$.
    Apply the above transformation to \cref{eq:sum} and 
    let $p_1 = p_2$ be the resulting equality.
    Then, $C_1^\vparam$ is equivalent to $C_2^\vparam$
    iff
    $p_1 = p_2$
    under the constraint that $v_{\theta_1},\ldots,v_{\theta_k} \in \unitc$.
\end{theorem}

\begin{example}\label{ex:roti3}
    Suppose we want to check the following equality:
    $
        e^{i\theta_1} \cdot 
        \phi^u(00) \cdot
        e^{i\theta_2} = 0
    $.
    We transform $\phi^u(00)$ into a fresh variable $\phi^u_{00}$
    and $e^{i\theta_1}$ and $e^{i\theta_2}$ into $v_{\theta_1}$ and $v_{\theta_2}$, respectively.
    This results in the following constrained \pit problem:
    $v_{\theta_1} \cdot \phi^u_{00} \cdot v_{\theta_2} = 0, \text{ for all } v_{\theta_1}, v_{\theta_2} \in \unitc \text{ and }  \phi^u_{00} \in \mathbb{C}$.
\end{example}

\section{Rewrite-Rule Synthesizer}\label{sec:synth}
We now present our rewrite-rule synthesizer.
The na\"ive approach is to enumerate pairs of circuits and check their equivalence---a quadratic explosion.
To avoid this quadratic explosion, we will
utilize a new probabilistic data structure in which circuits are inserted and stored in their respective equivalence classes. 
We call this data structure a \emph{polynomial identity filter} (\pif), because it uses the high-probability guarantees of Schwartz--Zippel to populate circuits into equivalence classes.

\subsection{The Polynomial Identity Filter (\pif)}
We will now define the polynomial identity filter (\pif).
Our goal is to design a data structure that groups 
polynomials into equivalence classes; 
when a new polynomial is \emph{inserted}, 
 it will assign it to the appropriate equivalence class.
The \pif directly builds upon the insights of Schwartz--Zippel.

 The key trick of the \pif is to randomly sample $\vec{\alpha}$
 \emph{only once} at initialization and use it to compare all inserted polynomials.
Building upon the high-probability guarantees of Schwartz--Zippel,
we ensure that all deduced equivalences are correct with a high probability.

 \paragraph{Initialization}
To define equivalence classes of polynomials, we will use a map $M$ from complex numbers to \emph{sets of polynomials}.
We initialize our polynomial identity filter as follows:
\begin{enumerate}
  \item Let $M$ map every complex number to the empty set.
  \item Let $\sspace \subset \mathbb{C}$ be a finite subset of the complex numbers.
  \item Sample $n$ independent values, $\alpha_1,\ldots,\alpha_n$, 
  from the uniform distribution over $\sspace$.
  \emph{These values are sampled once initially and used throughout the lifetime of the data structure.}
\end{enumerate}

\paragraph{Inserting a polynomial into \pif}
When a new polynomial $\poly$ is inserted into the \pif, we update the map
by adding $\poly$ to the set $M[\poly(\vec{\alpha})]$.
Intuitively, the set $M[\poly(\vec{\alpha})]$ is the set of all polynomials
that evaluate to the same complex number on the input $\vec{\alpha}$.

\paragraph{Correctness guarantees}
Suppose we insert $\ell$ polynomials into the \pif. 
Intuitively, the \pif implicitly applies the \pit algorithm from \cref{sec:pit} to all pairs of polynomials, i.e., $\leq \ell^2$ polynomial identity checks.
Since every identity test has a false positive probability (if the polynomials are not equal), the total failure probability of the \pif increases.
Luckily, the high-probability guarantees of Schwartz--Zippel still provide us with a pretty good failure probability.
Simply following the union bound, the failure probabilities add up, as formalized in the following theorem:

\begin{theorem}[\pif worst-case guarantees]\label{thm:pif}
  Suppose we insert $\ell$ mutually inequivalent polynomials into a new \pif.
  Let $d$ be the maximum degree of all $\ell$ polynomials.
  The probability that one of the cells of $M$ contains more 
  than one polynomial is at most $\ell^2d / |R|$.
\end{theorem}

\begin{proof}
Let $p_1, \ldots, p_\ell$ be mutually inequivalent polynomials over the same set of $n$ variables. If we insert $\ell$ polynomials into the \pif, it implicitly performs the following randomized computation:
\begin{enumerate}
  \item Sample $\alpha_1,\ldots,\alpha_n$ independently and uniformly from $R$.
  \item For every pair $\poly_i$ and $\poly_j$, where $i\neq j$, check if $\poly_i(\vec{\alpha}) = \poly_j(\vec{\alpha})$.
\end{enumerate}
(Note that step 2 is performed efficiently by computing each $\poly_i(\vec{\alpha})$ separately and inserting $\poly_i$  into $M[\poly_i(\vec{\alpha})]$.
All $\poly_i, \poly_j$ such that $\poly_i(\vec{\alpha}) = \poly_j(\vec{\alpha})$ are therefore inserted into the same cell of $M$.)

Following Schwartz--Zippel (\cref{theorem:pit}), for any pair of polynomials $\poly_i$ and $\poly_j$, where $i\neq j$, $\Pr[\poly_i(\vec{\alpha}) = \poly_j(\vec{\alpha})] \leq d / |R|$.
Therefore, following the union bound, the probability that one of the cells of $M$ contains more than one polynomial is
$$\Pr[\exists i \neq j \ldotp \poly_i(\vec{\alpha}) = \poly_j(\vec{\alpha})] \leq 
\sum_{i,j \in [1,\ell]\\, i\neq j} \Pr[\poly_i(\vec{\alpha}) = \poly_j(\vec{\alpha})] \leq 
\ell^2 d / |R|
$$
\end{proof}

\begin{example}
  If we insert $10^6$ mutually inequivalent polynomials of degree 10 into a new \pif, 
  and we use 64-bit integers for $\sspace$,
  then the probability of the \pif declaring a pair equivalent is $\sim 10^{-7}$.
\end{example}

\paragraph{Implementation considerations}
To further minimize failure probability, if necessary,
after populating the data structure, we can apply \pit (with freshly sampled values) to each pair of polynomials in each equivalence class.
An equivalence class will typically contain a small number of polynomials, allowing us to enumerate all pairs and reverify their equivalence.
To avoid floating-point errors, we can restrict our sample space $\sspace$
to rational numbers (see \cref{app:rational} for details).

\subsection{Rewrite-Rule Synthesizer}

We now have all the ingredients needed to describe our rewrite-rule synthesis technique.

\paragraph{Symbolic-circuit grammar}
\cref{fig:grammar} shows the grammar of symbolic circuits that we consider.
We fix finite sets of one- and two-qubit gates, symbolic gates, parameter variables ($\theta$), and constants.
We also allow for arithmetic expressions over parameters, e.g., $\theta_1 + \theta_2$.

\paragraph{The synthesis algorithm}
\cref{alg:synth}  synthesizes pairs of equivalent circuits.
It starts with a \pif instance containing the empty circuit.
Then, in a bottom-up-synthesis fashion, the algorithm enumerates circuits of increasing size, up to a bound $k$, and inserts them into the \pif.
We assume that all circuits are transformed into  polynomials, following \cref{thm:trans}, before inserting them into the \pif.
We also assume that the polynomials all share the same $\cvar_{\vec{x},\vec{y}}$ variables from \cref{thm:trans}.

For symbolic circuits, the algorithm considers every possible interpretation of the symbolic gates' state transformers ($f^u$).
Note that the space of interpretations is restricted to reversible functions because quantum operations are reversible.   

\begin{wrapfigure}{r}{.47\textwidth}
  \footnotesize
  \begin{minipage}{.47\textwidth}
  \begin{align*}
    C ~ \coloneq & ~ G_{1,1}^\param~i ~\mid~ G_{1,2}^\param~i ~\mid~ \ldots & \text{1-qubit gates}\\
    & \mid ~ G_{2,1}^\param~i~j ~\mid~ G_{2,2}^\param~i~j 
     ~\mid~ \ldots & \text{2-qubit gates}\\
    & \mid S_1 \mid S_2 \mid \ldots & \text{symbolic gates}\\
    & \mid C_1; C_2 & \text{sequential comp.}\\
    \param ~\coloneq & ~ \theta_1 \mid \theta_2 \mid \ldots \mid -\param \mid  c\param \mid \param + \param &\text{parameter expr.}\\
    i,j \in &~ [1,n] & \text{qubit indices}\\
    c \in &~ \{\pi,-\pi,\pi/2,\ldots\} & \text{constants}
  \end{align*}
  \end{minipage}
  \caption{Circuit synthesis grammar}\label{fig:grammar}
\end{wrapfigure}

After \cref{alg:synth} completes, we take each equivalence class in the \pif and generate a set of rules.
For every equivalent pair of circuits, $(C_1,C_2)$, where $C_2$ is smaller than $C_1$ (by number of gates), we generate the rewrite rule $C_1 \to C_2$.
We call these \emph{size-reducing} rules.
For equivalent pairs of the same size, we generate $C_1 \to C_2$ and $C_2 \to C_1$.
We call these \emph{size-preserving} rules.
For symbolic circuits, we construct rewrite rules where the two circuits have the same interpretation.

\paragraph{Pruning techniques}
To prune unnecessary rules, we adopt two techniques from Quartz \cite{xu2022quartz}: (1) Picking a representative circuit from each equivalence class to construct larger circuits with the grammar
(any equivalent circuit can be rewritten to the representative and vice versa). 
(2) Prune rules where both sides have common subcircuits. We also incorporate some new heuristics such as pruning rules where the left-hand side contains functions in parameter expressions, e.g., $\theta_1 + \theta_2$. The full list of additional
pruning we perform is described in \cref{sec:our_pruning}.

\begin{algorithm}[t]  \captionsetup{font=footnotesize} % set size of caption font

\caption{Circuit equivalence synthesizer}\label{alg:synth}\footnotesize
\begin{algorithmic}
\Procedure{\textsc{synth-eq}}{}
\State Construct a \pif instance and insert the empty circuit
\State Let $\mathcal{F}$ be the space of all  reversible functions in $\Zn \to \Zn$
\State Let $\mathcal{C}$ be all circuits with size up to some fixed bound, following  grammar in \cref{fig:grammar}
\For{$C \in \mathcal{C}$} \Comment{{\color{gray}in order of increasing size}}
\If{$C$ contains no symbolic gates}
     insert $C$ into the \pif
\Else
    \State Let $f_1, \ldots,f_l$ be the uninterpreted state transfomers in $C$
    \For{every interpretation $I$ of $f_1,\ldots,f_l$ from $\mathcal{F}$}
      \State insert $C(I)$ into the \pif 
    \EndFor
\EndIf
\EndFor
\EndProcedure
\end{algorithmic}
\end{algorithm}

\section{Circuit Optimizer}\label{sec:opt}
Given a circuit, we apply the synthesized rewrite rules to minimize some cost function: Commonly, this is the number of gates in the circuit because each gate, particularly two-qubit gates, introduces noise in the computation.
There are two critical challenges here:
\begin{enumerate}
  \item How do we apply symbolic rules that can match arbitrary subcircuits? While there are standard algorithms for finding patterns in a quantum circuit, there are no general techniques for finding patterns that satisfy a given constraint. 
  \item In what order to apply the rules? 
  Optimizers, like \voqc and \tket, employ a fixed schedule of optimizations chosen by the compiler designer.
\ours synthesizes tens of thousands of rules, and we simply cannot ask a developer to experiment with different schedules.\footnote{Equality saturation, as realized in the state-of-the-art library, egg~\cite{willsey2021egg}, cannot scale to large numbers of rules, especially \emph{multi-pattern} ones, and cannot apply symbolic rules natively.}

\end{enumerate}
To address these challenges, we present 
(1) an algorithm for matching and applying symbolic rewrite  rules, and
(2) a beam-search-based optimization algorithm.

\newcommand{\match}{\textsc{match}}
\newcommand{\matchsym}{\textsc{match-sym}}

\subsection{Rule-Matching Algorithm}

Given a quantum circuit $C$ and a rewrite rule $C_l \to C_r$, we want to find subcircuits of $C$ that match the pattern $C_l$, and rewrite them to $C_r$.
For non-symbolic rewrite rules, this is a standard process.
\begin{wrapfigure}{r}{.4\textwidth}
  \begin{subfigure}{.4\textwidth}\footnotesize
    \centering
      \begin{quantikz}[row sep = 1pt, column sep=5pt]
      & \gate{R_z^{\theta_1}} &  \gate[wires=2,style={fill=yellow!10}][.5cm]{S} & \qw & \qw\\
      &  \qw &  & \gate{R_z^{\theta_2}} & \qw
      \end{quantikz}
      \vspace{1em}
      \hrule
\vspace{1em}
      \begin{quantikz}[row sep = 1pt, column sep=5pt]
        & \gate{R_z^{\pi}} &  \ctrl{1}\gategroup[3,steps=5,style={fill=yellow!10, inner xsep=0pt},
        background,label style={label position=below,anchor=
        north,yshift=-0.2cm}]{{satisfies constraints on} $S$} & \qw & \targ{} & \ctrl{1} & \qw & \qw & \qw\\
        &  \qw & \targ{} &  \gate[style={fill=yellow!10}]{R_z^{\frac{\pi}{4}}} & \ctrl{-1} & \targ{} & \ctrl{1} & \gate{R_z^{\frac{\pi}{2}}} & \qw\\
        &  \qw & \qw &  \qw& \qw & \qw & \targ{} & \qw & \qw
      \end{quantikz}
  \end{subfigure}%
  \caption{Example of \matchsymb}\label{fig:matchs}
\end{wrapfigure}
First, a quantum circuit is represented as a directed-acyclic graph (\textsc{dag}), just like in our graphical representations of circuits in, e.g., \cref{fig:rules}.
% The \textsc{dag} explicitly represents gate parallelism.
%
Finding the pattern $C_l$ in $C$ boils down to the following problem:
Find a subgraph in $C$ that is \emph{isomorphic} to $C_l$.\footnote{additionally the subgraph has to be \emph{convex}.}
Since this is a well-known problem, we use $\match(C_l,C)$ to denote the procedure that returns \emph{all} subgraphs in $C$ that match the pattern $C_l$.

\paragraph{Matching symbolic patterns}
For simplicity, and without loss of generality, we consider symbolic rewrite rules that contain a single symbolic gate $S$.
We  fix a rule of the form $C_l; S; C_l' \to C_r;S;C_r'$, where the state transformer of $S$ is interpreted by $I$.
We want to formalize matching the pattern $C_l; S; C_l'$ in a circuit $C$.
We assume that   $\sem{S(I)}$ is of the form $\phi^u(\vec{x}) \ket{f(\vec{x})}$
and that the circuit $C$ has $n$ qubits.
The idea is that we will have to try every possible circuit that matches the path sum of $S$, as formalized in $\matchsym$:
\begin{equation}\label{eq:matchs}
\matchsym(C_l;S,C_l',\ C) = 
  \bigcup_{C_S \in \mathcal{S}} \match(C_l;C_S;C_l',\ C)
\end{equation}
where  $\mathcal{S} = \{C_S \mid C_S \text{ is a non-symbolic $n$-qubit circuit and }
\sem{C_S} \text{ is of the form } \phi(\vec{x}\ldots)\ket{f(\vec{x})\ldots}\}$.
Observe that the circuits $C_S$ can apply operations to more qubits than in $S$,
(as indicated by the $\ldots$).
As formalized in \cref{thm:match_sym}, this procedure preserves the correctness of the 
rewrite rule.

\begin{example}
  Consider the 2-qubit symbolic pattern in \cref{fig:matchs} (top),
  where $\sem{S(I)} \equiv \phi^u(x_1x_2) \ket{x_2x_1}$.
  The 3-qubit circuit in \cref{fig:matchs} (bottom) matches the symbolic pattern.
  The highlighted  subcircuit has a path sum of the form
  $e^{i(2x_2-1)\pi/4} \ket{x_2x_1\ldots}$, which matches $\sem{S(I)}$,
  because it swaps the first two qubits, $x_1$ and $x_2$.
\end{example}

\begin{theorem}[Soundness of \matchsym]\label{thm:match_sym}
  Given a symbolic rewrite rule of the form $C_l; S; C_l' \to C_r;S;C_r'$
  and $C_S \in \mathcal{S}$, $C_l; C_S; C_l' \equiv C_r;C_S;C_r'$.
\end{theorem}

\paragraph{Implementing \matchsym}
We implement $\matchsym$  by restricting $\mathcal{S}$ to the space of subcircuits of $C$.
For efficiency, we  limit $\mathcal{S}$ by (1) only considering subcircuits of $C$ over monomial gates, since $S$ is monomial, 
and (2)  limiting the search to subcircuits \emph{between} the set of subcircuits that match $C_l$ and $C_l'$.
Additionally (see \cref{sec:imp}) we limit the size of circuits in $\mathcal{S}$. 
Our approach for checking if a subcircuit is monomial is inspired by \citet{nam2018automated}'s
rotation-merging implementation.

\begin{algorithm}[t!]
  \captionsetup{font=footnotesize} % set size of caption font

\caption{Maximal beam search }\label{fig:algs}\footnotesize
    \begin{algorithmic}
    \Procedure{\maxbeam}{$C$}
    \State Create priority queue $Q$ of bounded size, and add $C$ to $Q$
    \State $C_\emph{best} \gets C$
    \While {$Q$ is not empty}
      \State dequeue circuit $C'$
      \If {$\cost(C') < \cost(C_\emph{best})$}
        \State $C_\emph{best} \gets C'$ 
      \EndIf 
      \For {every rewrite rule $R$}
        \State $C'_R \gets \textsc{apply-max}(R,C')$
        \If {$\cost(C'_R) \leq \cost(C_\emph{best})$ and $C'_R$ has not been seen before}
          \State add $C'_R$ to $Q$
        \EndIf 
      \EndFor
    \EndWhile
    \Return $C_\emph{best}$
    \EndProcedure
  \end{algorithmic}
\end{algorithm}

\subsection{Maximal Beam Search}
To find an optimal circuit, one needs to exhaustively consider every possible ordering of rewrite rule application. To limit the combinatorial explosion, the scheduling algorithm we propose,
\maxbeam (\cref{fig:algs}), 
limits the size of the search space in two ways:
(1) Instead of considering a single application of a rewrite rule in each step of the search, \maxbeam  
greedily considers \emph{maximal} applications of a rewrite rule. 
(2) \maxbeam is a \emph{beam search} through the space of rewrites.

\begin{definition}[Maximal matching set]
  Consider a rewrite rule $C_l \to C_r$
  and a circuit $C$.
  A \emph{maximal matching} set is a subset $\mathcal{M} \subseteq \match(C_l,C)$ such that
(1)   no pair of subcircuits $C_l',C_l'' \in \mathcal{M}$ overlap in $C$, and
(2) there is no $\mathcal{M}'$, where $\mathcal{M} \subset \mathcal{M}' \subseteq \match(C_l,C)$, that satisfies condition (1). 
The same definition applies to symbolic rules.
\end{definition}

\begin{figure}[t!]
  \centering
    \begin{subfigure}{.4\textwidth}
    \begin{quantikz}[row sep = 6pt,column sep = 10pt]
      & 
      \gate{R_z{^\pi}} \gategroup[1,steps=2,style={dashed, rounded corners, inner xsep=0pt},
      background,
      label style={xshift = -0.5cm}]{a}
      & \gate{R_z^{\frac{\pi}{2}}} \gategroup[1,steps=2,style={ rounded corners, inner ysep=10pt},
      background]{b}
      & \gate{R_z^{\frac{\pi}{3}}} \gategroup[1,steps=2,style={dashed,rounded corners, inner xsep=0pt},
      background,
      label style={xshift = 0.5cm}]{c} & \gate{R_z^{\frac{\pi}{4}}}  & \qw
    \end{quantikz}
  \end{subfigure}
    \begin{subfigure}{.4\textwidth}\centering
      \begin{quantikz}[row sep = 6pt,column sep=10pt]
        & \gate{R_z^{\theta_1}} & \gate{R_z^{\theta_2}} & \qw
        \end{quantikz}
        $\to$
        \begin{quantikz}[row sep = 6pt, column sep=10pt]
        & \gate{R_z^{\theta_1 + \theta_2}} & \qw
      \end{quantikz}
    \end{subfigure}
    \caption{Maximal match example}\label{fig:max}
  \end{figure}
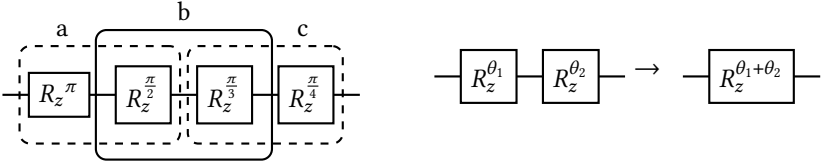

\begin{example}
  Consider the single-qubit circuit with a sequence of four rotations in \cref{fig:max} (left)
  and the rewrite rule that merges two rotations (right).
  The rule matches three subcircuits
  as shown by the three boxes.
Matches (a) and (b) overlap, as well as (b) and (c).
So $\applymax$ chooses a set of matches that do not overlap---e.g., the dotted ones---and applies the rewrite to them.
\end{example}

The \maxbeam algorithm begins with a priority queue of fixed size containing the input circuit $C$ that
we wish to optimize.
The priority queue uses a cost function, \cost.
The algorithm picks the next circuit from $Q$ and rewrites it.
For every rewrite rule $R$, it applies $R$ \emph{maximally} to the current circuit $C'$.
Specifically, the function $\applymax$ finds a maximal set of \emph{non-overlapping} matches for the rewrite rule $R$ in $C'$ and rewrites all the matches, producing a new circuit $C_R'$. 
In practice, we implement $\applymax$ greedily and do not try to find a \emph{maximum} matching set, only a maximal one.
\maxbeam can be terminated after a finite number of iterations or within some time limit.

\section{Implementation and Evaluation}\label{sec:imp}

\paragraph{Synthesized optimizers} 
We implemented \ours in ${\sim}\numprint{3700}$ lines of Java.
We evaluated \ours on four different gate sets:
(1) the standard gate set for \ibm computers,
(2) the gate set for Rigetti computers,
(3) the gate set for \emph{ion trap} computers (like IonQ \cite{ionqGateset}),
and (4) the gate set of~\citet{nam2018automated} (henceforth, Nam).
The \ibm and Rigetti gate sets support devices with superconducting qubits,
which are the largest quantum devices physically realized so far. Ion trap architectures 
are attractive due to their all-to-all qubit connectivity, which reduces the need for
expensive \emph{swaps}. The Nam gate set is interesting to study because it 
closely resembles the \emph{Clifford+T} universal gate set where Clifford gates 
can be efficiently simulated on a classical computer. However, 
unlike the other gate sets, the Nam gate set is not physically realized in any 
quantum hardware. 

\cref{table:rule_gen} summarizes the gate sets
and the rules synthesized.
For all gate sets, we limit rewrite rules to be over a maximum of 3 qubits and vary the maximum \emph{size} of a rule: the number of gates on either side.
We choose the largest size for which \ours can synthesize rules within ~3 minutes.
For example, for \ibm, in 72 seconds \ours  synthesizes 701 rules, out of which 48 rules are symbolic.
Failure probability is an upper bound on the \pif returning an incorrect rule (\cref{thm:pif}).
Observe how vanishingly small the failure probabilities are, e.g., $10^{-17}$ for \ibm.
\pagebreak

\paragraph{Research questions} We aim to answer the following research questions:

    (\textbf{Q1}) How does \ours compare to state-of-the-art optimizers?

    \textbf{(Q2)} How does \ours compare to superoptimization?

    \textbf{(Q3)} Which synthesized rewrite rules are useful?

\paragraph{Benchmarks} Throughout, we will use  a set of 33 benchmark circuits, comprised of those from prior work on optimization ~\cite[]{nam2018automated, amy2014tdepth, hietala2021verified, xu2022quartz}
and a new class of circuits. The benchmarks from prior work include arithmetic circuits and 
Toffoli gate networks. We added \emph{quantum approximate optimization
algorithm} (\qaoa) circuits that approximate the maximum cut in a 3-regular graph.
\qaoa is a promising and  near-term application because it can approximate \textsc{np}-hard combinatorial problems
on \textsc{nisq} machines without error correction.

\paragraph{Instantiation of \ours}
We use the total number of gates in a circuit as \ours's cost function ($\cost$ in \cref{fig:algs}).
We fix the priority queue size (in \cref{fig:algs}) to 8000 circuits.
For matching symbolic circuits, we limit the number of qubits and size of the $C_S$ circuits in \cref{eq:matchs} to 7 and 10, respectively.

\paragraph{Metrics}
To compare tools, the main metric we use is the number of two-qubit gates, because they have 
\emph{orders of magnitude} higher error rates compared to single-qubit gates.
For instance, the error rates for single- and two-qubit gates on the \ibm Toronto device are on the order of $10^{-4}$ and $10^{-2}$, respectively \cite{toronto} (further, some single-qubit gates, like $R_z$, are error-free as they are simulated classically). 
To further illustrate the importance of two-qubit gate reduction,
we use \emph{fidelity} results (success probability), which we statically estimate based on publicly available error rates from 
the \ibm Toronto \cite{toronto}, Rigetti Aspen-11 \cite{aspen11}, and IonQ Aria \cite{aria} devices. Fidelity is the probability that none of the gates in a circuit cause an error. For a circuit $G_1;\ldots;G_n$, its fidelity is $\prod_i (1 - \text{ error rate of } G_i)$.

\begin{table*}[t!]
  \caption{Rewrite-rule synthesis results}
  \centering
  \setlength{\tabcolsep}{4.5pt}
  \scriptsize
    \begin{tabular}{l l c c c c c r c }
      \toprule
      Gate set & Gates & \# Qubits & Size & \# Possible Rules & \# Rules & \# Symbolic Rules & Failure Prob. & Time (s) \\
      \midrule
      \ibm    & $\uone^\theta$, $\utwo^{\theta_1, \theta_2}$, $\uthree^{\theta_1, \theta_2, \theta_3}$, \cnot & 3 & 4 & \multicolumn{1}{r}{$1.5 \times 10^{13}$}   & \multicolumn{1}{r}{\numprint{701}}   & \multicolumn{1}{r}{48}        & $10^{-17}$      & \multicolumn{1}{r}{72} \\ 
      Nam     & $\h$, $\x$, $\rz^\theta$, \cnot                                                                   & 3 & 6 & \multicolumn{1}{r}{$5.4 \times 10^{18}$} & \multicolumn{1}{r}{\numprint{14544}} & \multicolumn{1}{r}{\numprint{2365}}& $10^{-12}$ & \multicolumn{1}{r}{135} \\ 
      Rigetti & $\rx^\pi$, $\rx^{\pi/2}$, $\rx^{-\pi/2}$, $\rz^\theta$, \cz                     & 3 & 5 & \multicolumn{1}{r}{$1.4 \times 10^{16}$}  & \multicolumn{1}{r}{\numprint{2242}}  & \multicolumn{1}{r}{809}   & $10^{-14}$          & \multicolumn{1}{r}{70} \\
      Ion     & $\rx^\theta$, $\ry^\theta$, $\rz^\theta$, $\rxx^\theta$                                       & 3 & 3 & \multicolumn{1}{r}{$1.6 \times 10^{11}$}      & \multicolumn{1}{r}{\numprint{1519}}  & \multicolumn{1}{r}{24}  & $10^{-19}$            & \multicolumn{1}{r}{15} \\
      \bottomrule
    \end{tabular}
    \label{table:rule_gen}
\end{table*}
\subsection*{Q1: How does \ours compare to state-of-the-art optimizers?}
\paragraph{Experimental setup} We compared \ours to four state-of-the-art optimizers: \ibm Qiskit \cite{Qiskit}, Quilc \cite{quilc},
\tket \cite{tket}, and \voqc \cite{hietala2021verified}. 
The first three are used in industrial toolkits; \voqc is a formally verified and very effective optimizer.%
\footnote{We excluded the \citet{nam2018automated} optimizer because it is proprietary and the existing data was not obtained from 
running on decomposed input circuits nor does it include the added benchmarks. We also excluded PyZX  \cite{kissinger2019pyzx} because it works well for reducing T gate count, which
is useful for future fault-tolerant machines, but can often increase total gate count. A comparison of \ours against PyZX with respect to T gate reduction is in \cref{app:tcount}.}
For each  benchmark, we set a time limit of 1 hour (we discuss running time of \ours in Q3).
To ensure a fair comparison of the optimization phases of the various tools,
we provide all tools with the same decomposed input circuit in the target gate set.

We use \emph{S-curves} to present the results.
For each benchmark circuit, we compute the quantity
$$
\frac{\text{\# of gates with tool } X - \text{\# of gates with \ours}}{\text{\# of gates in unoptimized circuit}}
$$
and present the benchmarks in increasing order.
Positive values imply that \ours outperforms tool $X$.
We compute the following quantity for fidelity:\footnote{We do not use the fidelity of the original circuit in the denominator here because it can be extremely small.}
$
\frac{ \text{fidelity with \ours\xspace } - \text{ fidelity with tool } X}{\text{maximum of fidelity with \ours and with tool }X}.
$ 

\begin{figure*}[t!]
  \centering
  \includegraphics[width=0.9\textwidth]{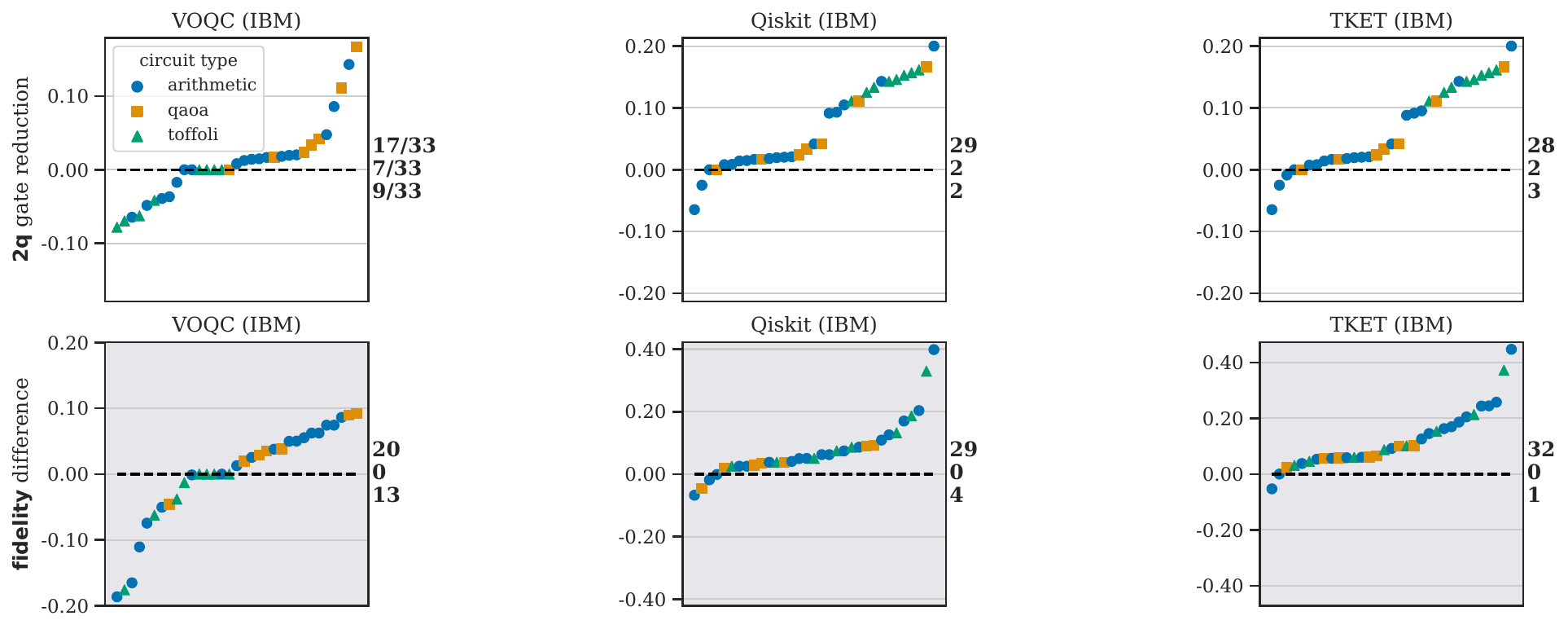}
  \caption
  { Comparison against state-of-the-art optimizers on \ibm.
  Each graph is annotated to the right with the number of circuits where \ours outperforms, matches, and underperforms (top-to-bottom)
  the other tool.} 
  \label{fig:rq1_ibm}
\end{figure*}

\paragraph{\ibm} 
\cref{fig:rq1_ibm} shows the S-curves for the \ibm gate set.
Consider, for instance, the top middle S-curve, which compares \ours to \ibm Qiskit.
For the majority of the benchmarks, 29/33, \ours outperforms Qiskit in two-qubit gate reduction---all the benchmarks above the dashed horizontal (0) line.
We see similar results with \tket.
\ours can outperform \voqc in 17/33 benchmarks, and exactly match its performance on 7/33 benchmarks.
The fidelity graphs, bottom row, depict a very similar story, indicating the close correspondence between two-qubit-gate count and fidelity.

\paragraph{Nam}
Results on Nam are in the appendix because they resemble the results for the \ibm gate set.

\paragraph{Rigetti and Ion} For the Rigetti and Ion gate sets, 
we compare against Quilc and Qiskit, respectively.
Implemented by Rigetti, Quilc is specialized for optimizing the Rigetti gate set.
To our knowledge, these are the only publicly available compilers that apply to those two gate sets.
We also compare against \tket for the Rigetti gate set but we note that the optimized circuits \tket produces \emph{do not}
adhere to the allowed angles for $\rx$ gates and therefore are not valid. 
See appendix for the results, which resemble the results for Quilc.
 
\begin{figure*}[t!]
  \centering
  \begin{subfigure}[b]{0.48\textwidth}
    \caption{Quilc on Rigetti gate set} 
    \centering
    \includegraphics[width=\textwidth]{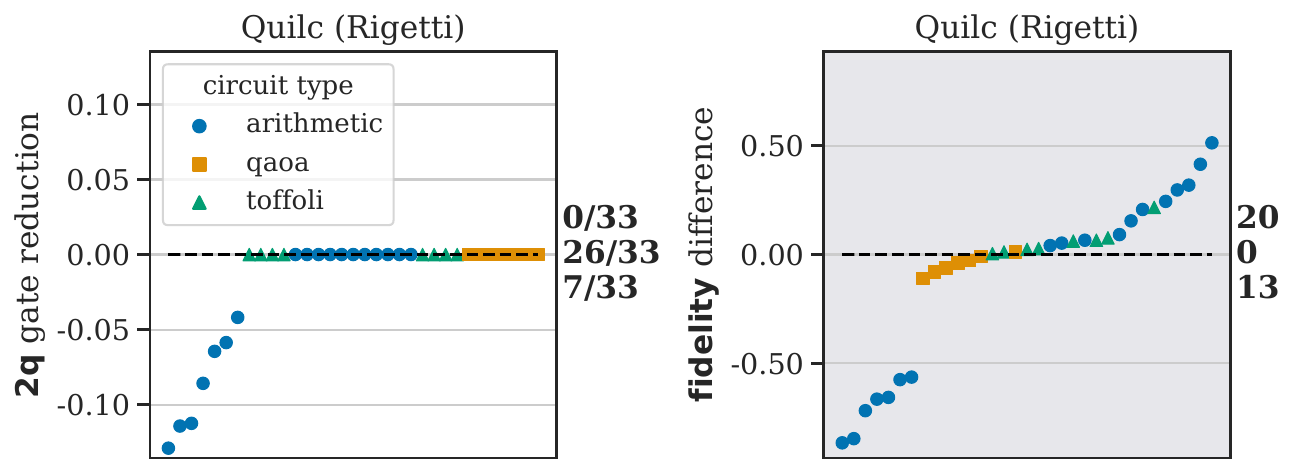}
  \end{subfigure}
  \begin{subfigure}[b]{0.48\textwidth}
    \caption{Qiskit on Ion gate set} 
    \centering
    \includegraphics[width=\textwidth]{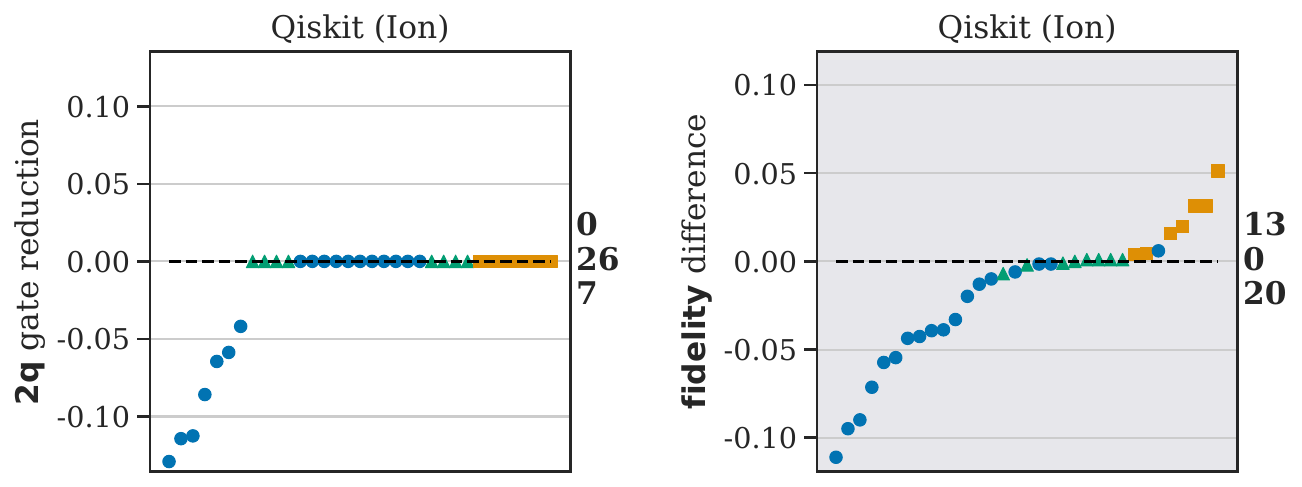}
  \end{subfigure}

  \caption
  { Comparison against state-of-the-art optimizers on Rigetti and Ion trap.} 
  \label{fig:rq1_rigetti_ion}
\end{figure*}

None of the tools are able to reduce two-qubit 
gate count for the majority of the benchmarks as shown in \cref{fig:rq1_rigetti_ion}, where these benchmarks lay on the dashed line.
However, we see reduction in single-qubit gates (see appendix), which is reflected in fidelity, as single-qubit gates errors dominate for all but 7/33 benchmarks.
 \ours is able to  outperform  Quilc on a majority of the benchmarks for the Rigetti gate set (20/33). 
For the Ion gate set, we see an opposite story: \ours 
outperforms Qiskit on 13/33 benchmarks and underperforms it on  20/33.

We investigated why 
Quilc and Qiskit are sometimes able to achieve reduction in two-qubit gates and isolated it to a powerful optimization that resynthesizes arbitrary two-qubit circuits~\cite{cross2019validating}. We cannot fully capture such optimizations and leave it as an avenue for future work.

\paragraph{\emph{Q1 summary}} \textbf{\ours is able to significantly outperform or match state-of-the-art optimizers on a majority of the benchmarks across all
gate sets with respect to two-qubit gate reduction. The results are similar for fidelity except on the Ion gate set where \ours only outperforms or matches Qiskit on 39\% of the benchmarks.}

\subsection*{Q2: How does \ours compare to superoptimization?}

\paragraph{Experimental setup} 
Next, we compare against the Quartz \emph{superoptimizer} \cite{xu2022quartz}.
Quartz is comprised of two phases that run in sequence: (1) The \emph{preprocessing} phase: a manually written set of optimizations that decompose a circuit to the target gate set and applies rotation merging and other domain specific optimizations---e.g., Hadamard and $\cz$ cancellation for Rigetti. 
(2) The \emph{search} phase: applies automatically synthesized (non-symbolic) rewrite rules and applies them to a circuit by enumerating different orderings.
The synthesized rules are verified with an \textsc{smt} solver,
but the manually written preprocessing phase is not verified.

We compare against Quartz along two dimensions: (1) the time to synthesize rules and (2) the quality of the 
synthesized rules. Both tools were alotted 1 hour of optimization time, 32GB of RAM, and 1 \textsc{CPU} core per benchmark. 
% Because Quartz does not support input circuits with 
% parameterized gates, we manually added three placeholder gates and their decompositions for each extra rotation needed for
% the QAOA benchmarks. 
We do not compare against Quartz 
for the Ion gate set, which they do not currently support.
To fairly compare against Quartz, we distinguish between two variants of the tool:
(1) Quartz, the full tool with the two phases, and (2) Quartz-NoPP, which is Quartz without the preprocessing phase---just the synthesized rules.

\begin{figure*}[t!]
  \centering
  \includegraphics[width=0.9\textwidth]{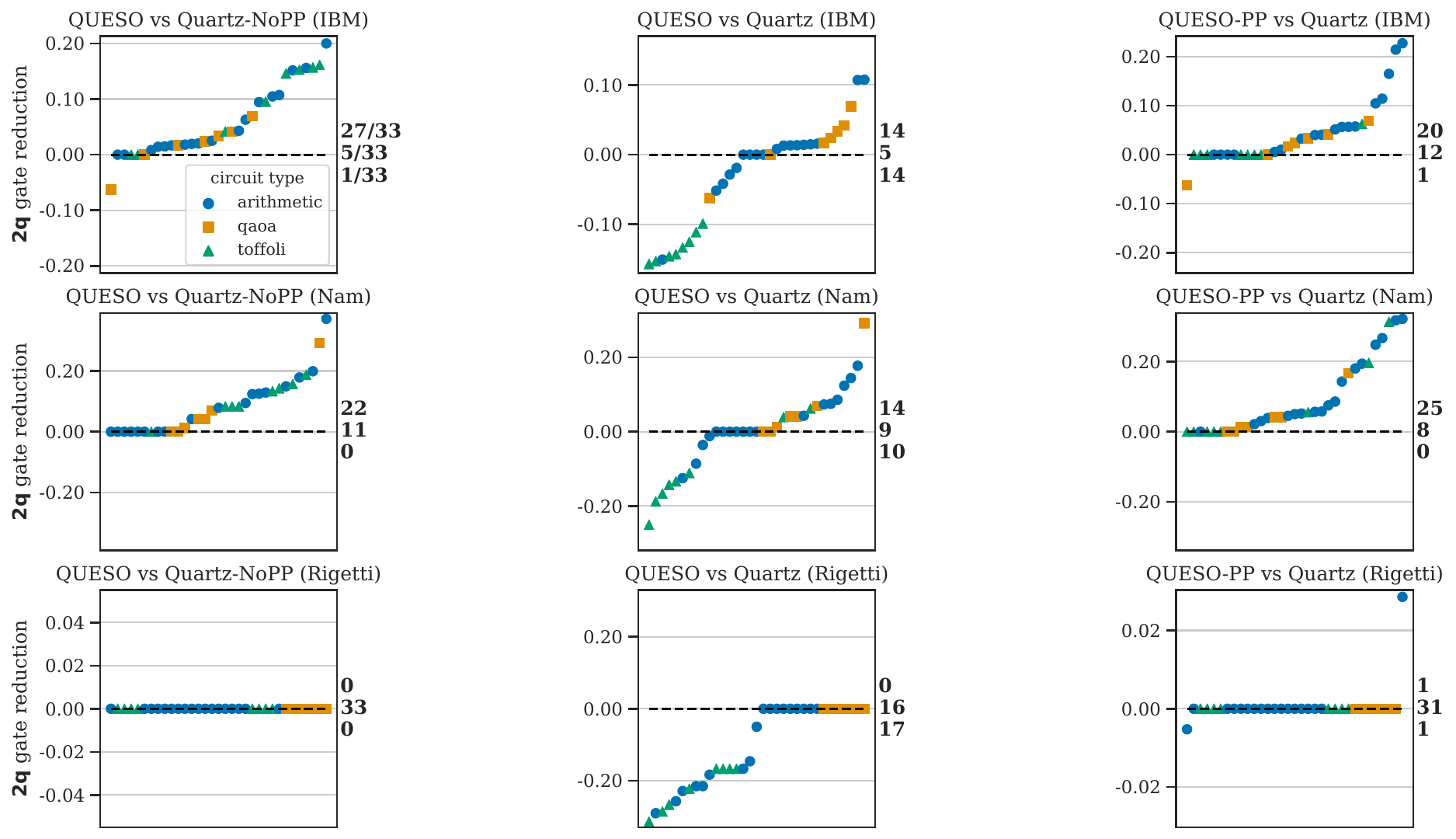}
  \caption{\textbf{1 hour timeout} comparison against Quartz:
  (left column) Quartz without the preprocessing phase;
  (middle column) Quartz with both phases;
  (right column) \ours with Quartz's preprocessing phase vs Quartz with both phases.} 
  \label{fig:rq2_2q}
\end{figure*}

\begin{figure*}[t!]
  \centering
  \includegraphics[width=0.9\textwidth]{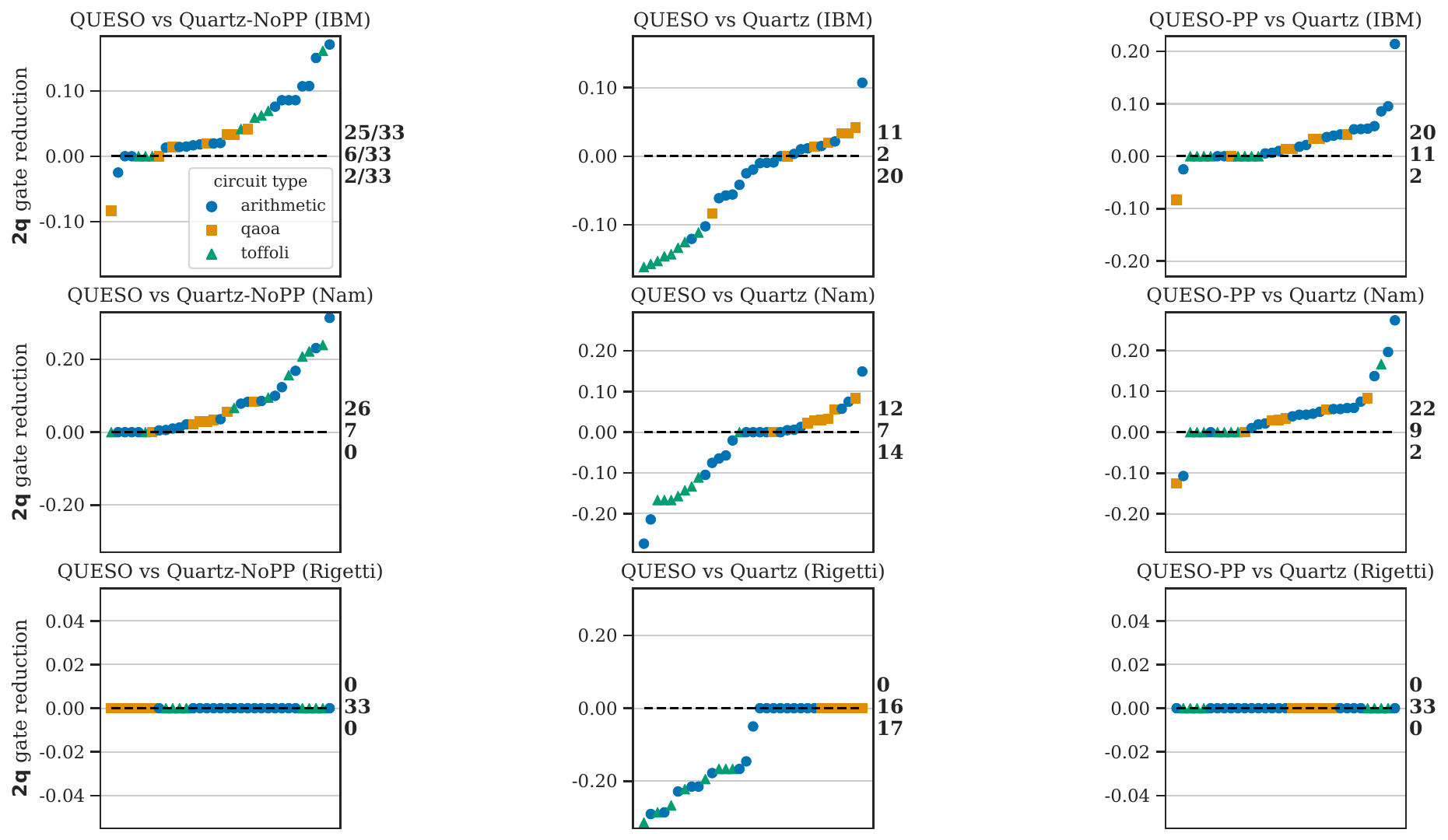}
  \caption{\textbf{24 hour timeout} comparison against Quartz} 
  \label{fig:rq2_2q_24hr}
\end{figure*}

\paragraph{Results} 
First, we observe that \ours is able to synthesize rules an order of magnitude faster than Quartz:
\ours synthesizes rules in 70-135 seconds, whereas Quartz takes up to \numprint{2303} seconds for rules of the same size. 
For example, \ours synthesizes rules for the \ibm gate set with up to 3 qubits and size 4 in 72 seconds while Quartz takes \numprint{2193} seconds.
Note that comparing rule-synthesis speed directly is challenging because \ours synthesizes more expressive, symbolic rules. We attribute our 
fast rule synthesis to using a \pif for equivalence checking rather than an \textsc{smt} solver.

\cref{fig:rq2_2q} shows the results of the comparison to Quartz.
The first column shows that \ours is able to significantly outperform Quartz with preprocessing disabled (Quartz-NoPP) on most benchmarks on \ibm and Nam gate sets. For Rigetti, both \ours and Quartz-NoPP cannot eliminate any two-qubit gates. 
These results demonstrates the power of our synthesized rules compared to Quartz's.

\cref{fig:rq2_2q} (middle column) shows the results against full Quartz, i.e., when enabling the hand-crafted preprocessing phase.
Note that in this case we provide Quartz with circuits pre-decomposition---i.e., with Toffoli gates.
For \ours, we use the same 
decomposition for each Toffoli gate, whereas Quartz's preprocessing greedily picks which decomposition to use.
The results on \ibm and Nam show \ours and Quartz are close in performance;
for Rigetti, thanks to the domain-specific optimizations in the preprocessing phase, Quartz is able to eliminate two-qubit gates in half of the benchmarks.

To further understand the effects of the preprocessing phase, \ourspp is the result of running \ours on the output of Quartz's preprocessing phase.
As the third column of \cref{fig:rq2_2q} shows, on \ibm and Nam, \ourspp outperforms full Quartz, and matches it on Rigetti.
These results further amplify the power of our symbolic rules in comparison with Quartz's.

The results are similar when running both tools for the full 24 hour timeout used in Quartz's original evaluation as shown in \cref{fig:rq2_2q_24hr}. 
The results for every one hour interval are included in the appendix. 
We observe that after 4 to 5 hours, Quartz (with preprocessing) catches up to \ours (without preprocessing) on the \ibm gate set on 6 of the benchmarks.
With the longer timeout, Quartz overall performs slightly better than before on the \ibm and Nam gate sets and remains the same
on the Rigetti gate set. The notable exception is the comparison against \ours and Quartz-NoPP on the Nam gate set where \ours benefits from the longer timeout.

\paragraph{\emph{Q2 Summary}} \textbf{\ours synthesizes rules for the same size and gate set 
up to 30x times faster than Quartz. When comparing synthesized rules, \ours outperforms or matches Quartz 
on 97\% of the benchmarks across all gate sets.}

\subsection*{Q3: Which synthesized rewrite rules are useful?}
We explore this question at two levels of granularity:
(A) Is there a subset of the rewrite rules that is sufficient for producing optimal circuits?
(B) Which classes of rules are useful for optimization?
We present results for the representative \ibm gate set.

\paragraph{(A) Results}
We collect the set of all rewrite rules that result in the best circuit that \ours can discover within 1 hour of execution time.
Across all benchmarks, we observe that a fixed subset of only 35 rules out of \numprint{701} rules are used to reach the best solution.
This implies that we can run \ours with a subset of the rules and achieve similar results in a significantly smaller amount of time.
We envision, for instance, that \ours can be \emph{finetuned} on a given class of problems to collect all relevant rules and discard the unnecessary ones.

\begin{figure*}[t!]
  \centering
  \includegraphics[width=\textwidth]{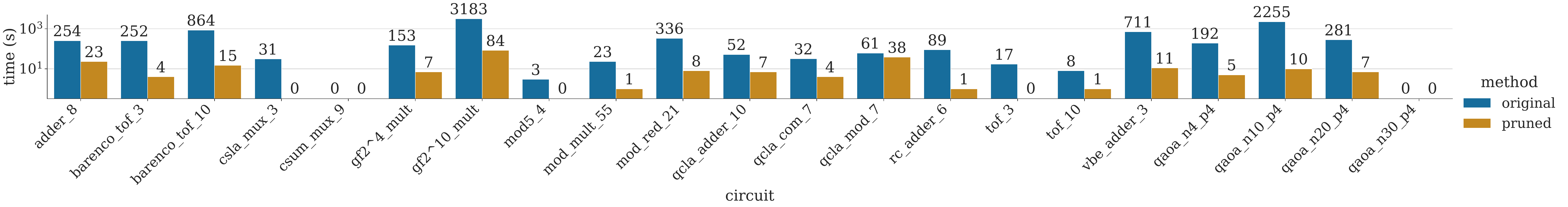}
  \caption{\textbf{Log-scale} comparison of \ours running time on \ibm with pruned (35) vs original (701) rewrite rules.} 
  \label{fig:rq3_time_to_best}
\end{figure*}

To understand the time savings, we run \ours with the \emph{pruned} set of 35 rules 
and ask: how long does it take for it to reach a circuit of equal cost (in terms of two-qubit gate count) to that found by \ours running on the full set of rules for 1 hour.
\cref{fig:rq3_time_to_best}
 shows the results for a representative subset of the 33 benchmarks.
We observe a drastic reduction in runtime. 
For instance, on the \textsf{qaoa\_n10\_p4} benchmark,
it takes \ours \numprint{2255} seconds to converge to the best circuit that can be found within 1 hour,
but it takes the pruned version only 10 seconds to arrive at a circuit
with the same number of two-qubit gates.
Overall, we observe runtime reductions of up to 225x, making \ours run in a few seconds to a minute on the majority of the benchmarks.

\begin{figure*}[t!]
  \centering
  \includegraphics[width=0.9\textwidth]{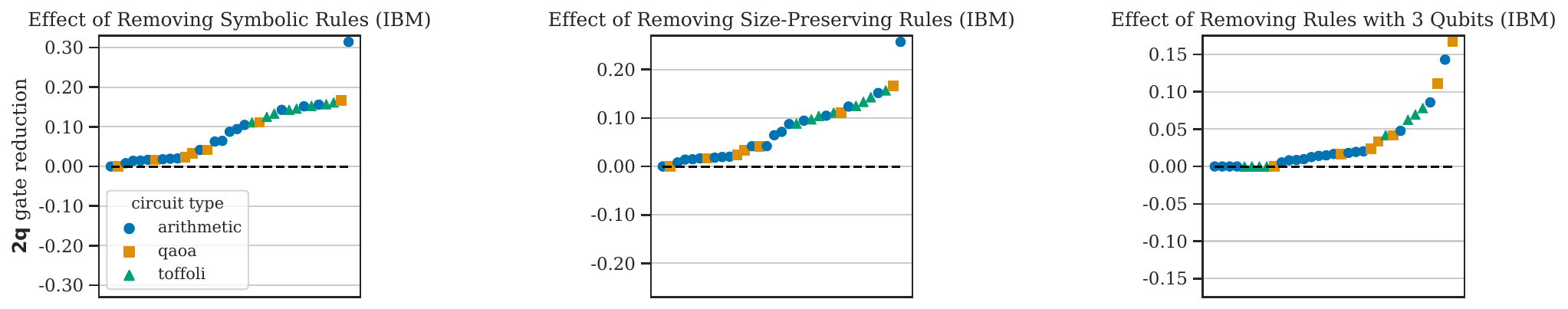}
  \caption{Effect of different types of rules. Points $> 0$ indicate that removing the rules is detrimental.} 
  \label{fig:rq3_toggles}
\end{figure*}

\paragraph{(B) Rule classes}
Next, we study the effect of symbolic rules, size-preserving rules, and rules over 3-qubit circuits.
\cref{fig:rq3_toggles} 
shows the S-curves comparing \ours versus \ours without a subset of the rules.
The graphs show that removing each type of rule  significantly affects
the performance---almost all the points are above the dashed line. Most importantly,
this shows that being able to synthesize and apply symbolic rules is critical. We also observe that greedily applying only size-reducing rules, 
results in a worse solution in almost all the benchmarks. Finally, we see a slightly less dramatic effect when 
removing rules with 3 qubits, i.e., restricting \ours to rules with up to 2 qubits, but a majority of the benchmarks still rely on these rules.

\paragraph{\emph{Q3 Summary}} \textbf{Restricting \ours to the small subset of rules used decreases
the time to reach the best solution by up to 225x. Symbolic rules significantly contribute 
to \ours's performance and the ability to synthesize and apply them is critical. }

\section{Related Work} 

\paragraph{Rewrite-rule synthesis}
Most quantum-circuit compilers use hand-crafted optimizations 
for a given gate set
\cite{Qiskit,tket,quilc,hietala2021verified,certiq,nam2018automated}.
Our approach is most similar to Quartz \cite{xu2022quartz} (and its predecessor~\cite{quanto}) and related work on rewrite-rule synthesis for computation graphs~\cite{jia2019taso} and \textsc{llvm}.
\ours differs from Quartz along two dimensions: expressivity and speed.
\ours can learn symbolic rules, unlike Quartz. 
These symbolic rules enable optimizations similar
to \citet{nam2018automated}'s rotation merging, which Quartz applies as an unverified hand-crafted preprocessing step.
Our synthesis phase is much faster due to the application of Schwartz--Zippel as opposed to \textsc{smt}-based verification.
Quartz implements a preprocessing pass to optimize the circuit before
beginning to apply rewrite rules using a cost-based backtracking search.
In contrast, we rely only on learned rules; as our goal is 
not superoptimization, we apply rules  greedily, allowing our approach to find smaller circuits faster. ZX calculus-based optimizers use graphical rewrite rules~\cite{ZX1,ZX2,kissinger2019pyzx}. 
However, PyZX's full optimization pass involves a subroutine \cite{pyzxdocs} that only supports circuits with Clifford + T gates
and does not support arbitrary rotation, making it rigid and not 
compatible with existing hardware platforms that execute gates with arbitrary rotation. \ours is designed to leverage hardware gates to unlock broad optimization opportunities and enable flexibility to adapt to changes in the basis gates.     

\paragraph{Verified optimizers}
Compilers are hard to get right~\cite{DBLP:conf/issta/SunLZS16}, and much progress has been made in building verified classical compilers~\cite{compcert,cakeML}.
This includes verified optimizing compilers, using interactive theorem proving~\cite{compcertPeephole, compcertPoly, compcertMiddle, cakeMLfp}
and automated techniques like \textsc{smt}-solving~\cite{alive,alive2,cobalt}.
In the quantum realm, 
similar efforts have included reversible circuit compilers verified in F*~\cite{reqwire,reverc}, the optimizer \textsc{voqc} \cite{hietala2021verified}, which has been formally verified in Coq,
and Giallar \cite{Giallar} for verification of Qiskit optimizations.
Our work, in contrast to the above, automatically synthesizes probabilistically verified rewrite rules,
relying on novel verification insights for this problem domain and a probabilistic data structure. 
Leveraging \pit for equivalence-checking has been applied in other areas such as 
program analysis \cite{gulwani2003} and machine learning \cite{wang2021}.

\paragraph{Circuit resynthesis}
There are quantum-circuit optimizations that \ours cannot discover.
The most interesting is that of 
\citet{cross2019validating}. This optimization finds maximal disjoint blocks of
gates that operate on a given control and target of a $\cnot$ in the block.
For each two-qubit block, the optimization computes a unitary operation and
resynthesizes a subcircuit for the block using either exact techniques \cite{KAK, KAK2}
or approximation. Other similar optimizations include \textsc{quest}~\cite{patelQUEST},
which performs approximate resynthesis of circuits to reduce their $\cnot$ count,
and several exact resynthesis techniques for reducing the $\cnot$ counts in circuits
\cite{CNOTSDP, Meuli2018, Davis2020}.

\section{Conclusions and Future Work}
We have described a technique for automatically generating quantum-circuit optimizers by synthesizing symbolic rewrite rules.
Our results demonstrate the remarkable ability of our synthesized optimizers to outperform or rival state-of-the-art optimizers.
For future work, we would like to explore (1) learning-based techniques for scheduling rewrite rules to speed up optimization, and (2) enhancements of symbolic rules to capture more sophisticated optimizations like those in Quilc.

\section*{Acknowledgements}
We thank the anonymous reviewers and our shepherd, Joseph Tassarotti, for their insightful feedback.
We also thank Justin Hsu and Thomas Reps for their input during the writing process. 
We are grateful to Martin Diges, Mingkuan Xu, and the CHTC team 
for their assistance in our experimentation as well as Max Willsey for providing guidance during
our exploration of egg.

This work is supported by \textsc{nsf} grants \#1652140 and \#2212232 and awards from Meta and Amazon. 
This research is also partially supported by the OVCRGE at the University of Wisconsin–Madison with funding 
from the Wisconsin Alumni Research Foundation.
Lauren Pick is supported by \textsc{nsf} grant \#2127309 to the Computing Research Association for the CIFellows Project.

\section*{Software Availability}

Our artifact is publicly available on Zenodo \cite{quesoArtifact}. It contains the \ours source code, evaluation infrastructure, and documentation.

\bibliography{references}

\newpage
\newpage
\appendix
\section{Quantum Hardware}\label{app:hardware}

\vspace{0.15in}

{\noindent \textbf{What are different quantum hardware platforms?} Physically, a quantum bit (qubit) can be realized using superconducting, semiconductor,  or atomic qubit devices. For example, \ibm, Rigetti, and Google use superconducting architecture, whereas IonQ and Honeywell use atomic qubits. The design tradeoffs between different quantum hardware platforms are summarized in~\cite{linke2017experimental}.   

\vspace{0.15in}

{\noindent \textbf{How are gates performed on hardware?}} Quantum hardware has two parts  -- (1) qubit devices and  (2) a control computer. The control computer manipulates qubit states by sending control pulses to the qubit devices. For superconducting qubits, gates are performed by using the sequence of microwave control pulses, whereas shining laser pulses manipulate atomic qubits. The control computer can support different gates by changing the sequence or re-shaping these control pulses. Typically the pulse sequences and shapes are stored in the FPGAs, which drive the microwave signal generators or the lasers. We can add a gate to the basis by reprogramming the control FPGA.

\vspace{0.15in}
	
{\noindent \textbf{Why are two-qubit gates significantly error prone?}} Two qubit gates entangle quantum bits using complex and relatively long control pulses to qubit devices. A qubit can decay to the lowest possible energy state during gate operations. Furthermore, the longer the gate duration, the exponentially higher the chances of qubit decay. As a result, the two-qubit gate error is about 10x higher, about 1\%}, on most industrial quantum computers~\cite{googledatasheet,toronto,aria,aspen11}.  Furthermore, two-qubit gates corrupt both the qubit devices involved in the operation and impose crosstalk errors on neighboring qubit devices, significantly degrading the output quality~\cite{murali2020software,das2021adapt}.

\vspace{0.15in}

{\noindent \textbf{What are "perfect" virtual gates?}} On most quantum hardware platforms, $\rz$ gates are implemented in software virtually~\cite{mckay2017efficient}. \textit{These virtual gates have zero latency and perfect fidelity}. The insight behind the virtual gate is quite elegant. $\rz$ gates affect the phase of the qubit state, which can be performed by simply adding a phase offset to the subsequent $\rx$ or $\ry$ gate pulses, such that the subsequent gate not only performs the intended operation but also adds the phase required by the $\rz$ gate. If the subsequent gate is another $\rz$ gate or sequence of $\rz$ gates, then the phase offset corresponding to the sum of all rotation angles in the sequence is used. Due to this hardware optimization, reducing the $\rz$ gate at the compiler level does not yield any fidelity benefits. Therefore, \textit{to evaluate the effectiveness of \ours, we count physical gates and omit ``virtual'' $\rz$ gates that are perfect and do not degrade fidelity on real quantum hardware.}

\vspace{0.15in}

{\noindent \textbf{Why are basis gates changing?}} On conventional computers and accelerators, instruction sets are rigid, as changing the instruction set involves physically redesigning the chip, whereas quantum hardware is highly reconfigurable. Engineers leverage this flexibility and continuously tune gates to enable higher gate fidelities. For example, all \ibm platforms recently modified their basis gates to support less flexible but more noise-tolerant gates by upgrading their control computer firmware. With scaling quantum hardware, basis gates are expected to change as hardware is evolving rapidly with the innovation in qubit devices and qubit control. If we design compiler optimizations for fixed basis gates, we will need to hand-tune the optimization every time we change the basis gates. Recent work has also shown that it can be beneficial to allow basis gates to differ across qubits based on calibration data to minimize errors \cite{lin2022chooseBasis}.

\section{Sampling in the rational domain}\label{app:rational}
To sample complex numbers on the unit circle, we note that 
every  rational number $r$
results in a unique rational coordinate on the unit circle:
$$\left(\frac{r^2 - 1}{1+r^2}, \frac{2r}{1+r^2}\right)$$
Specifically, $r$ is the slope of the line between $(1,0)$
and the point above.

Therefore, if we uniformly sample $r$ from any finite subset of $\mathbb{Q}$
and apply the above formula, we have a complex number on the unit circle:
\begin{itemize}
    \item Let $X \subset \mathbb{Q}$.
    \item Sample $r$ uniformly from $X$.
    \item Return $\frac{r^2 - 1}{1+r^2} + \frac{2r}{1+r^2}i$
\end{itemize}

\begin{theorem}
The above procedure uniformly samples from the set 
$$\left\{\frac{r^2 - 1}{1+r^2} + \frac{2r}{1+r^2}i \mid r \in X\right\}$$
\end{theorem}
\begin{proof}
    Follows from the fact that each $r \in X$
    produces a unique number on the unit circle,
    $\frac{r^2 - 1}{1+r^2} + \frac{2r}{1+r^2}i$.
    Since $r$ is sampled uniformly from $X$,
    the output of the algorithm is uniformly distributed
    over the set $\left\{\frac{r^2 - 1}{1+r^2} + \frac{2r}{1+r^2}i \mid r \in X\right\}$.
\end{proof}

\section{More implementation details}

\subsection{Pruning techniques}
\label{sec:our_pruning}
\begin{figure*}[h]
    \centering
    \includegraphics[width=0.66\textwidth]{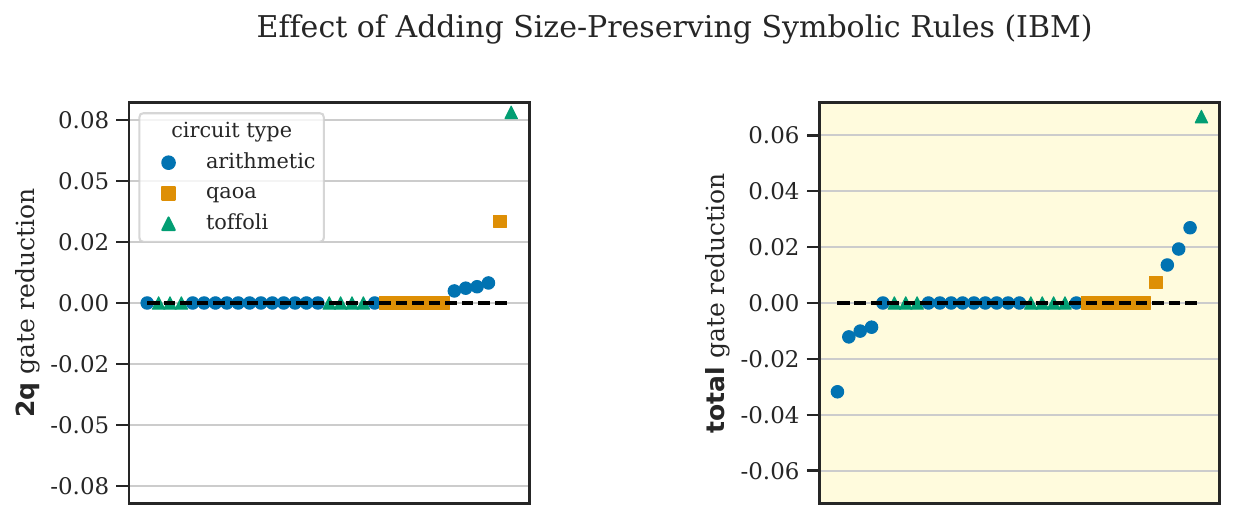}
    \caption{Comparison between running \ours using a baseline set of rules and adding size-preserving symbolic rules.} 
    \label{fig:rq3_add_size_preserve_symb}
\end{figure*}

In addition to the pruning adopted from Quartz described in \cref{sec:synth} we 
prune rules that satisfy the following:
\begin{itemize}
    \item Either side is a disconnected graph. The way we enumerate circuits syntactically results in some rules that are unnecessary in the quantum setting such as the rule saying two gates on different qubits can commute.
    \item The pattern to search for contains a parameter with an arithmetic expression such as $\theta_1+\theta_2$. Unless the parameter also appears in the replacement, the search would need to decompose an angle into a sum of two angles to find a match.
    \item The symbolic parameters in the replacement are not a subset of the symbolic parameters in the pattern. For example, this is possible when the symbolic parameters on both sides sum to zero.  
    \item The qubits in the replacement are not a subset of the qubits in the pattern. This can happen in symbolic rules if the state transformer swaps the state of two qubits.
    \item A symbolic rule where the subcircuit before or after the symbolic gate in the pattern is empty. This is for efficiency and limits the search for a matching symbolic subcircuit to cases where it is more constrained. Additionally, rules of this form are almost always size-preserving rules, which \cref{fig:rq3_add_size_preserve_symb} shows do not help in most cases. 
\end{itemize}

\subsection{Experimental setup}
\paragraph{Instantiation of \ours} \cref{table:rule_gen_angles} shows the different parameters included in the grammar when synthesizing
each gate set. For the \ibm gate set, each parameter is only allowed to be used once in a circuit. We limit circuits with symbolic gates to have
at most 2 qubits. We found that using symbolic rules 
of size at most 3 resulted in the best performance. The maximum sizes of non-symbolic rules used are shown in \cref{table:rule_gen}.
For the Ion gate set, we use a cost function that excludes $\rz$ gates from the total gate count.

\begin{table*}[h!]
    \caption{Parameters allowed in synthesis}
    \centering
    \footnotesize
      \begin{tabular}{l l}
        \toprule
        Gate set & Parameters \\
        \midrule
        \ibm & $\theta_1$, $\theta_2$, $\theta_3$, $\theta_1 + \theta_2$, $\theta_1 + \theta_2 + \theta_3$\\ 
        Nam & $\theta_1$, $\theta_2$, $\theta_1 + \theta_2$\\ 
        Rigetti & $\theta_1$, $\theta_2$, $\theta_1 + \theta_2$, $-\theta_1$\\\
        Ion & $\theta_1$, $\theta_2$, $\theta_1 + \theta_2$, $-\theta_1$, $\pi$, $\frac{\pi}{2}$\\
        \bottomrule
      \end{tabular}
      \label{table:rule_gen_angles}
  \end{table*}

\paragraph{Invoking other tools} We invoked \voqc, \tket, Quilc, and Qiskit using their Python interfaces. We ran Quartz 
using the same parameters from \cite{xu2022quartz}. Because Quartz behaves nondeterministically 
for the \textsf{mod5\_4} benchmark on the Nam gate set, we report an average of 7 runs.

\paragraph{Hardware}
All benchmarks were executed on a cluster of Intel\textsuperscript{\textregistered} Xeon\textsuperscript{\textregistered}, AMD EPYC\texttrademark \xspace, and AMD Opteron\texttrademark \xspace CPUs clocked an average of 2.4GHz.

\paragraph{Calculating fidelity}
We use the following publically reported gate fidelities for each device where $f_1$ and $f_2$ are the fidelities for single and two-qubit gates, respectively:
\begin{description}
    \item[\ibm Toronto \cite{toronto}] $f_1=0.999606$, $f_2=0.98719$
    \item[Rigetti Aspen-11 \cite{aspen11}] $f_1=0.998$, $f_2=0.902$
    \item[IonQ Aria \cite{aria}] $f_1=0.9995$, $f_2=0.996$
\end{description}

\section{Proofs}

\subsection{\cref{lemma:eq1}}
Follows directly from the definition of equivalence.

\subsection{\cref{lemma:poly}}
By contradiction:
Suppose \cref{eq:sum} does not hold but $C_1$ and $C_2$ are equivalent.
By \cref{lemma:eq1}, we know that $\psi_1^{\vec{x}}(\vec{y},\param) = \psi_2^{\vec{x}}(\vec{y},\param)$,
but there must be  a value $c$ such that $c \cdot \psi_1^{\vec{x}}(\vec{y},\param) \neq c \cdot \psi_2^{\vec{x}}(\vec{y},\param) $. Contradiction.

Conversely, suppose that \cref{eq:sum} holds but $C_1$ and $C_2$ are not equivalent. 
By \cref{lemma:eq1}, this means that there is a valuation of the variables $\vec{x},\vec{y},...$ such that $\psi_1^{\vec{x}}(\vec{y},\param) \neq \psi_2^{\vec{x}}(\vec{y},\param)$.
Then, for $\cvar_{\vec{x},\vec{y}} \neq 0$, we have $\cvar_{\vec{x},\vec{y}}\psi_1^{\vec{x}}(\vec{y},\param) \neq \cvar_{\vec{x},\vec{y}} \psi_2^{\vec{x}}(\vec{y},\param)$.
Hence, \cref{eq:sum} does not hold.

\subsection{\cref{thm:trans}}

We will prove this by way of the contrapositive. The idea is that the transformation induces a bijective correspondence between counterexamples to equivalence. Suppose first that $\circuit_1 \not\equiv \circuit_2$, where the amplitudes for 
$\circuit_1$ are given by $$\psi_1^a(\vec{b}, \vparam) = c_{\vec{a},\vec{b}}\prod_{i=1}^p (\phi^u_i(\vec{a},\vec{b}))^{m_i}\prod_{j=1}^k (e^{i\theta_j})^{n_i}$$ and the amplitudes for $\circuit_2$ are given by 
$$\psi_2^a(\vec{b}, \vparam) = c'_{\vec{a},\vec{b}}\prod_{i=1}^p (\phi^u_i(\vec{a},\vec{b}))^{m'_i}\prod_{j=1}^k (e^{i\theta_j})^{n'_i}$$
 This means that there is some $\vec{a},\vec{b} \in \Zn$, interpretation of the amplitude transformers $\hat{\phi}^u_1, \ldots, \hat{\phi}^u_p$ and value of the parameters $\hat{\theta}_1 \ldots \hat{\theta}_k$ such that 
\begin{align*}
    c_{\vec{a},\vec{b}}\prod_{i=1}^p (\hat{\phi}^u_i(\vec{a},\vec{b}))^{m_i}\prod_{j=1}^k (e^{i\hat{\theta}_j})^{n_i} \neq c'_{\vec{a},\vec{b}}\prod_{i=1}^p (\hat{\phi}^u_i(\vec{a},\vec{b}))^{m'_i}\prod_{j=1}^k (e^{i\hat{\theta}_j})^{n'_i}
\end{align*}
Therefore, there is a point such that we have the following inequality between polynomials. 

$$\sum_{\vec{x},\vec{y} \in \Zn} c_{\vec{x},\vec{y}} \cvar_{\vec{x},\vec{y}} \prod_{i=1}^p (\phi^u_{i,\vec{x},\vec{y}})^{m_i}\prod_{j=1}^k (\cvar_{\theta_j})^{n_i} \neq  \sum_{\vec{x},\vec{y} \in \Zn} c'_{\vec{x},\vec{y}} \cvar_{\vec{x},\vec{y}} \prod_{i=1}^p (\phi^u_{i, \vec{x},\vec{y}})^{m'_i}\prod_{j=1}^k (\cvar_{\theta_j})^{n'_i}$$

Namely, set $\phi^u_{i,\vec{a},\vec{b}} = \hat{\phi}^u_i(\vec{a},\vec{b})$, $v_{\theta_j} = e^{i\hat{\theta}_j}$, and $\cvar_{\vec{a},\vec{b}} = 1$ and all other variables to 0.

The other direction is similar. Given a point $\hat{\phi}^u_{i,\vec{a},\vec{b}}$, $\hat{\cvar}_{\theta_j}$, and $\hat{\cvar}_{\vec{a},\vec{b}}$ such that the polynomials are not equal,
choose some $\vec{a}, \vec{b}$ such that the corresponding terms evaluate to different values at this point. Then define an interpretation of the amplitude transformers by $\phi_i^u(\vec{a}, \vec{b}) = \hat{\phi}^u_{i,\vec{a},\vec{b}}$. Recall that each $\hat{\cvar}_{\theta_j}$ is in $\unitc$. Let $\theta_j = \hat{\theta}_j$ where $\hat{\theta}_j$ is the unique value in $[0, 2\pi)$ satisfying $\cos \hat{\theta}_j = Re(\hat{\cvar}_{\theta_j})$ and $\sin \hat{\theta}_j = Im(\hat{\cvar}_{\theta_j})$.

\subsection{\cref{thm:match_sym}}

Let $\sem{S(I)}$ be of the form $\phi^u(\vec{x}) \ket{f(\vec{x})}$ where $n = |\vec{x}|$. That is, it has an interpreted 
state transformer but an uninterpreted amplitude transformer and operates over $n$ qubits. Let $\sem{C_S}$ be of the form 
$\phi(\vec{x}\ldots) \ket{f(\vec{x}\ldots)}$ where both the state and amplitude transformers are interpreted and $C_S$  
operates over $k = |\vec{x}\ldots| \geq n$ qubits. Consider the symbolic rewrite rule $C_l; S; C_l' \to C_r;S;C_r'$.
From \cref{alg:synth}, this rule would only be synthesized if $C_l; S; C_l' \equiv C_r;S;C_r'$. 
Recall that any circuit that matches the state transformer of $S$ will satisfy this equivalence because 
the amplitude transformer is uninterpreted and could therefore be anything.
Observe that (1) a state transformer need only match $S$ for the $n$ qubits in $S$ for the equivalence to hold
because $C_l$, $C_l'$, $C_r$, and $C_r'$ only operate on at most those $n$ qubits
and (2) the equivalence for the $k-n$ qubits in $C_S$ but not $S$ have their equivalence trivially preserved
because only $C_S$ operates on those qubits. Combining (1) and (2), we get $C_l; C_S; C_l' \equiv C_r;C_S;C_r'$.

\section{Experimental results continued}

\subsection{How does \ours compare to other optimizers in reducing T gate count?}
\label{app:tcount}

\begin{figure*}[t!]
  \centering
  \includegraphics[width=0.9\textwidth]{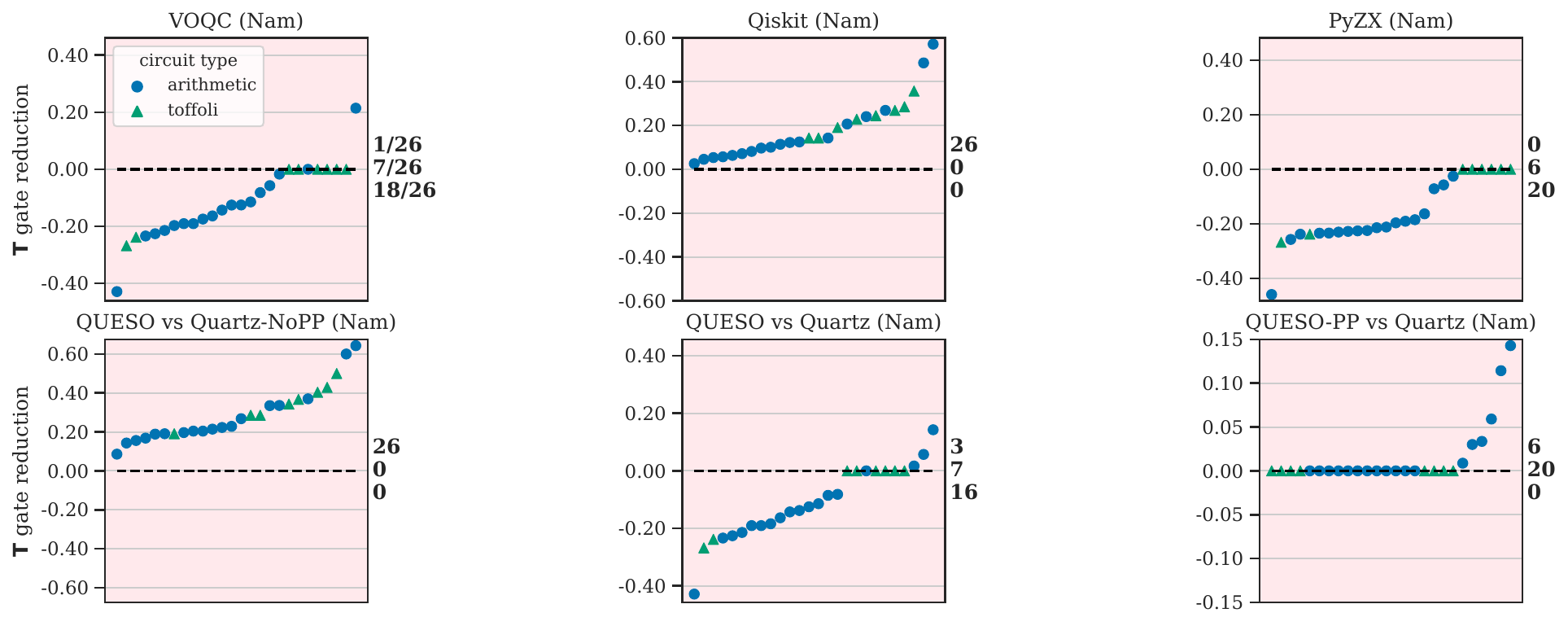}
  \caption{Comparison against other optimizers with respect to T gate reduction on the Nam gate set.} 
  \label{fig:rq4_tcount}
\end{figure*}

\paragraph{Results} We compare against other optimizers with respect to T gate reduction using the Nam gate set. 
Although T gates can be efficiently simulated in the NISQ setting, they are the primary source of bottlenecks in fault tolerant quantum computing.
\cref{fig:rq4_tcount} shows the results against other optimizers. Notably, we include PyZX now because its primary goal is to reduce T gate count. 
\ours is not able to outperform other optimizers using a dedicated rotation merging pass despite being able to synthesize a symbolic rule capturing the
essence of rotation merging. 

\paragraph{\emph{Summary}} \textbf{With respect to T gate reduction, \ours outperforms or matches optimizers that do not use a dedicated rotation merging pass and underperforms optimizers that do.}

\begin{figure*}[h]
    \centering
    \includegraphics[width=\textwidth]{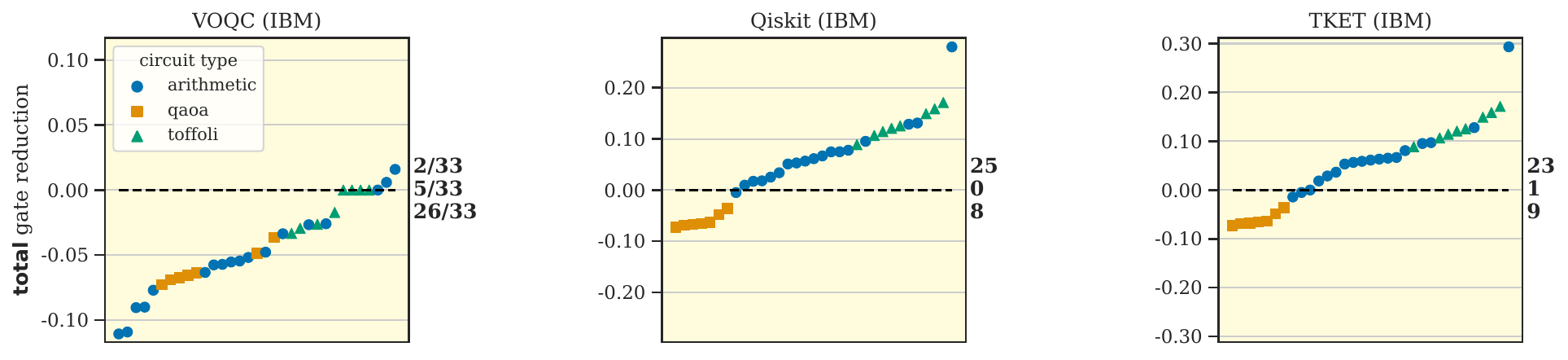}
    \caption{ Comparison against state-of-the-art optimizers for the \ibm gate set with respect to total gate count reduction.} 
    \label{fig:rq1_ibm_total}
\end{figure*}

\begin{figure*}[h]
    \centering
    \begin{subfigure}[b]{0.66\textwidth}
        \caption{Two-qubit gate reduction} 
        \centering
        \includegraphics[width=\textwidth]{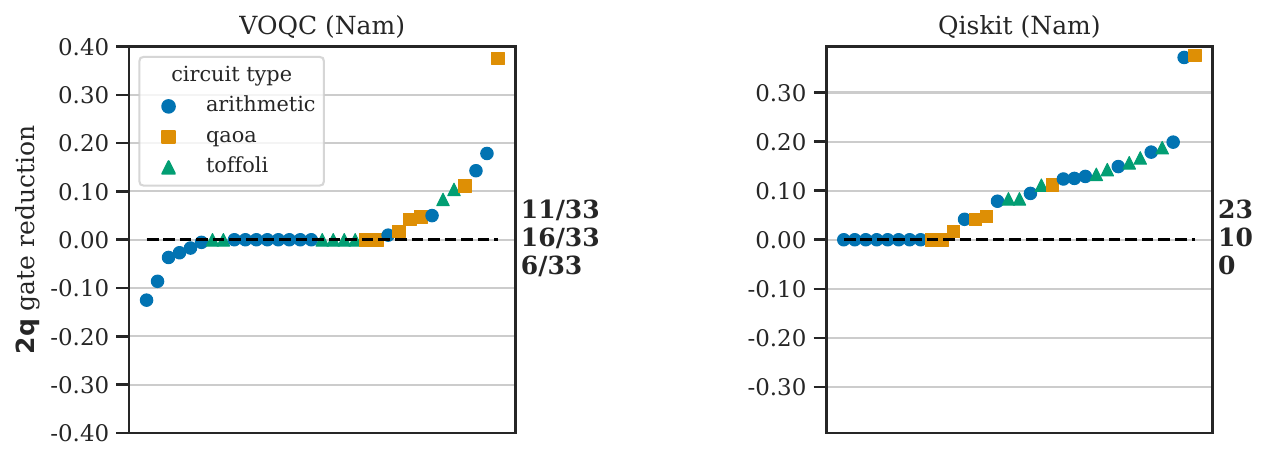}
    \end{subfigure}

    \begin{subfigure}[b]{0.66\textwidth}
        \caption{Total gate reduction} 
        \centering
        \includegraphics[width=\textwidth]{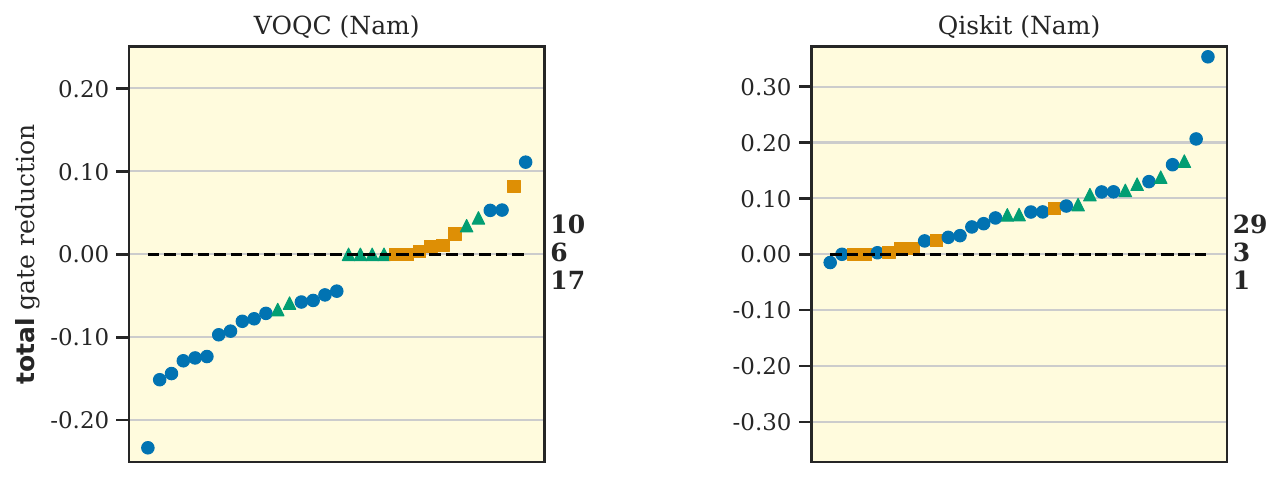}
    \end{subfigure}
    \caption{ Comparison against state-of-the-art optimizers for the Nam gate set. }
    \label{fig:rq1_nam}
\end{figure*}

\begin{figure*}[h]
    \centering
    \includegraphics[width=\textwidth]{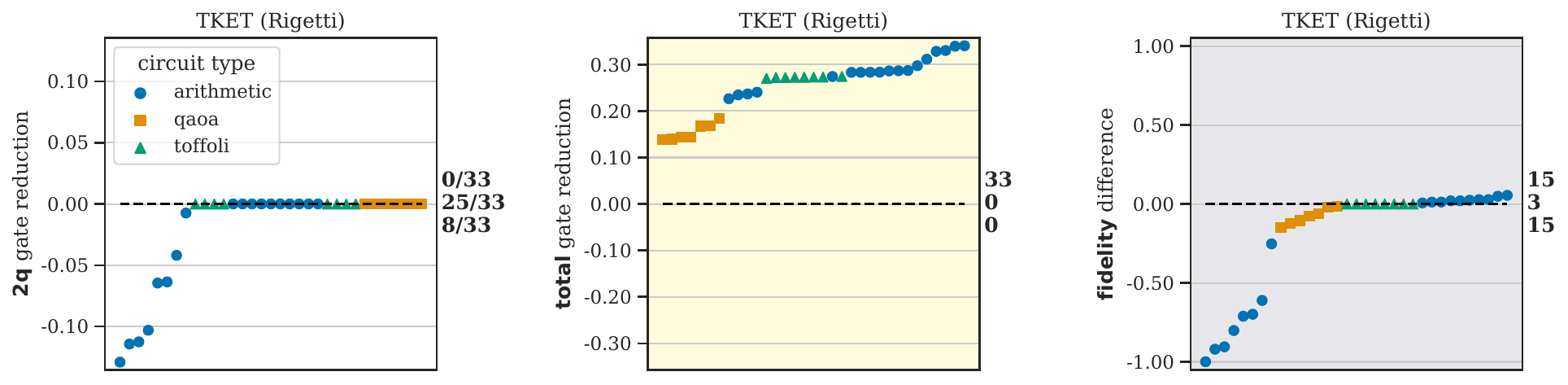}
    \caption{\tket on Rigetti gate set} 
    \label{fig:rq1_rigetti_tket}
\end{figure*}

\begin{figure*}[h]
    \centering
    \includegraphics[width=0.66\textwidth]{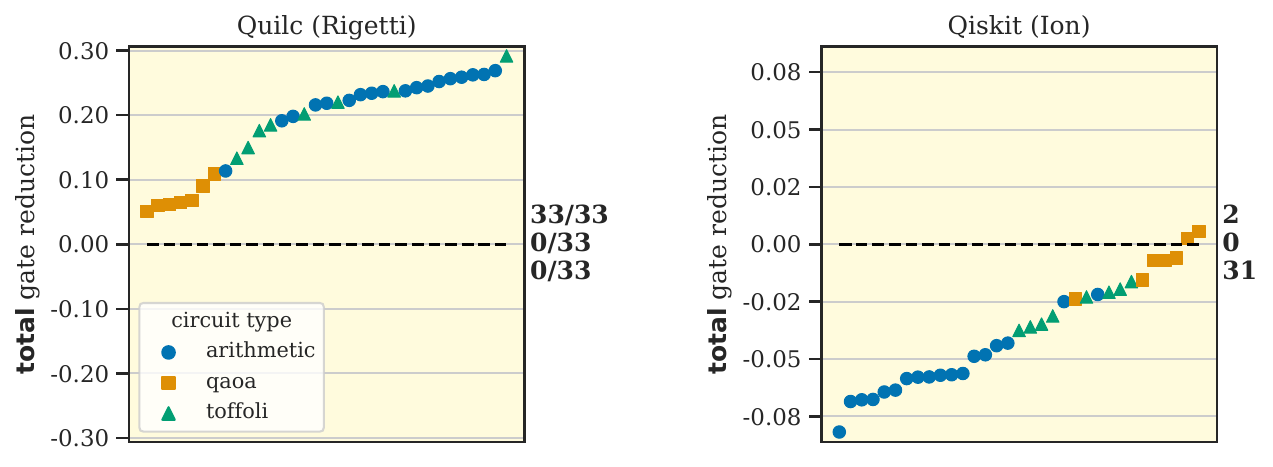}
    \caption{Total gate reduction on Rigetti gate set compared to Quilc and Ion gate set compared to Qiskit. The cost function for the Ion gate set is excluding $\rz$ gates. } 
    \label{fig:rq1_rigetti_quilc_ion_total}
\end{figure*}

\begin{figure*}[h]
    \centering
    \includegraphics[width=0.33\textwidth]{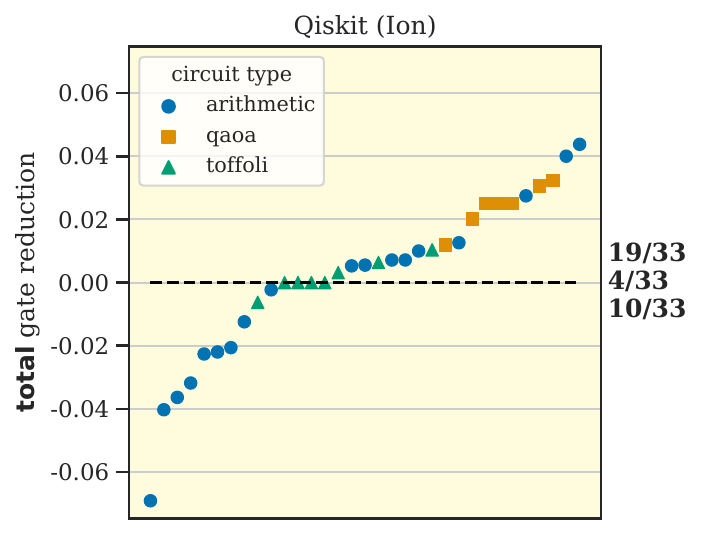}
    \caption{Total gate reduction on Ion gate set with cost function including $\rz$ gates.} 
    \label{fig:rq1_ion_normal}
\end{figure*}

\begin{figure*}[h]
    \centering
    \includegraphics[width=\textwidth]{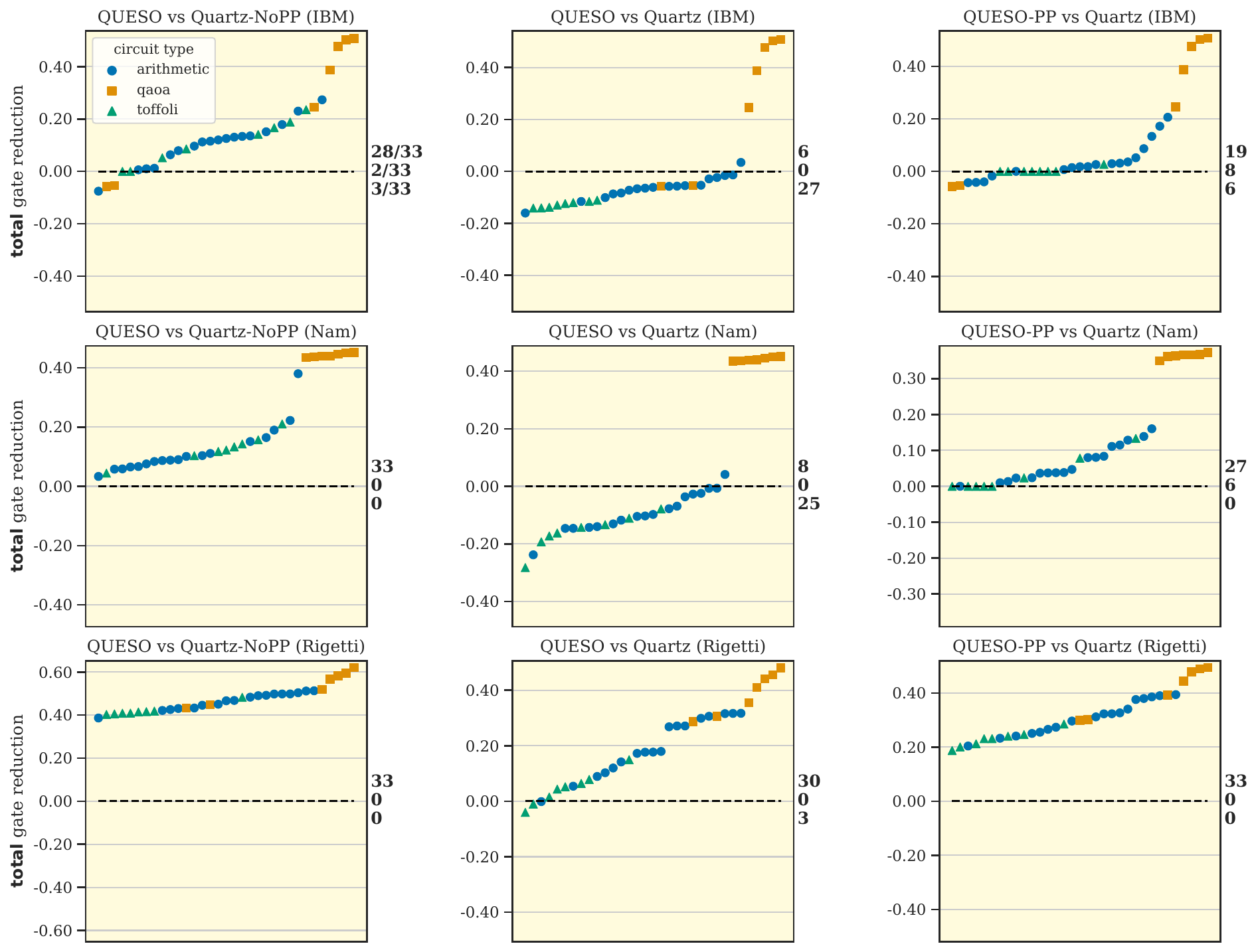}
    \caption{\textbf{1 hour timeout} comparison against Quartz on reduction in total gate count with preprocessing disabled and enabled. 
    \ourspp is our tool run on the result of Quartz's preprocessing.} 
    \label{fig:rq2_total}
\end{figure*}

\begin{figure*}[h]
    \centering
    \includegraphics[width=\textwidth]{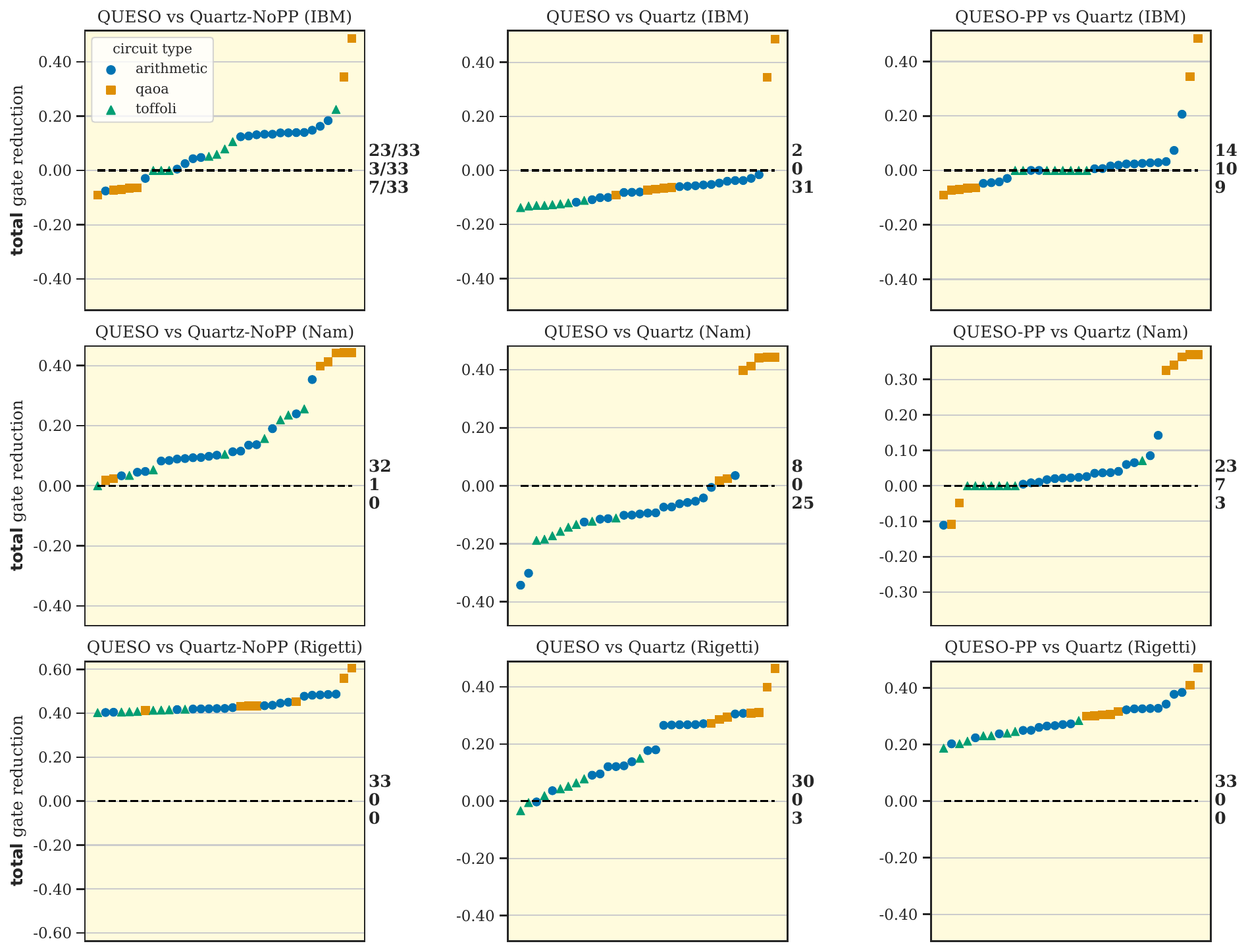}
    \caption{\textbf{24 hour timeout} comparison against Quartz on reduction in total gate count with preprocessing disabled and enabled. 
    \ourspp is our tool run on the result of Quartz's preprocessing.} 
    \label{fig:rq2_total_24hr}
\end{figure*}

\begin{figure*}[h]
    \centering
    \includegraphics[width=\textwidth]{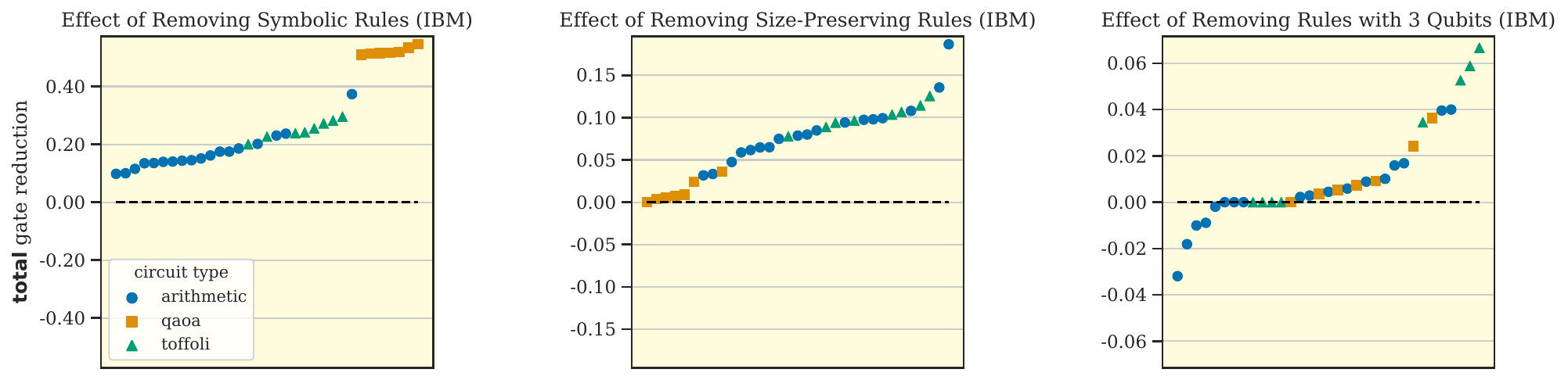}
    \caption{Comparison between running \ours using a baseline set of rules and various subsets for total gate reduction. } 
    \label{fig:rq3_toggles_total}
\end{figure*}

\begin{figure*}[h]
    \centering
    \includegraphics[width=\textwidth]{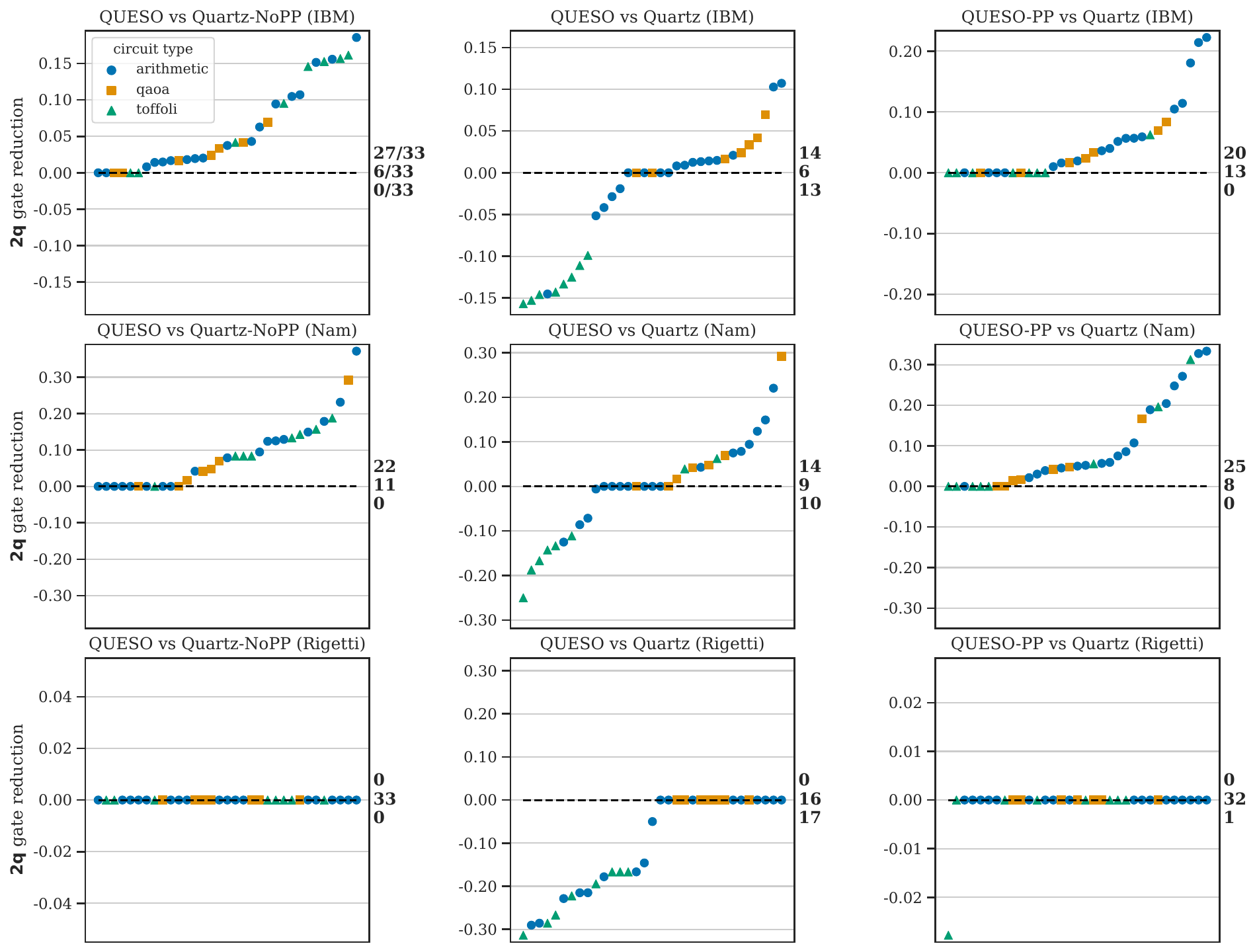}
    \caption{\textbf{2 hour timeout} comparison against Quartz.} 
    \label{fig:rq2_2hr}
\end{figure*}

\begin{figure*}[h]
    \centering
    \includegraphics[width=\textwidth]{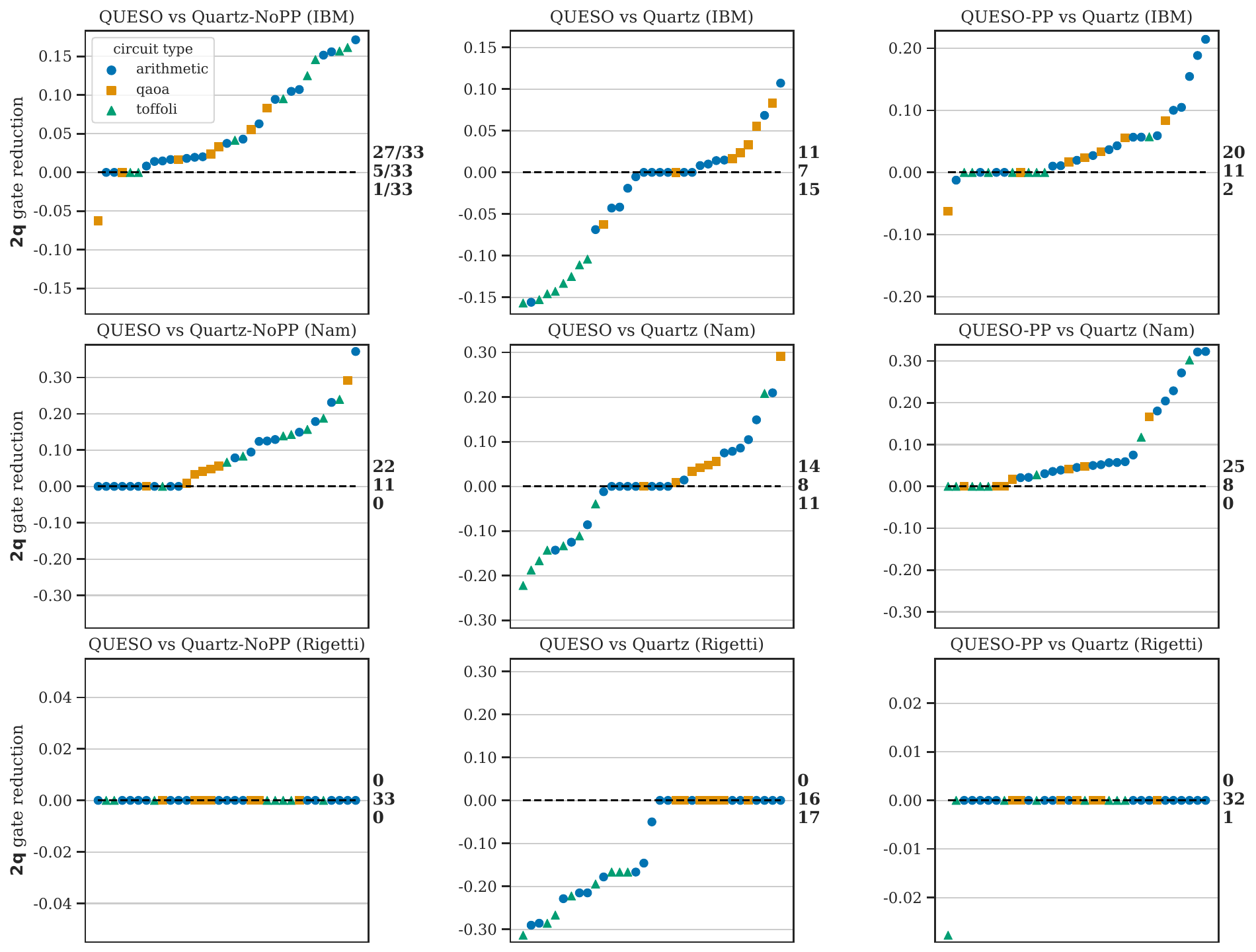}
    \caption{\textbf{3 hour timeout} comparison against Quartz.} 
    \label{fig:rq2_3hr}
\end{figure*}

\begin{figure*}[h]
    \centering
    \includegraphics[width=\textwidth]{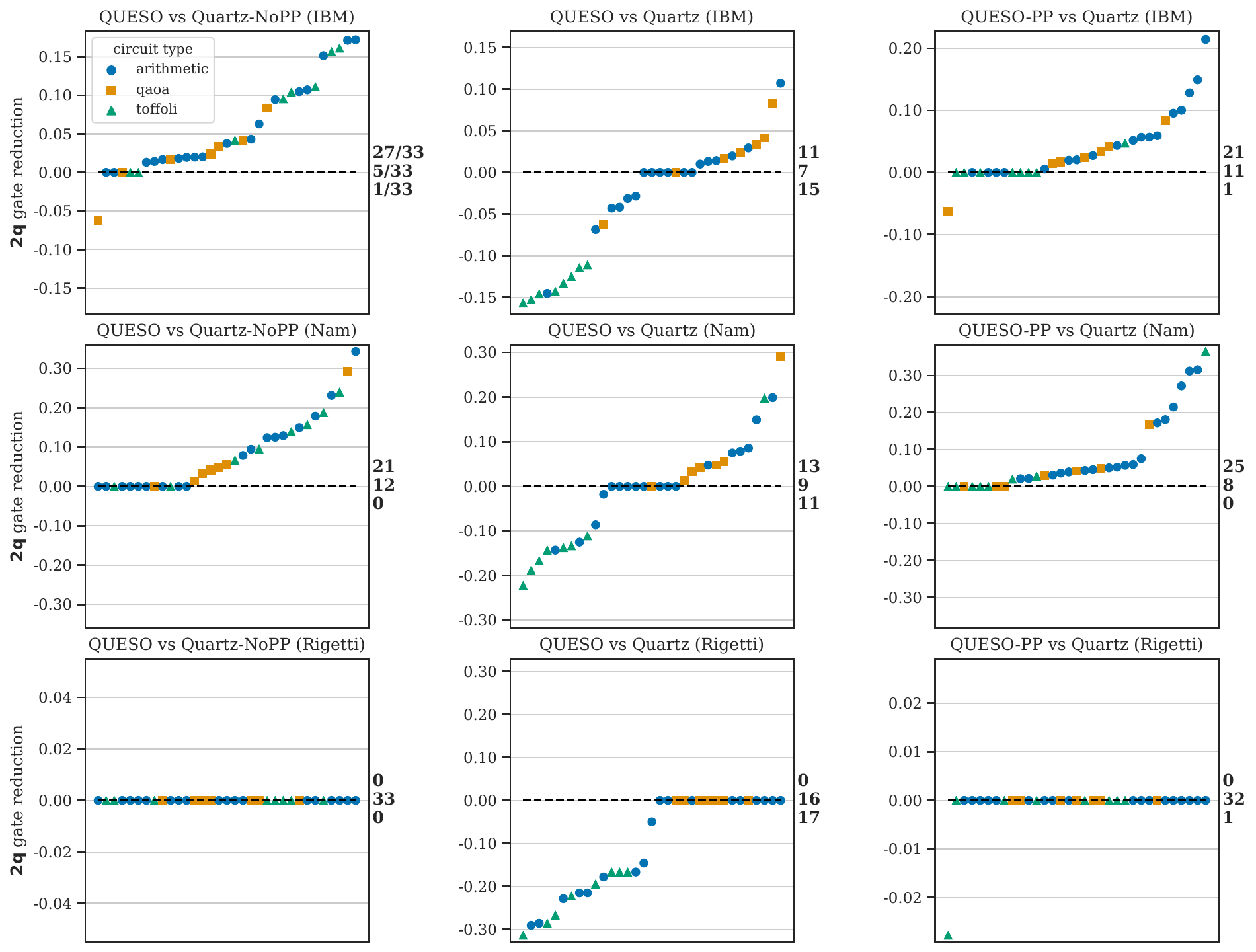}
    \caption{\textbf{4 hour timeout} comparison against Quartz.} 
    \label{fig:rq2_4hr}
\end{figure*}

\begin{figure*}[h]
    \centering
    \includegraphics[width=\textwidth]{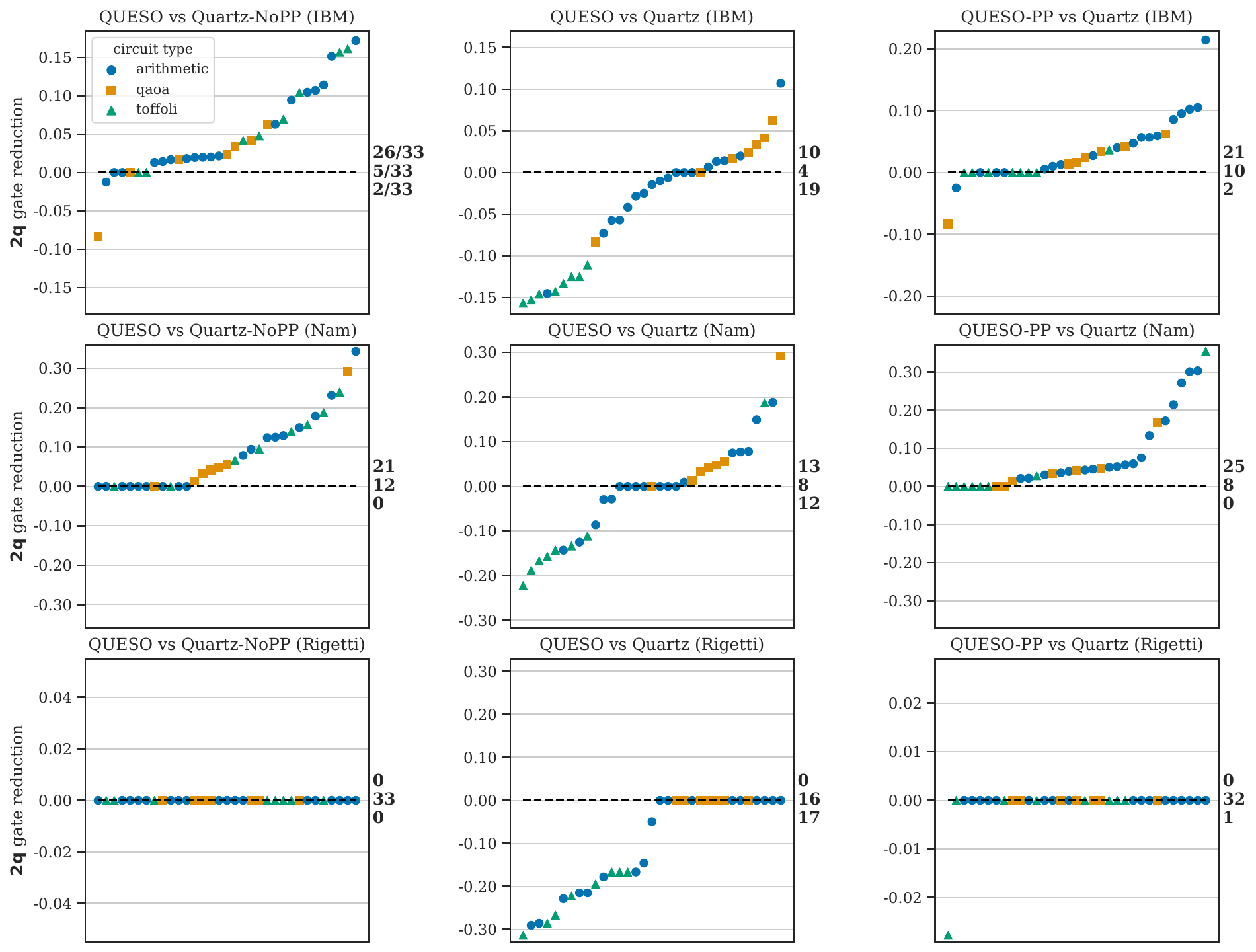}
    \caption{\textbf{5 hour timeout} comparison against Quartz.} 
    \label{fig:rq2_5hr}
\end{figure*}

\begin{figure*}[h]
    \centering
    \includegraphics[width=\textwidth]{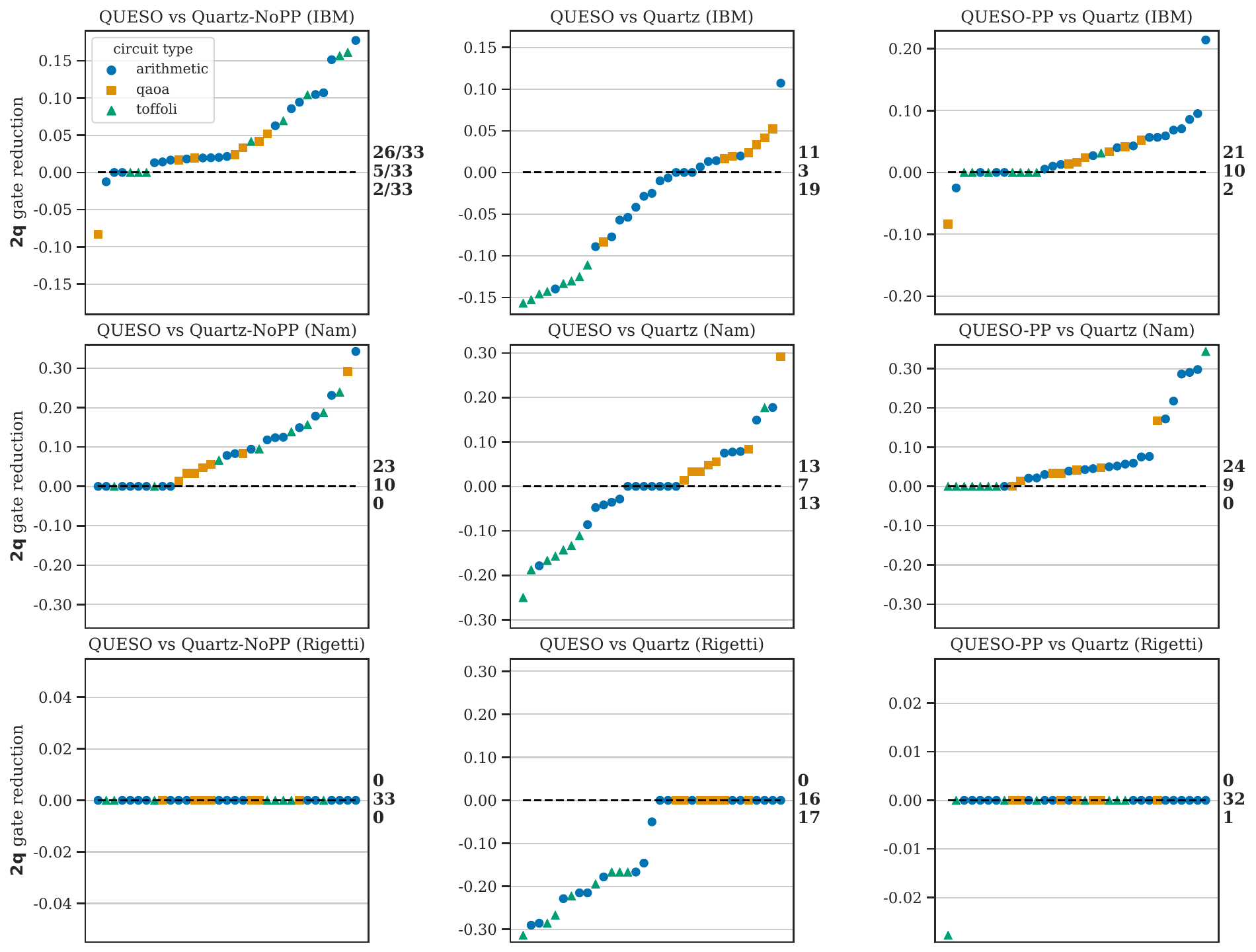}
    \caption{\textbf{6 hour timeout} comparison against Quartz.} 
    \label{fig:rq2_6hr}
\end{figure*}

\begin{figure*}[h]
    \centering
    \includegraphics[width=\textwidth]{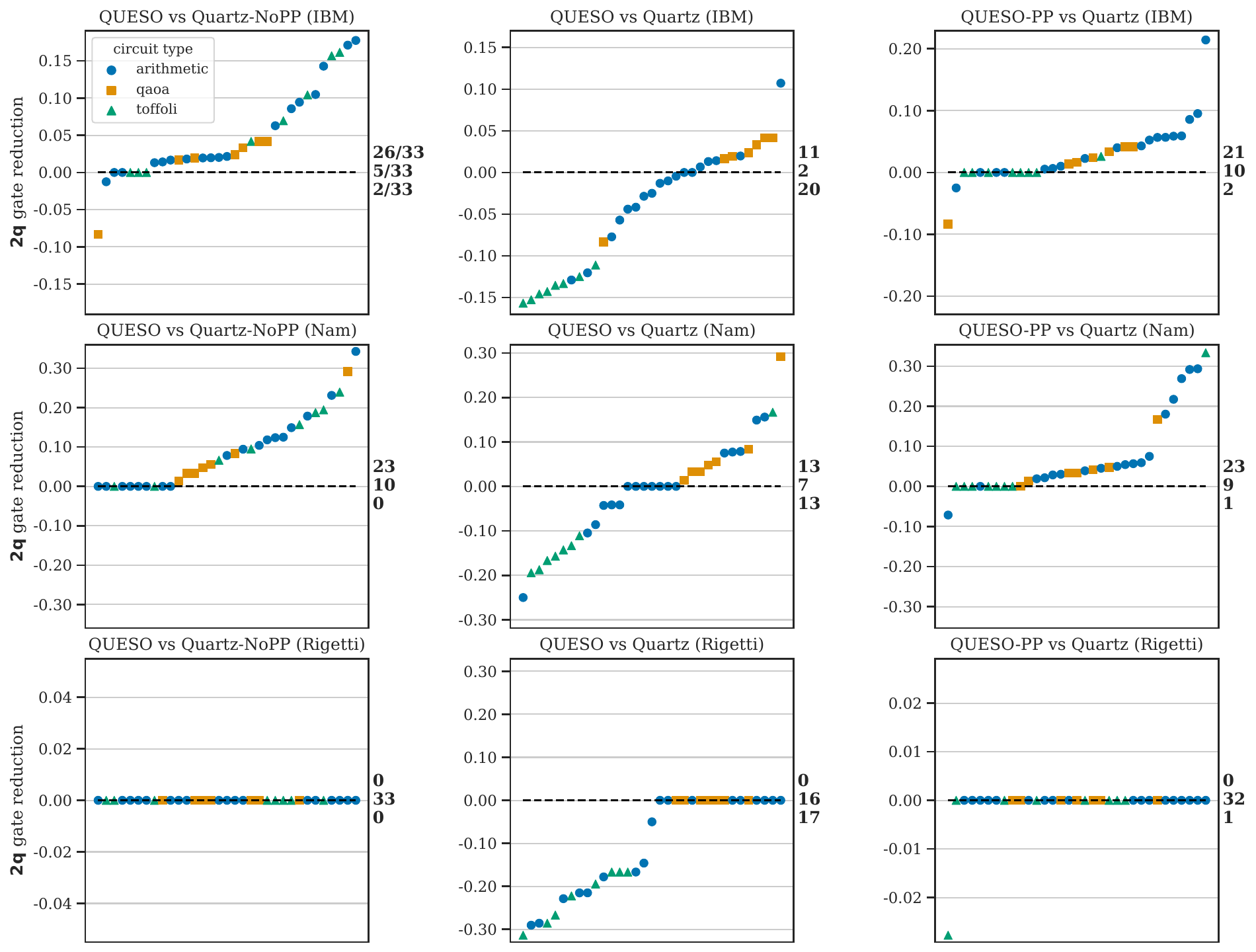}
    \caption{\textbf{7 hour timeout} comparison against Quartz.} 
    \label{fig:rq2_7hr}
\end{figure*}

\begin{figure*}[h]
    \centering
    \includegraphics[width=\textwidth]{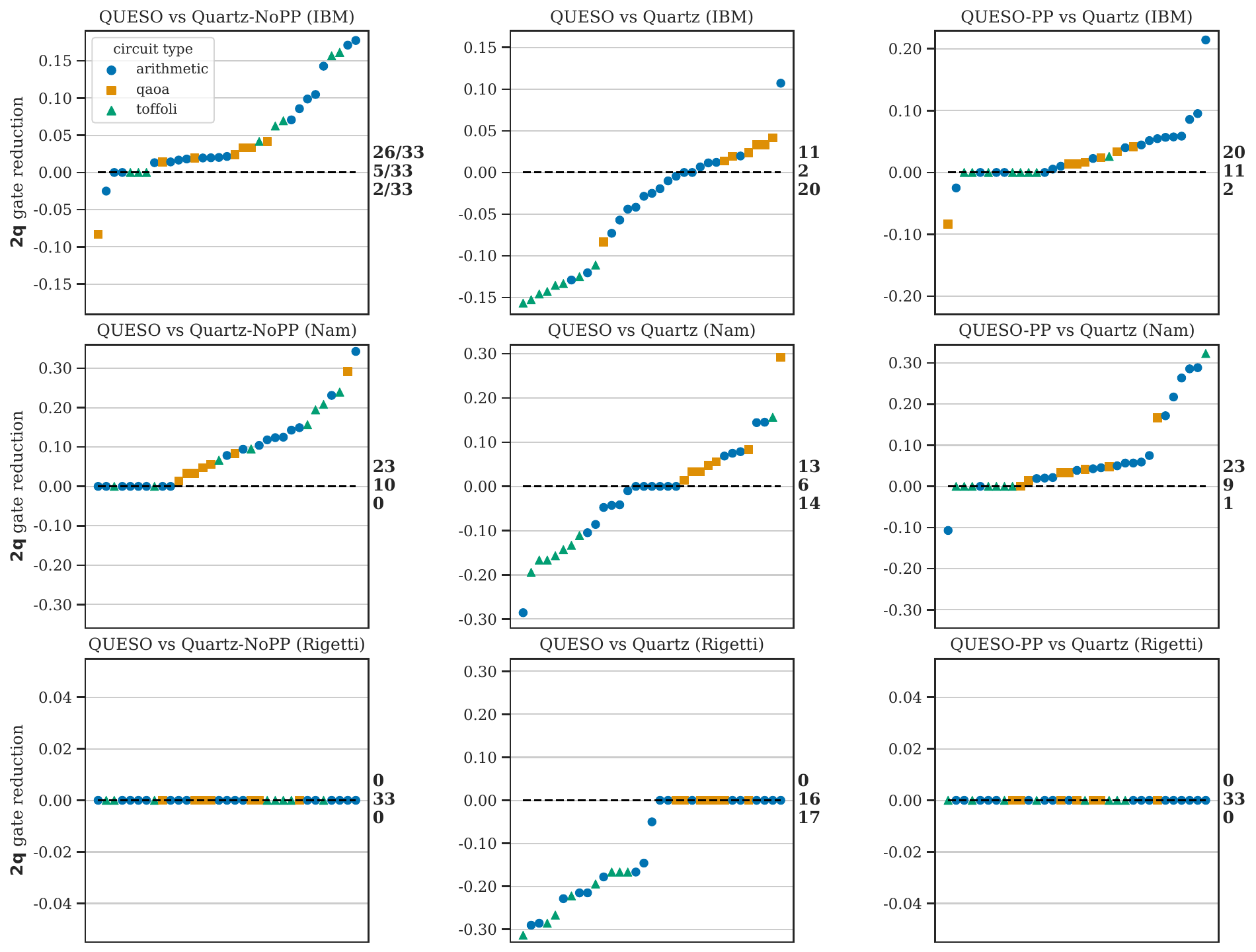}
    \caption{\textbf{8 hour timeout} comparison against Quartz.} 
    \label{fig:rq2_8hr}
\end{figure*}

\begin{figure*}[h]
    \centering
    \includegraphics[width=\textwidth]{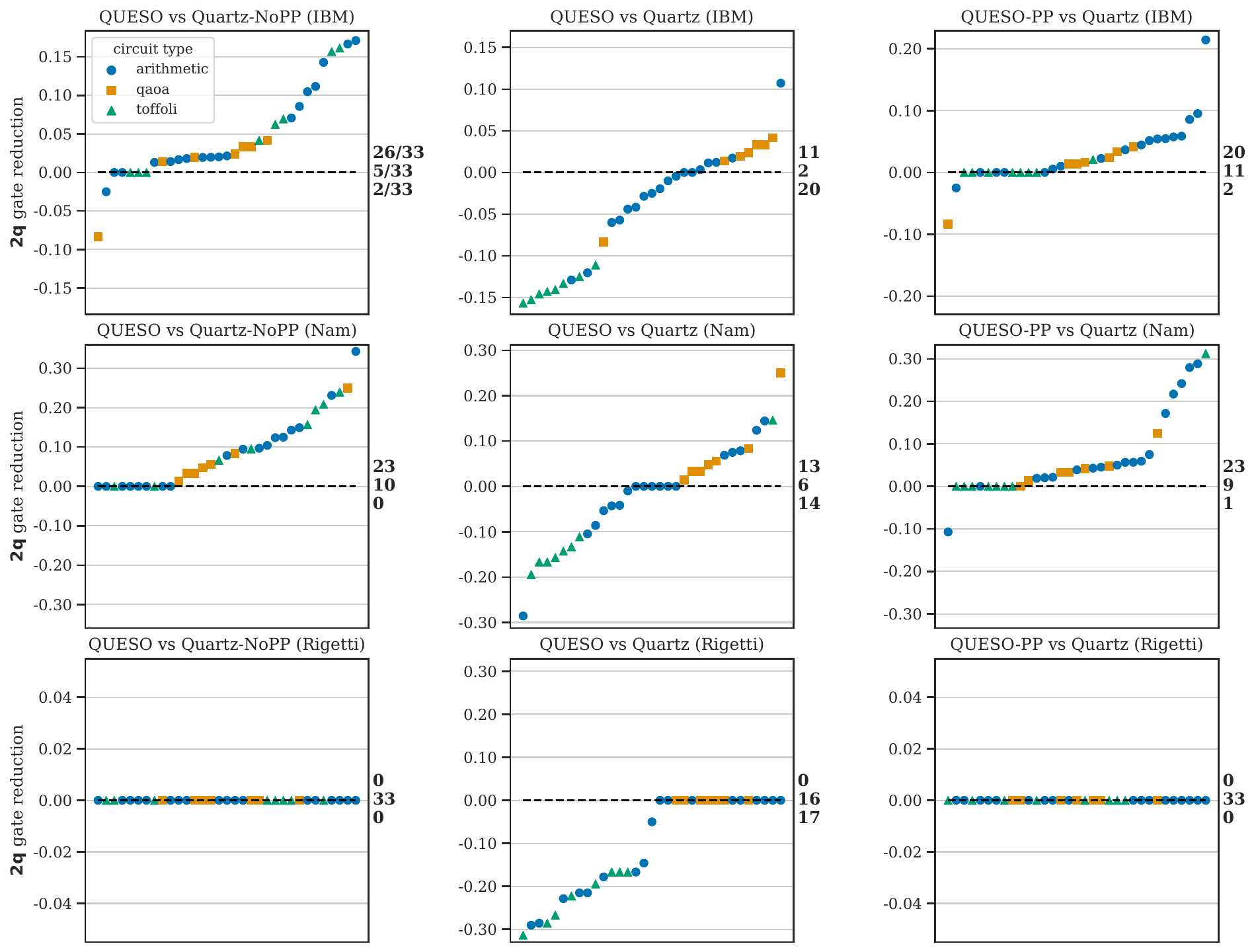}
    \caption{\textbf{9 hour timeout} comparison against Quartz.} 
    \label{fig:rq2_9hr}
\end{figure*}

\begin{figure*}[h]
    \centering
    \includegraphics[width=\textwidth]{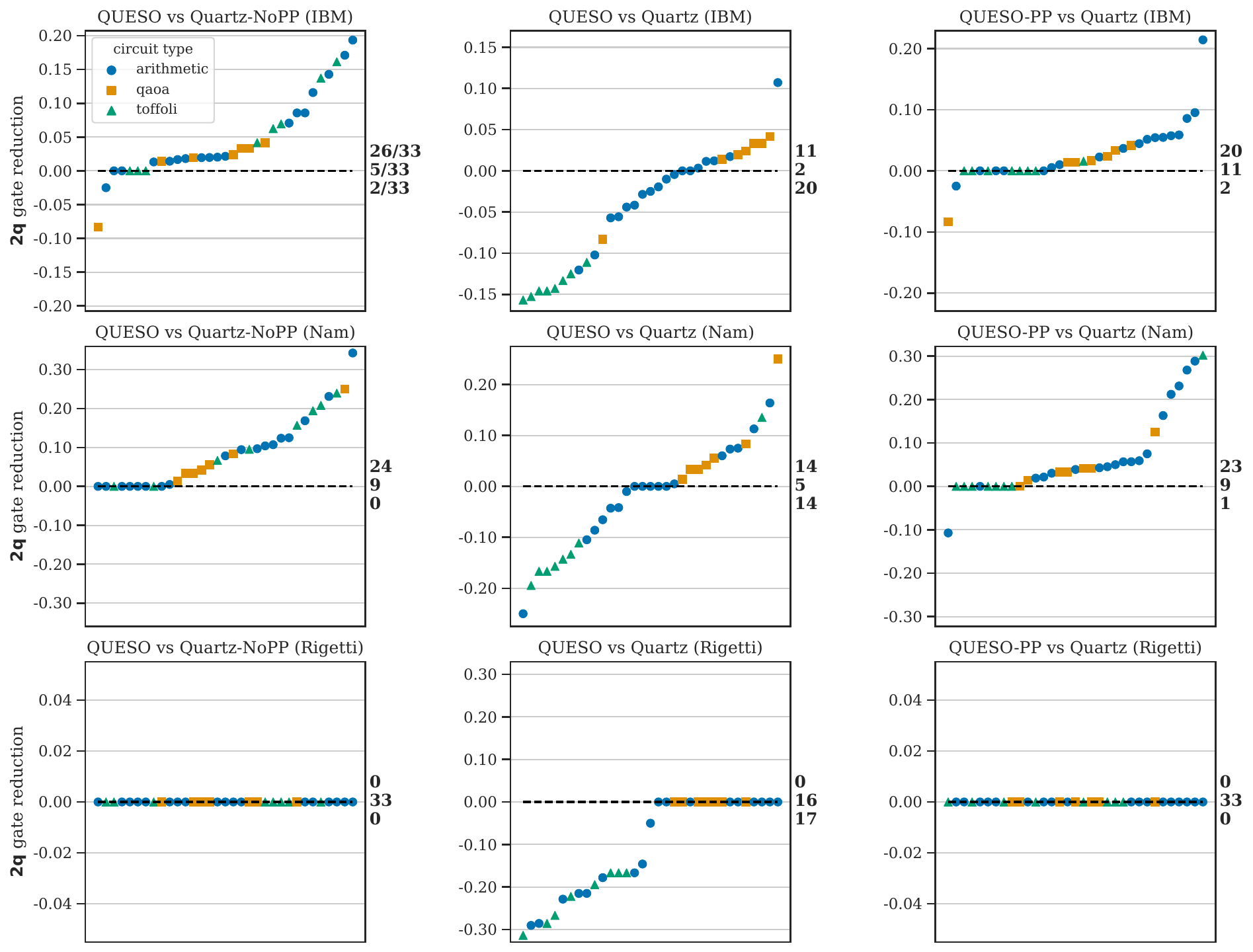}
    \caption{\textbf{10 hour timeout} comparison against Quartz.} 
    \label{fig:rq2_10hr}
\end{figure*}

\begin{figure*}[h]
    \centering
    \includegraphics[width=\textwidth]{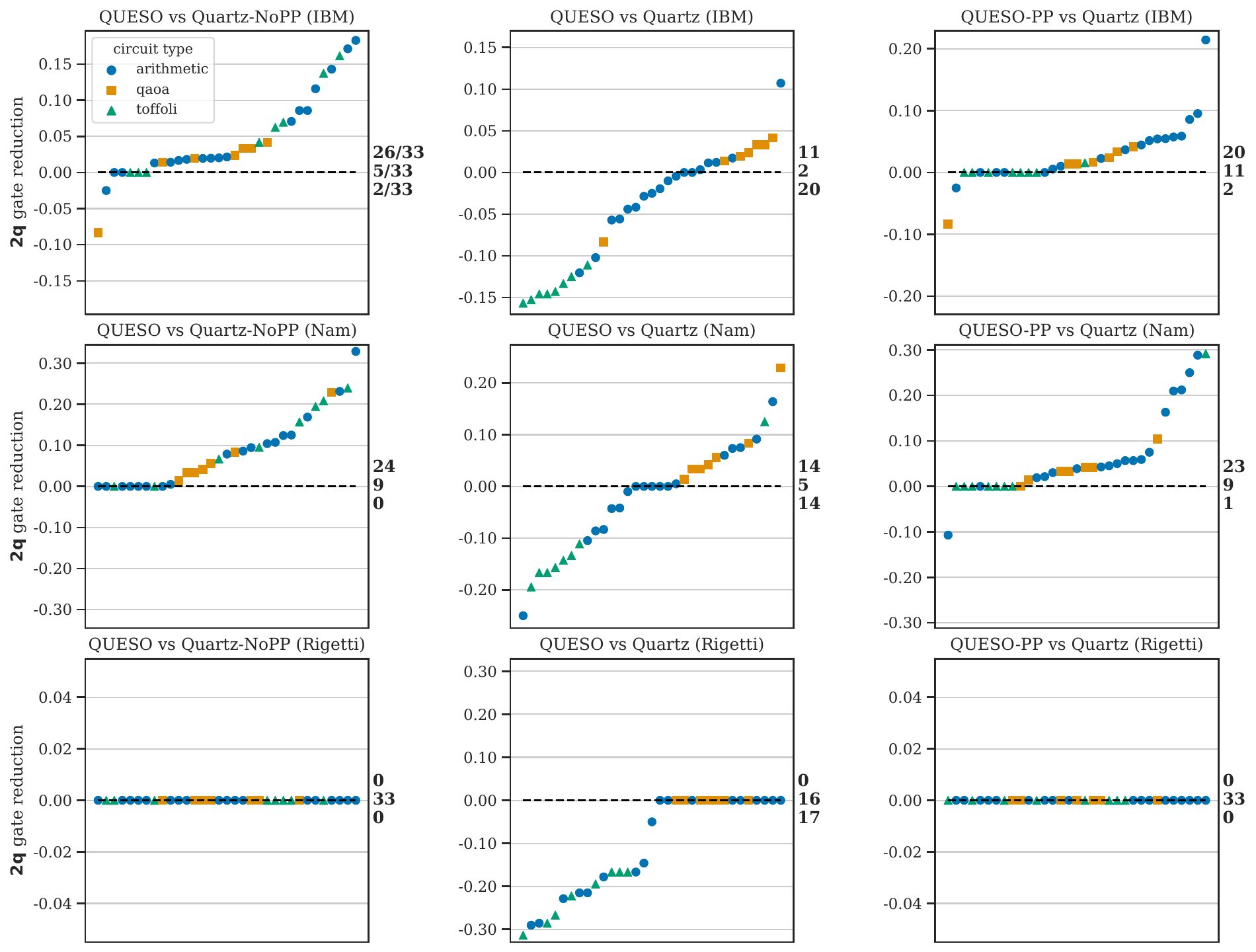}
    \caption{\textbf{11 hour timeout} comparison against Quartz.} 
    \label{fig:rq2_11hr}
\end{figure*}

\begin{figure*}[h]
    \centering
    \includegraphics[width=\textwidth]{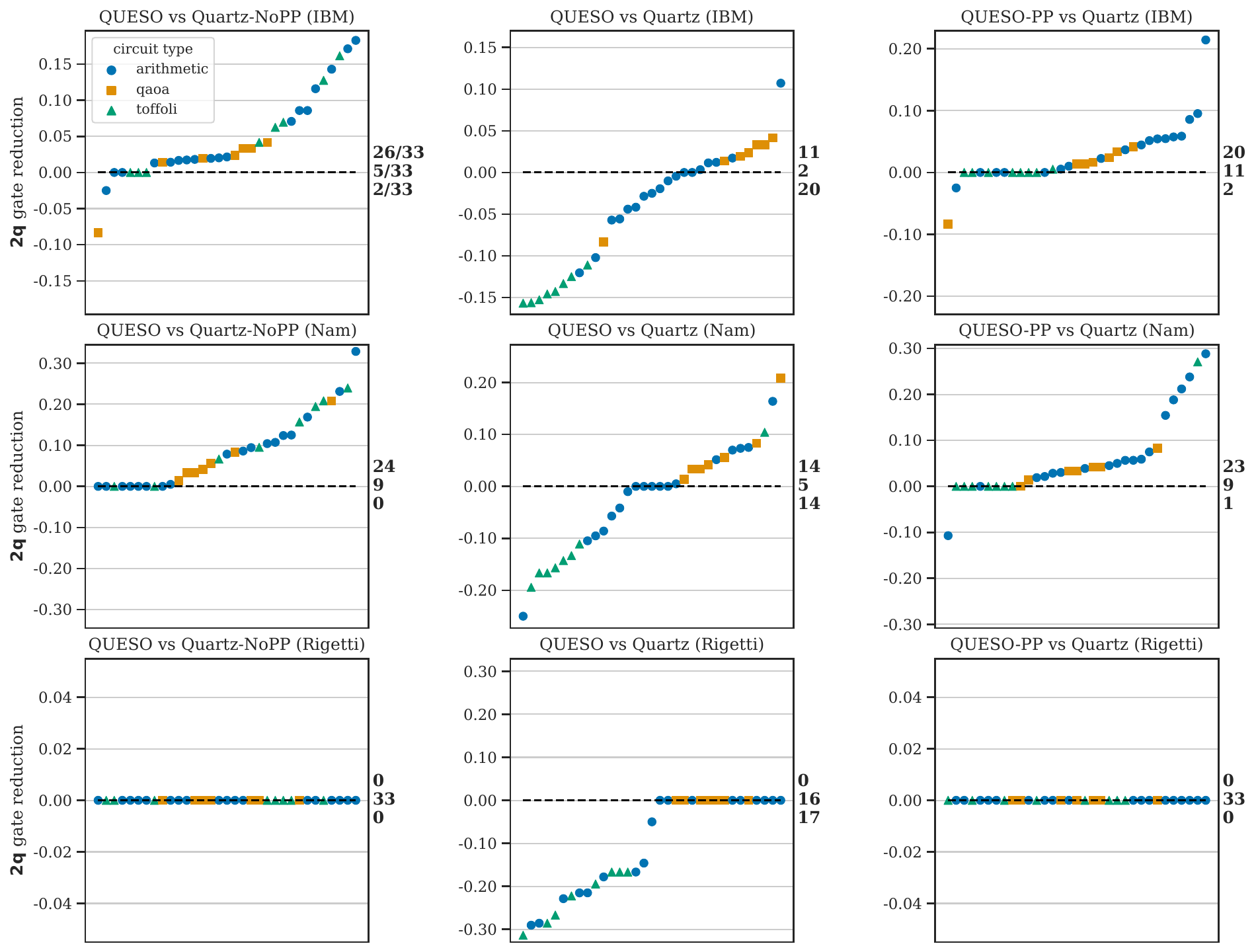}
    \caption{\textbf{12 hour timeout} comparison against Quartz.} 
    \label{fig:rq2_12hr}
\end{figure*}

\begin{figure*}[h]
    \centering
    \includegraphics[width=\textwidth]{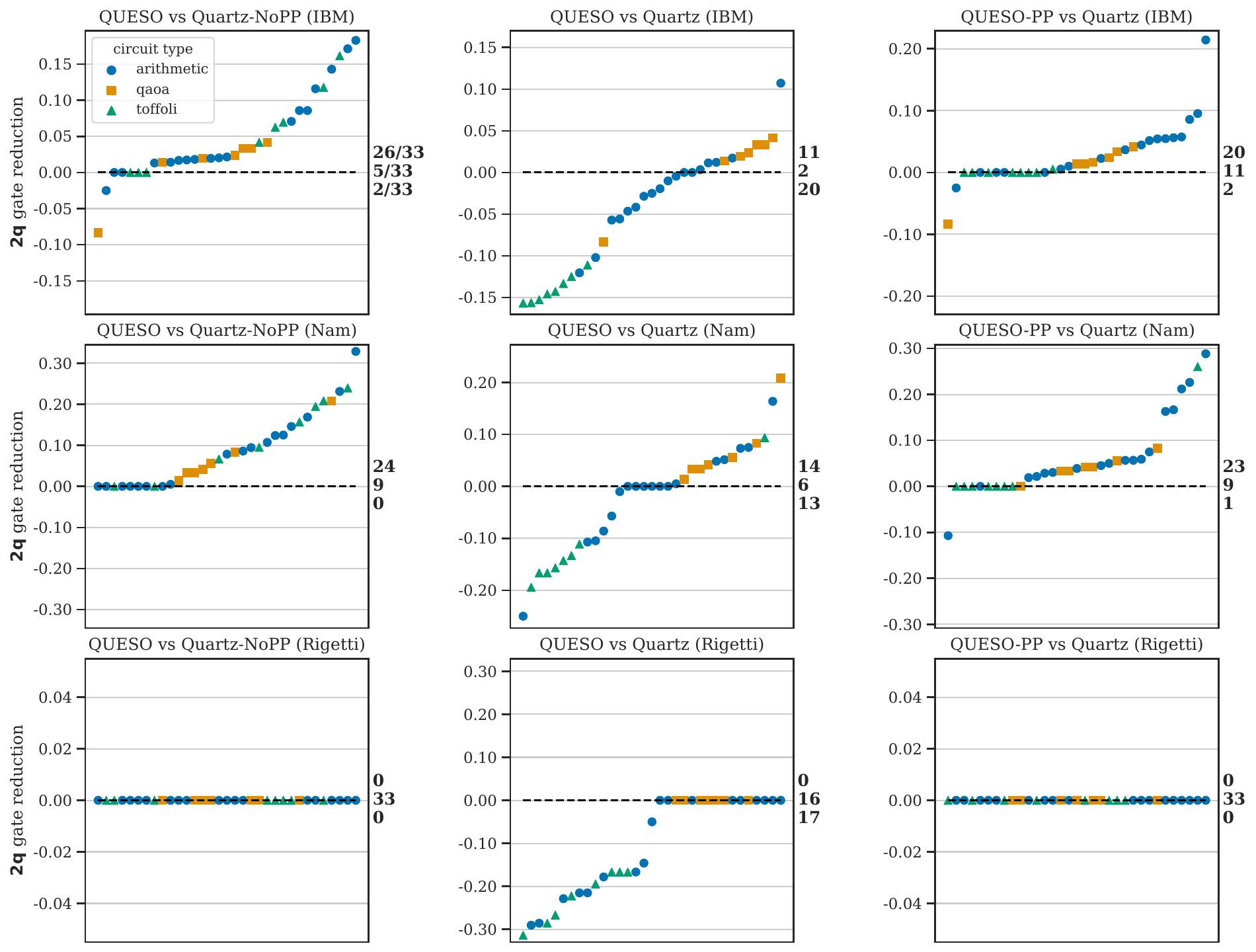}
    \caption{\textbf{13 hour timeout} comparison against Quartz.} 
    \label{fig:rq2_13hr}
\end{figure*}

\begin{figure*}[h]
    \centering
    \includegraphics[width=\textwidth]{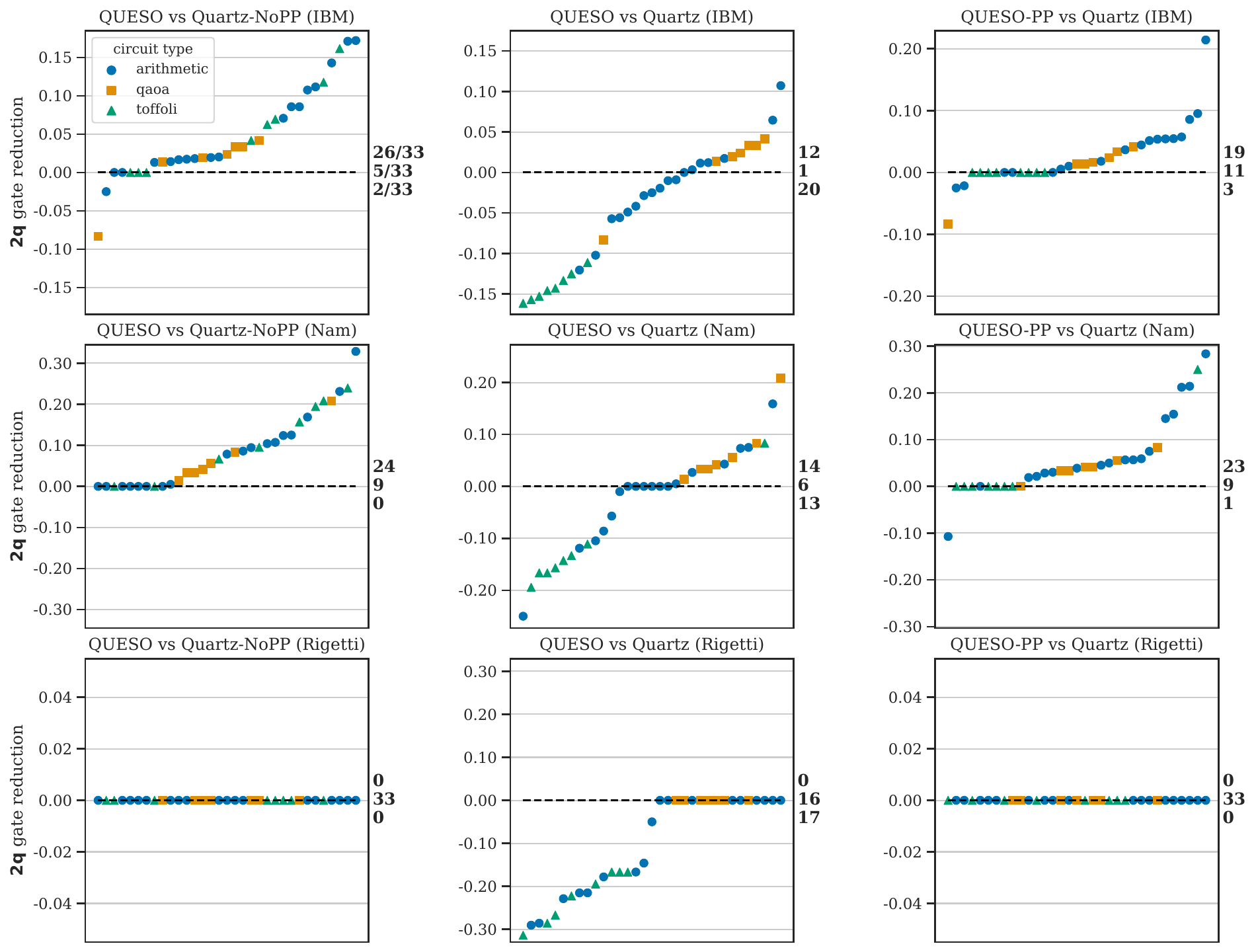}
    \caption{\textbf{14 hour timeout} comparison against Quartz.} 
    \label{fig:rq2_14hr}
\end{figure*}

\begin{figure*}[h]
    \centering
    \includegraphics[width=\textwidth]{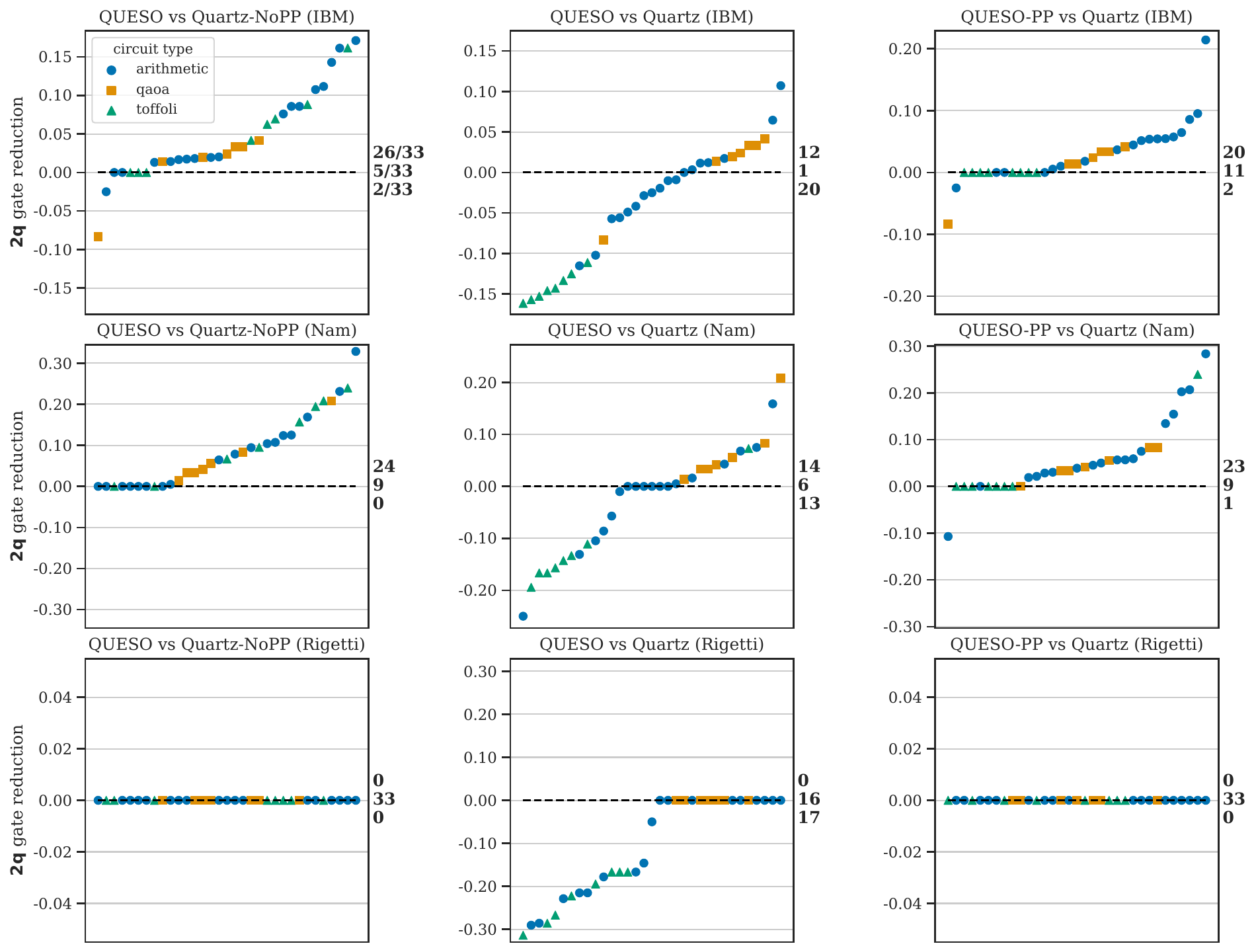}
    \caption{\textbf{15 hour timeout} comparison against Quartz.} 
    \label{fig:rq2_15hr}
\end{figure*}

\begin{figure*}[h]
    \centering
    \includegraphics[width=\textwidth]{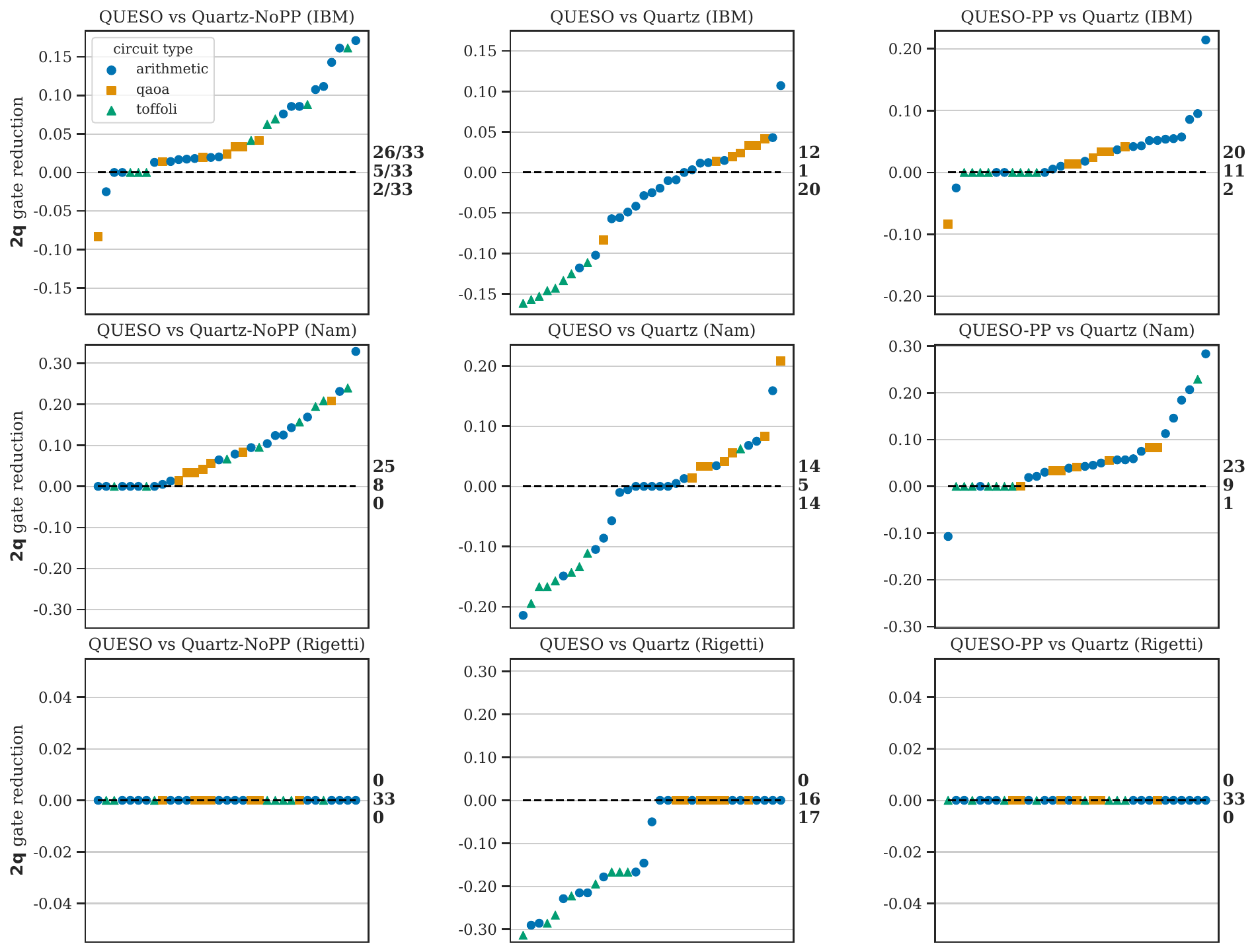}
    \caption{\textbf{16 hour timeout} comparison against Quartz.} 
    \label{fig:rq2_16hr}
\end{figure*}

\begin{figure*}[h]
    \centering
    \includegraphics[width=\textwidth]{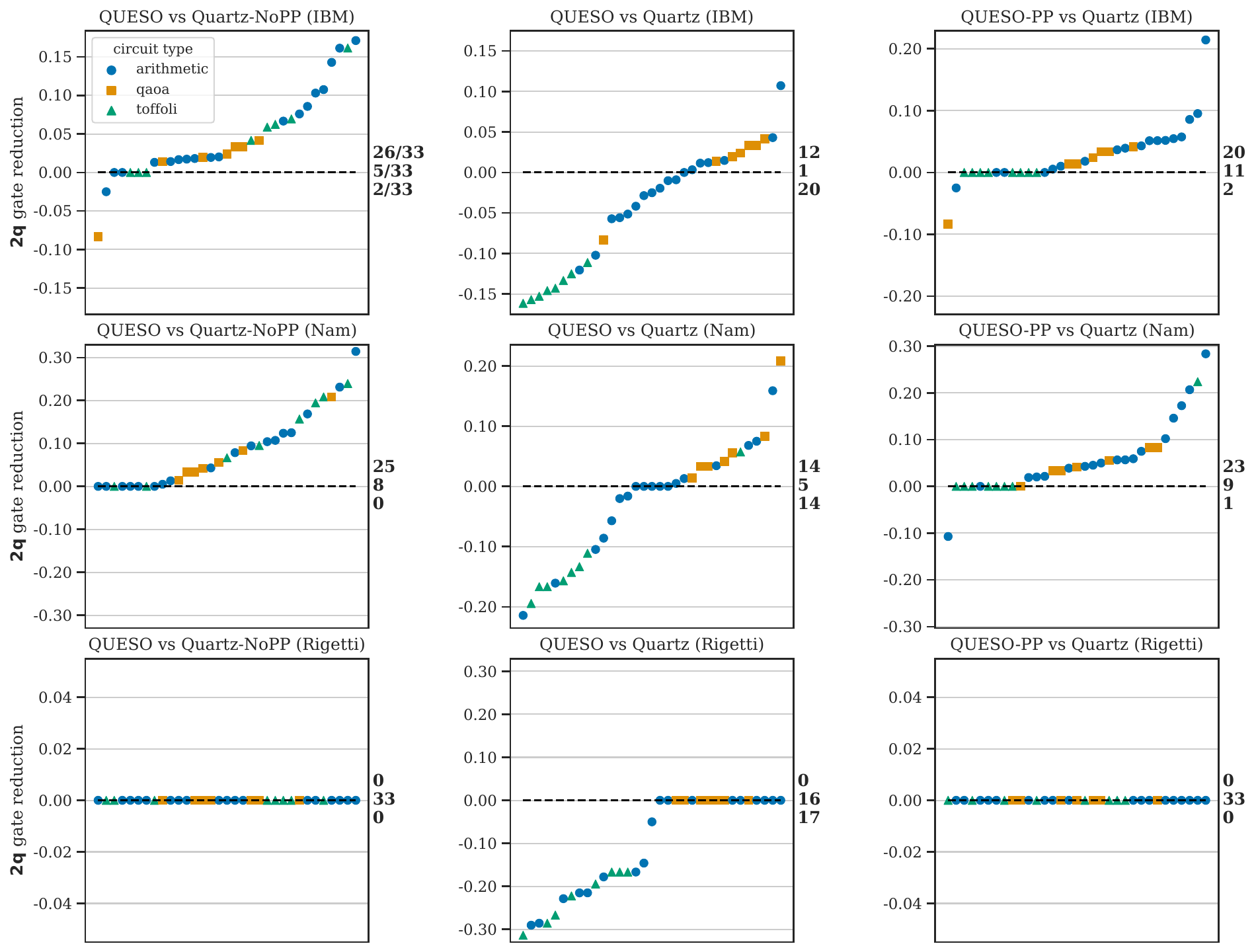}
    \caption{\textbf{17 hour timeout} comparison against Quartz.} 
    \label{fig:rq2_17hr}
\end{figure*}

\begin{figure*}[h]
    \centering
    \includegraphics[width=\textwidth]{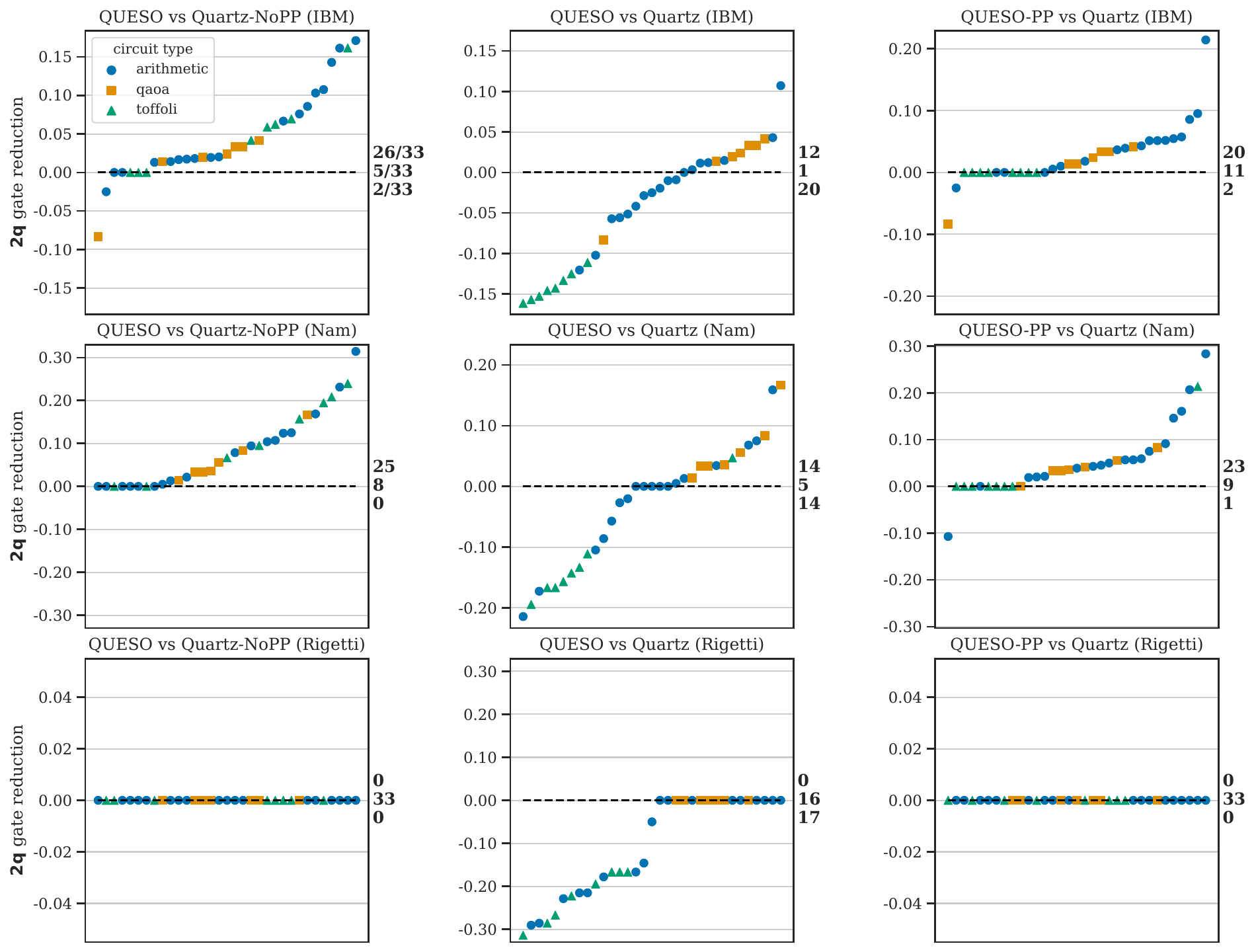}
    \caption{\textbf{18 hour timeout} comparison against Quartz.} 
    \label{fig:rq2_18hr}
\end{figure*}

\begin{figure*}[h]
    \centering
    \includegraphics[width=\textwidth]{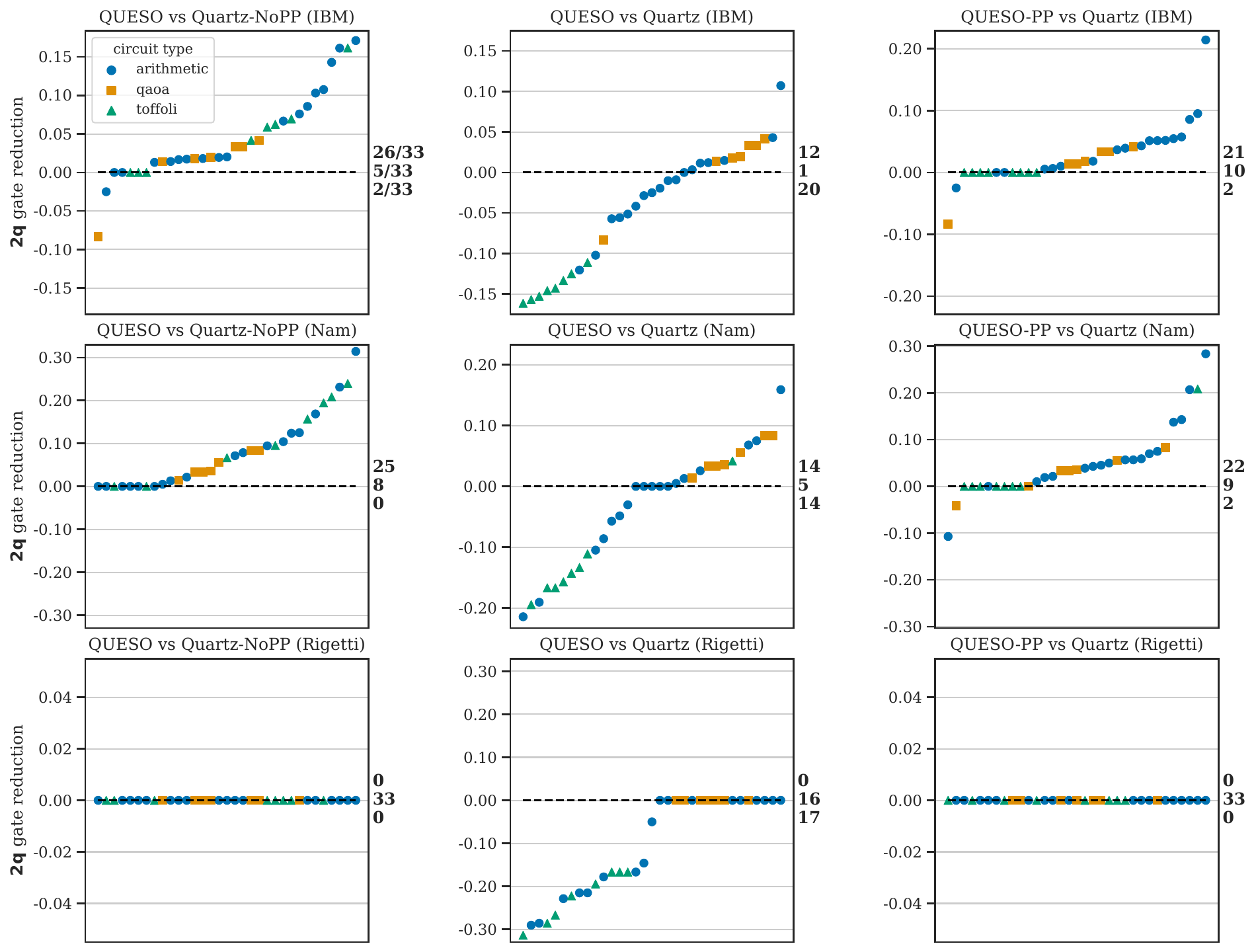}
    \caption{\textbf{19 hour timeout} comparison against Quartz.} 
    \label{fig:rq2_19hr}
\end{figure*}

\begin{figure*}[h]
    \centering
    \includegraphics[width=\textwidth]{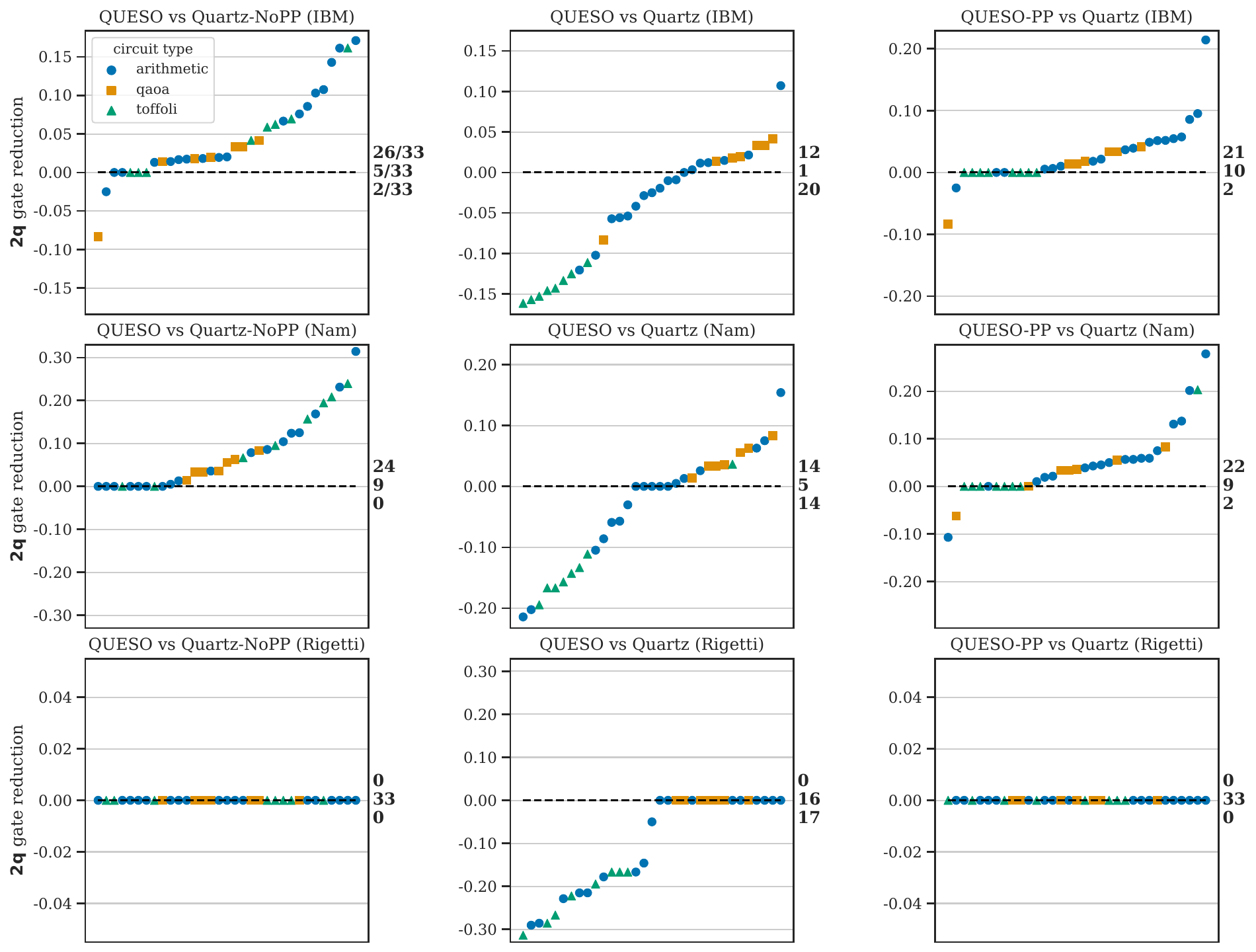}
    \caption{\textbf{20 hour timeout} comparison against Quartz.} 
    \label{fig:rq2_20hr}
\end{figure*}

\begin{figure*}[h]
    \centering
    \includegraphics[width=\textwidth]{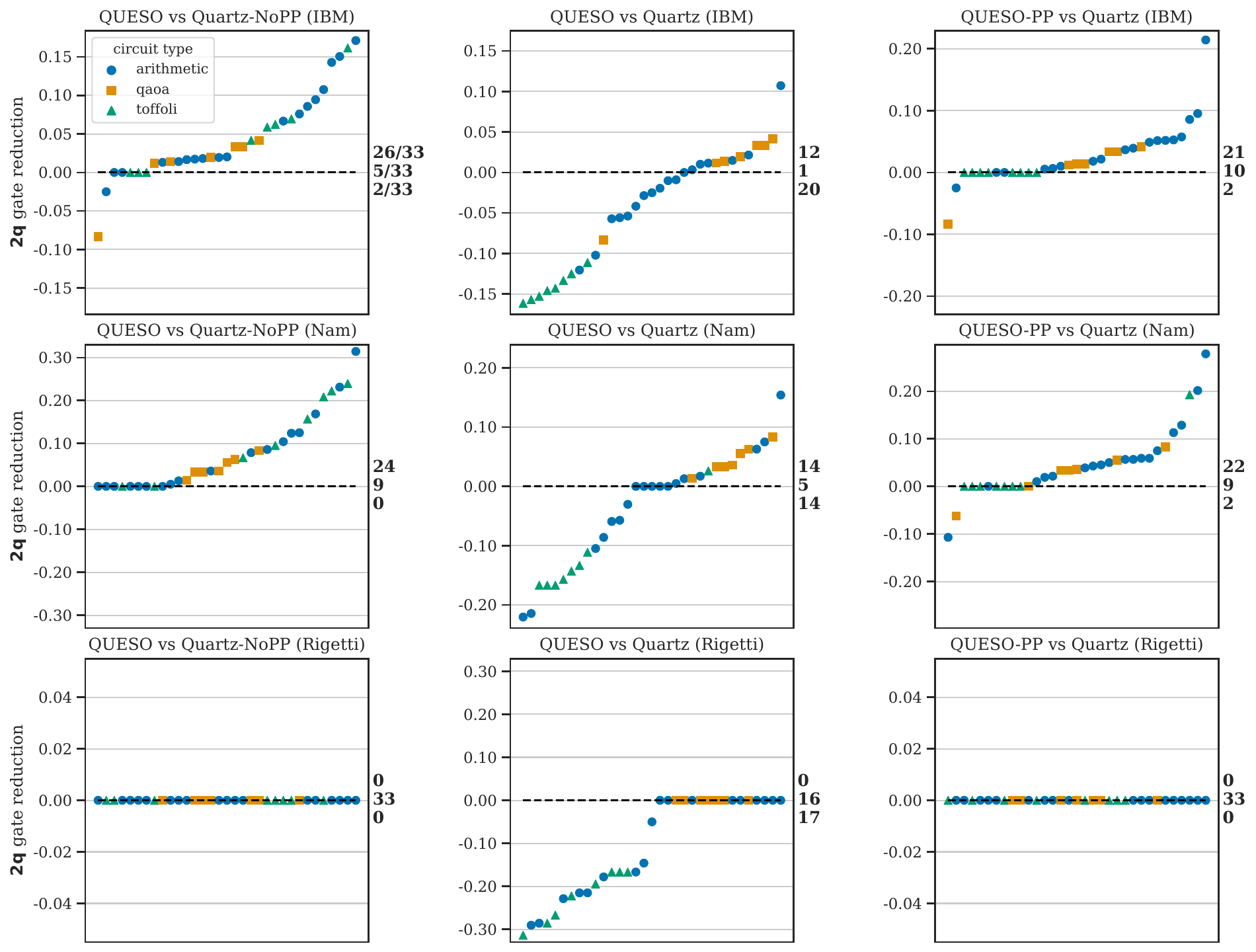}
    \caption{\textbf{21 hour timeout} comparison against Quartz.} 
    \label{fig:rq2_21hr}
\end{figure*}

\begin{figure*}[h]
    \centering
    \includegraphics[width=\textwidth]{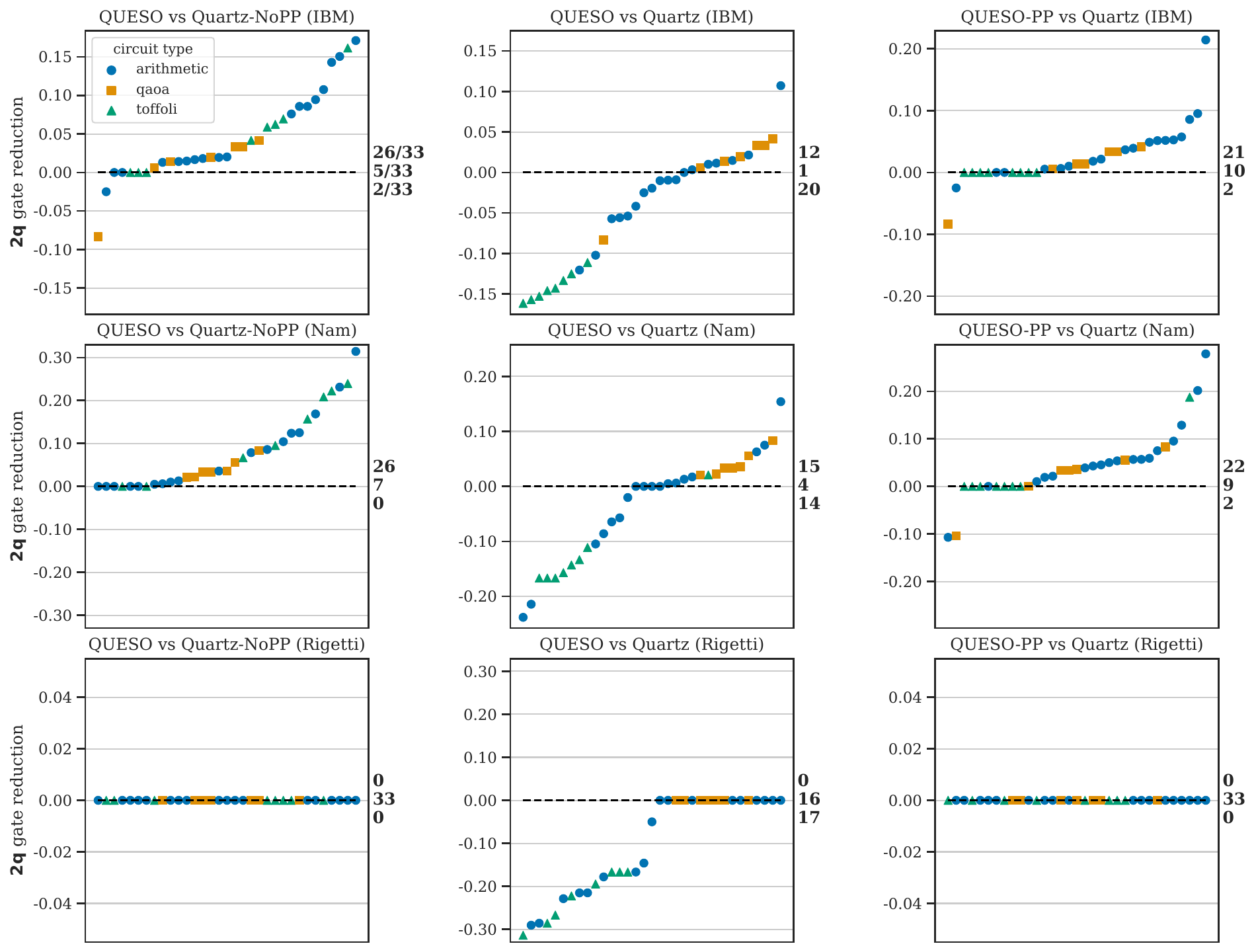}
    \caption{\textbf{22 hour timeout} comparison against Quartz.} 
    \label{fig:rq2_22hr}
\end{figure*}

\begin{figure*}[h]
    \centering
    \includegraphics[width=\textwidth]{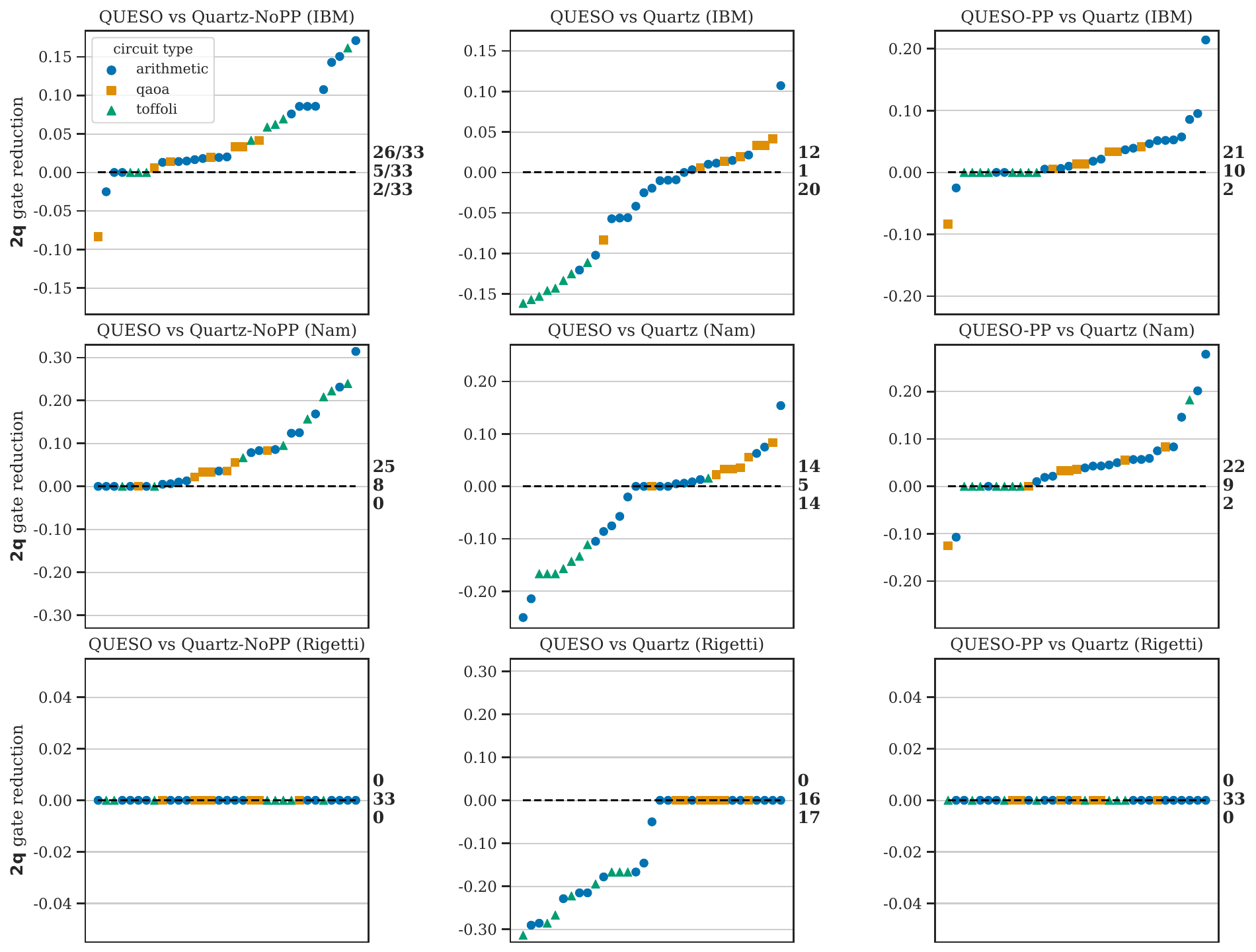}
    \caption{\textbf{23 hour timeout} comparison against Quartz.} 
    \label{fig:rq2_23hr}
\end{figure*}

\FloatBarrier
\section{Experimental results in tabular form}

\subsection{Total gate count 1 hour}
%& \multicolumn{8}{c}{\bfseries Total Gate Count} \\
\begin{table}[h]
    \caption{\textbf{Total} gate count results for \ibm gate set with \textbf{1 hour timeout}}
    \begin{filecontents*}{plot_generation/all_results_ibm.csv}
Circuit,Original,VOQC,Qiskit,TKET,Quartz,Quartz-NoPP,QUESO,QUESO-PP
adder\_8,900,644,805,802,718,894,687,563
barenco\_tof\_3,58,43,51,51,36,47,44,36
barenco\_tof\_4,114,80,100,100,67,102,83,67
barenco\_tof\_5,170,117,149,149,98,154,122,98
barenco\_tof\_10,450,302,394,394,265,423,317,253
csla\_mux\_3,170,148,156,157,143,148,147,146
csum\_mux\_9,420,354,382,372,343,383,378,361
gf2\string^4\_mult,225,185,206,206,179,218,191,179
gf2\string^5\_mult,347,282,318,319,281,338,291,276
gf2\string^6\_mult,495,399,454,454,394,483,426,385
gf2\string^7\_mult,669,536,614,614,535,657,573,523
gf2\string^8\_mult,883,691,804,806,695,877,759,672
gf2\string^9\_mult,1095,870,1006,1009,874,1092,969,842
gf2\string^10\_mult,1347,1068,1238,1240,1059,1345,1215,1017
mod5\_4,63,53,58,58,51,56,52,38
mod\_mult\_55,119,99,106,103,91,94,103,96
mod\_red\_21,278,206,227,230,206,244,222,196
qcla\_adder\_10,521,403,469,464,398,498,430,371
qcla\_com\_7,443,292,398,384,270,408,341,267
qcla\_mod\_7,884,664,786,776,708,878,720,590
rc\_adder\_6,200,153,170,170,160,173,171,168
tof\_3,45,36,40,40,31,36,36,31
tof\_4,75,58,66,66,49,58,58,49
tof\_5,105,80,92,92,67,89,80,67
tof\_10,255,190,222,222,157,226,190,157
vbe\_adder\_3,150,94,136,138,84,135,94,71
qaoa\_n4\_p4,220,92,92,92,88,88,100,100
qaoa\_n6\_p4,330,138,138,138,135,135,154,154
qaoa\_n8\_p4,440,184,184,184,320,320,212,212
qaoa\_n10\_p4,550,230,230,230,479,479,266,266
qaoa\_n14\_p4,770,322,322,322,741,741,374,374
qaoa\_n20\_p4,1100,460,460,460,1089,1089,536,536
qaoa\_n30\_p4,1650,690,690,690,1647,1647,810,810
\end{filecontents*}
\csvreader[
        tabular = l|r r r r r r r r, 
        table foot = \hline, 
        table head = \bfseries Circuit & \bfseries Original & \bfseries \voqc & \bfseries Qiskit & \bfseries \tket & \bfseries Quartz & \bfseries Quartz-NoPP & \bfseries \ours & \bfseries \ourspp \\\hline
    ]{plot_generation/all_results_ibm.csv}{}
    {\csvcoli & \csvcolii & \csvcoliii & \csvcoliv & \csvcolv & \csvcolvi & \csvcolvii & \csvcolviii & \csvcolix}
\end{table}

\begin{table}[h]
    \caption{\textbf{Total} gate count results for Rigetti gate set with \textbf{1 hour timeout}}
    \begin{filecontents*}{plot_generation/all_results_rigetti.csv}
Circuit,Original,Quilc,Quartz,Quartz-NoPP,QUESO,QUESO-PP
adder\_8,4412,3212,2767,4401,2142,1460
barenco\_tof\_3,268,193,148,242,134,86
barenco\_tof\_4,528,418,272,480,264,160
barenco\_tof\_5,788,553,386,716,394,228
barenco\_tof\_10,2088,1540,960,2052,1044,569
csla\_mux\_3,870,608,654,810,418,370
csum\_mux\_9,1848,1421,1100,1800,935,670
gf2\string^4\_mult,1059,748,796,963,512,454
gf2\string^5\_mult,1633,1197,1252,1571,809,696
gf2\string^6\_mult,2329,1741,1867,2300,1154,992
gf2\string^7\_mult,3147,2371,2560,3132,1564,1347
gf2\string^8\_mult,4213,3057,3340,4204,2081,1741
gf2\string^9\_mult,5149,3796,4205,5143,2579,2197
gf2\string^10\_mult,6333,4724,5179,6329,3172,2685
mod5\_4,305,225,197,279,143,116
mod\_mult\_55,545,408,361,503,267,212
mod\_red\_21,1208,845,738,1081,614,430
qcla\_adder\_10,2535,1769,1722,2510,1267,932
qcla\_com\_7,2048,1462,1131,2010,1020,638
qcla\_mod\_7,4186,3001,2762,4170,2022,1409
rc\_adder\_6,1010,750,606,920,485,353
tof\_3,207,135,135,188,104,76
tof\_4,345,218,199,315,172,114
tof\_5,483,325,271,441,240,155
tof\_10,1173,797,631,1071,580,360
vbe\_adder\_3,740,452,367,680,368,216
qaoa\_n4\_p4,712,332,472,576,268,259
qaoa\_n6\_p4,1068,501,712,864,385,390
qaoa\_n8\_p4,1424,664,1085,1319,579,527
qaoa\_n10\_p4,1780,805,1445,1724,714,655
qaoa\_n14\_p4,2492,1171,2146,2461,1010,914
qaoa\_n20\_p4,3560,1646,2997,3547,1428,1297
qaoa\_n30\_p4,5340,2375,4579,5333,2014,1966
\end{filecontents*}
\csvreader[
        tabular = l|r r r r r r, 
        table foot = \hline, 
        table head = \bfseries Circuit & \bfseries Original & \bfseries Quilc & \bfseries Quartz & \bfseries Quartz-NoPP & \bfseries \ours & \bfseries \ourspp \\\hline
    ]{plot_generation/all_results_rigetti.csv}{}
    {\csvcoli & \csvcolii & \csvcoliii & \csvcoliv & \csvcolv & \csvcolvi & \csvcolvii }
\end{table}

\begin{table}[h]
    \caption{\textbf{Total} gate count results for Ion gate set with \textbf{1 hour timeout}}
    \begin{filecontents*}{plot_generation/all_results_ion.csv}
Circuit,Original,Qiskit,QUESO,QUESO-RZ-Obj
adder\_8,2616,1381,1387,1507
barenco\_tof\_3,160,91,92,96
barenco\_tof\_4,316,176,175,187
barenco\_tof\_5,472,261,258,278
barenco\_tof\_10,1252,686,673,733
csla\_mux\_3,510,277,263,306
csum\_mux\_9,1120,664,656,692
gf2\string^4\_mult,635,355,347,386
gf2\string^5\_mult,981,543,536,599
gf2\string^6\_mult,1401,777,763,858
gf2\string^7\_mult,1895,1041,1031,1163
gf2\string^8\_mult,2533,1381,1483,1542
gf2\string^9\_mult,3105,1701,1765,1911
gf2\string^10\_mult,3821,2095,2359,2354
mod5\_4,181,101,100,109
mod\_mult\_55,325,178,165,192
mod\_red\_21,728,381,397,422
qcla\_adder\_10,1503,805,839,908
qcla\_com\_7,1226,654,693,725
qcla\_mod\_7,2494,1278,1309,1424
rc\_adder\_6,594,314,288,327
tof\_3,123,71,71,73
tof\_4,205,116,116,120
tof\_5,287,161,161,167
tof\_10,697,386,386,402
vbe\_adder\_3,440,229,245,265
qaoa\_n4\_p4,448,148,139,155
qaoa\_n6\_p4,672,219,211,235
qaoa\_n8\_p4,896,328,299,323
qaoa\_n10\_p4,1120,386,358,394
qaoa\_n14\_p4,1568,566,518,562
qaoa\_n20\_p4,2240,772,716,788
qaoa\_n30\_p4,3360,1171,1087,1191
\end{filecontents*}
\csvreader[
        tabular = l|r r r r, 
        table foot = \hline, 
        table head = \bfseries Circuit & \bfseries Original & \bfseries Qiskit & \bfseries \ours & \bfseries \oursrz \\\hline
    ]{plot_generation/all_results_ion.csv}{}
    {\csvcoli & \csvcolii & \csvcoliii & \csvcoliv & \csvcolv }
\end{table}

\begin{table}[h]
    \caption{\textbf{Total} gate count results for Nam gate set with \textbf{1 hour timeout}}
    \begin{filecontents*}{plot_generation/all_results_nam.csv}
Circuit,Original,VOQC,Qiskit,Quartz,Quartz-NoPP,QUESO,QUESO-PP
adder\_8,900,682,869,730,900,752,605
barenco\_tof\_3,58,50,56,38,54,48,38
barenco\_tof\_4,114,95,109,68,114,90,68
barenco\_tof\_5,170,140,162,102,170,150,98
barenco\_tof\_10,450,365,427,322,450,395,262
csla\_mux\_3,170,158,168,156,164,149,148
csum\_mux\_9,420,308,420,306,420,406,252
gf2\string^4\_mult,225,192,213,180,219,202,177
gf2\string^5\_mult,347,291,327,281,343,308,273
gf2\string^6\_mult,495,410,465,399,493,450,381
gf2\string^7\_mult,669,549,627,541,667,611,516
gf2\string^8\_mult,883,705,819,703,883,832,682
gf2\string^9\_mult,1095,885,1023,877,1093,1020,835
gf2\string^10\_mult,1347,1084,1257,1060,1345,1257,1009
mod5\_4,63,56,62,40,63,49,33
mod\_mult\_55,119,90,117,94,115,108,94
mod\_red\_21,278,214,261,228,272,230,196
qcla\_adder\_10,521,438,512,448,521,467,406
qcla\_com\_7,443,314,428,345,441,357,274
qcla\_mod\_7,884,723,853,725,884,786,651
rc\_adder\_6,200,157,195,154,200,182,152
tof\_3,45,40,44,35,42,40,35
tof\_4,75,65,73,55,75,65,55
tof\_5,105,90,102,75,105,90,75
tof\_10,255,215,247,195,255,215,175
vbe\_adder\_3,150,101,146,92,150,93,80
qaoa\_n4\_p4,220,124,124,205,205,106,128
qaoa\_n6\_p4,330,186,186,327,327,178,206
qaoa\_n8\_p4,440,248,248,437,437,244,276
qaoa\_n10\_p4,550,310,310,549,549,310,350
qaoa\_n14\_p4,770,434,434,769,769,426,482
qaoa\_n20\_p4,1100,620,620,1099,1099,616,696
qaoa\_n30\_p4,1650,930,930,1650,1650,930,1050
\end{filecontents*}
\csvreader[
        tabular = l|r r r r r r r, 
        table foot = \hline, 
        table head = \bfseries Circuit & \bfseries Original & \bfseries \voqc & \bfseries Qiskit & \bfseries Quartz & \bfseries Quartz-NoPP & \bfseries \ours & \bfseries \ourspp \\\hline
    ]{plot_generation/all_results_nam.csv}{}
    {\csvcoli & \csvcolii & \csvcoliii & \csvcoliv & \csvcolv & \csvcolvi & \csvcolvii & \csvcolviii }
\end{table}

\FloatBarrier
\subsection{Two-qubit gate count 1 hour}

\begin{table}[h]
    \caption{\textbf{Two-qubit} gate count results for \ibm gate set with \textbf{1 hour timeout}}
    \begin{filecontents*}{plot_generation/all_results_ibm_2q.csv}
Circuit,Original,VOQC,Qiskit,TKET,Quartz,Quartz-NoPP,QUESO,QUESO-PP
adder\_8,409,331,385,383,391,409,347,298
barenco\_tof\_3,24,20,24,24,18,22,21,18
barenco\_tof\_4,48,38,48,48,34,48,41,34
barenco\_tof\_5,72,56,72,72,50,72,61,50
barenco\_tof\_10,192,146,192,192,142,192,161,130
csla\_mux\_3,80,74,71,71,73,75,73,73
csum\_mux\_9,168,168,168,168,168,168,168,168
gf2\string^4\_mult,99,99,99,99,97,99,97,96
gf2\string^5\_mult,154,154,154,154,153,154,151,148
gf2\string^6\_mult,221,221,221,221,220,221,217,211
gf2\string^7\_mult,300,300,300,300,299,300,295,287
gf2\string^8\_mult,405,405,405,402,405,405,399,382
gf2\string^9\_mult,494,494,494,494,494,494,487,466
gf2\string^10\_mult,609,609,609,609,609,609,604,574
mod5\_4,28,28,28,28,27,27,24,21
mod\_mult\_55,48,46,48,48,44,46,46,44
mod\_red\_21,105,99,105,104,92,105,94,81
qcla\_adder\_10,233,207,213,209,199,233,211,187
qcla\_com\_7,186,148,174,174,129,186,157,128
qcla\_mod\_7,382,344,366,366,364,382,358,301
rc\_adder\_6,93,81,81,81,87,91,87,87
tof\_3,18,16,18,18,14,16,16,14
tof\_4,30,26,30,30,22,26,26,22
tof\_5,42,36,42,42,30,40,36,30
tof\_10,102,86,102,102,70,102,86,70
vbe\_adder\_3,70,54,62,62,46,62,48,38
qaoa\_n4\_p4,48,48,48,48,37,37,40,40
qaoa\_n6\_p4,72,72,72,72,69,69,64,64
qaoa\_n8\_p4,96,96,96,96,96,96,92,92
qaoa\_n10\_p4,120,120,120,120,120,120,116,116
qaoa\_n14\_p4,168,168,168,168,168,168,164,164
qaoa\_n20\_p4,240,240,240,240,240,240,236,236
qaoa\_n30\_p4,360,360,360,360,360,360,360,360
\end{filecontents*}
\csvreader[
        tabular = l|r r r r r r r r, 
        table foot = \hline, 
        table head = \bfseries Circuit & \bfseries Original & \bfseries \voqc & \bfseries Qiskit & \bfseries \tket & \bfseries Quartz & \bfseries Quartz-NoPP & \bfseries \ours & \bfseries \ourspp \\\hline
    ]{plot_generation/all_results_ibm_2q.csv}{}
    {\csvcoli & \csvcolii & \csvcoliii & \csvcoliv & \csvcolv & \csvcolvi & \csvcolvii & \csvcolviii & \csvcolix}
\end{table}

\begin{table}[h]
    \caption{\textbf{Two-qubit} gate count results for Rigetti gate set with \textbf{1 hour timeout}}
    \begin{filecontents*}{plot_generation/all_results_rigetti_2q.csv}
Circuit,Original,Quilc,Quartz,Quartz-NoPP,QUESO,QUESO-PP
adder\_8,409,385,321,409,409,321
barenco\_tof\_3,24,24,20,24,24,20
barenco\_tof\_4,48,48,40,48,48,40
barenco\_tof\_5,72,72,60,72,72,60
barenco\_tof\_10,192,192,160,192,192,160
csla\_mux\_3,80,71,76,80,80,76
csum\_mux\_9,168,168,168,168,168,168
gf2\string^4\_mult,99,99,99,99,99,99
gf2\string^5\_mult,154,154,154,154,154,154
gf2\string^6\_mult,221,221,221,221,221,221
gf2\string^7\_mult,300,300,300,300,300,300
gf2\string^8\_mult,405,405,405,405,405,405
gf2\string^9\_mult,494,494,494,494,494,494
gf2\string^10\_mult,609,609,609,609,609,609
mod5\_4,28,28,28,28,28,28
mod\_mult\_55,48,48,40,48,48,40
mod\_red\_21,105,105,81,105,105,81
qcla\_adder\_10,233,213,199,233,233,199
qcla\_com\_7,186,174,132,186,186,132
qcla\_mod\_7,382,366,312,382,382,314
rc\_adder\_6,93,81,73,93,93,73
tof\_3,18,18,14,18,18,14
tof\_4,30,30,22,30,30,22
tof\_5,42,42,30,42,42,30
tof\_10,102,102,70,102,102,70
vbe\_adder\_3,70,62,52,70,70,50
qaoa\_n4\_p4,48,48,48,48,48,48
qaoa\_n6\_p4,72,72,72,72,72,72
qaoa\_n8\_p4,96,96,96,96,96,96
qaoa\_n10\_p4,120,120,120,120,120,120
qaoa\_n14\_p4,168,168,168,168,168,168
qaoa\_n20\_p4,240,240,240,240,240,240
qaoa\_n30\_p4,360,360,360,360,360,360
\end{filecontents*}
\csvreader[
        tabular = l|r r r r r r, 
        table foot = \hline, 
        table head = \bfseries Circuit & \bfseries Original & \bfseries Quilc & \bfseries Quartz & \bfseries Quartz-NoPP & \bfseries \ours & \bfseries \ourspp \\\hline
    ]{plot_generation/all_results_rigetti_2q.csv}{}
    {\csvcoli & \csvcolii & \csvcoliii & \csvcoliv & \csvcolv & \csvcolvi & \csvcolvii }
\end{table}

\begin{table}[h]
    \caption{\textbf{Two-qubit} gate count results for Ion gate set with \textbf{1 hour timeout}}
    \begin{filecontents*}{plot_generation/all_results_ion_2q.csv}
Circuit,Original,Qiskit,QUESO,QUESO-RZ-Obj
adder\_8,409,385,409,409
barenco\_tof\_3,24,24,24,24
barenco\_tof\_4,48,48,48,48
barenco\_tof\_5,72,72,72,72
barenco\_tof\_10,192,192,192,192
csla\_mux\_3,80,71,80,80
csum\_mux\_9,168,168,168,168
gf2\string^4\_mult,99,99,99,99
gf2\string^5\_mult,154,154,154,154
gf2\string^6\_mult,221,221,221,221
gf2\string^7\_mult,300,300,300,300
gf2\string^8\_mult,405,405,405,405
gf2\string^9\_mult,494,494,494,494
gf2\string^10\_mult,609,609,609,609
mod5\_4,28,28,28,28
mod\_mult\_55,48,48,48,48
mod\_red\_21,105,105,105,105
qcla\_adder\_10,233,213,233,233
qcla\_com\_7,186,174,186,186
qcla\_mod\_7,382,366,382,382
rc\_adder\_6,93,81,89,93
tof\_3,18,18,18,18
tof\_4,30,30,30,30
tof\_5,42,42,42,42
tof\_10,102,102,102,102
vbe\_adder\_3,70,62,70,70
qaoa\_n4\_p4,48,48,48,48
qaoa\_n6\_p4,72,72,72,72
qaoa\_n8\_p4,96,96,96,96
qaoa\_n10\_p4,120,120,120,120
qaoa\_n14\_p4,168,168,168,168
qaoa\_n20\_p4,240,240,240,240
qaoa\_n30\_p4,360,360,360,360
\end{filecontents*}
\csvreader[
        tabular = l|r r r r, 
        table foot = \hline, 
        table head = \bfseries Circuit & \bfseries Original & \bfseries Qiskit & \bfseries \ours & \bfseries \oursrz \\\hline
    ]{plot_generation/all_results_ion_2q.csv}{}
    {\csvcoli & \csvcolii & \csvcoliii & \csvcoliv & \csvcolv }
\end{table}

\begin{table}[h]
    \caption{\textbf{Two-qubit} gate count results for Nam gate set with \textbf{1 hour timeout}}
    \begin{filecontents*}{plot_generation/all_results_nam_2q.csv}
Circuit,Original,VOQC,Qiskit,Quartz,Quartz-NoPP,QUESO,QUESO-PP
adder\_8,409,337,409,407,409,348,298
barenco\_tof\_3,24,22,24,16,22,20,16
barenco\_tof\_4,48,44,48,30,48,39,30
barenco\_tof\_5,72,66,72,48,72,66,44
barenco\_tof\_10,192,176,192,188,192,176,128
csla\_mux\_3,80,74,80,76,80,70,70
csum\_mux\_9,168,168,168,166,168,168,112
gf2\string^4\_mult,99,99,99,99,99,99,96
gf2\string^5\_mult,154,154,154,154,154,154,148
gf2\string^6\_mult,221,221,221,221,221,221,211
gf2\string^7\_mult,300,300,300,300,300,300,285
gf2\string^8\_mult,405,405,405,405,405,405,384
gf2\string^9\_mult,494,494,494,494,494,494,466
gf2\string^10\_mult,609,609,609,609,609,609,574
mod5\_4,28,28,28,22,28,23,18
mod\_mult\_55,48,40,48,40,48,46,40
mod\_red\_21,105,93,105,105,105,92,79
qcla\_adder\_10,233,207,233,231,233,211,189
qcla\_com\_7,186,148,186,182,186,149,123
qcla\_mod\_7,382,338,382,380,382,352,306
rc\_adder\_6,93,73,93,73,93,81,71
tof\_3,18,16,18,14,16,16,14
tof\_4,30,26,30,22,30,26,22
tof\_5,42,36,42,30,42,36,30
tof\_10,102,86,102,90,102,86,70
vbe\_adder\_3,70,54,70,47,70,44,41
qaoa\_n4\_p4,48,48,48,44,44,30,36
qaoa\_n6\_p4,72,72,72,69,69,64,68
qaoa\_n8\_p4,96,96,96,96,96,92,92
qaoa\_n10\_p4,120,120,120,120,120,120,120
qaoa\_n14\_p4,168,168,168,167,167,160,160
qaoa\_n20\_p4,240,240,240,239,239,236,236
qaoa\_n30\_p4,360,360,360,360,360,360,360
\end{filecontents*}
\csvreader[
        tabular = l|r r r r r r r, 
        table foot = \hline, 
        table head = \bfseries Circuit & \bfseries Original & \bfseries \voqc & \bfseries Qiskit & \bfseries Quartz & \bfseries Quartz-NoPP & \bfseries \ours & \bfseries \ourspp \\\hline
    ]{plot_generation/all_results_nam_2q.csv}{}
    {\csvcoli & \csvcolii & \csvcoliii & \csvcoliv & \csvcolv & \csvcolvi & \csvcolvii & \csvcolviii }
\end{table}

\FloatBarrier
\subsection{Total gate count 24 hour}
\begin{table}[h]
    \caption{\textbf{Total} gate count results for \ibm gate set with \textbf{24 hour timeout}}
    \begin{filecontents*}{plot_generation/all_results_ibm_24hr.csv}
Circuit,Original,VOQC,Qiskit,TKET,Quartz,Quartz-NoPP,QUESO,QUESO-PP
adder\_8,900,644,805,802,585,840,675,560
barenco\_tof\_3,58,43,51,51,36,47,44,36
barenco\_tof\_4,114,80,100,100,67,91,82,67
barenco\_tof\_5,170,117,149,149,98,138,120,98
barenco\_tof\_10,450,302,394,394,253,411,310,253
csla\_mux\_3,170,148,156,157,139,139,144,144
csum\_mux\_9,420,354,382,372,341,377,375,361
gf2\string^4\_mult,225,185,206,206,178,217,187,178
gf2\string^5\_mult,347,282,318,319,275,336,288,275
gf2\string^6\_mult,495,399,454,454,388,483,417,385
gf2\string^7\_mult,669,536,614,614,532,656,557,521
gf2\string^8\_mult,883,691,804,806,693,866,743,672
gf2\string^9\_mult,1095,870,1006,1009,868,1078,925,842
gf2\string^10\_mult,1347,1068,1238,1240,1051,1328,1161,1016
mod5\_4,63,53,58,58,51,55,52,38
mod\_mult\_55,119,99,106,103,91,94,103,96
mod\_red\_21,278,206,227,230,205,230,218,196
qcla\_adder\_10,521,403,469,464,379,479,407,364
qcla\_com\_7,443,292,398,384,268,388,316,265
qcla\_mod\_7,884,664,786,776,598,818,702,581
rc\_adder\_6,200,153,170,170,151,168,163,160
tof\_3,45,36,40,40,31,36,36,31
tof\_4,75,58,66,66,49,58,58,49
tof\_5,105,80,92,92,67,80,80,67
tof\_10,255,190,222,222,157,205,190,157
vbe\_adder\_3,150,94,136,138,82,113,94,71
qaoa\_n4\_p4,220,92,92,92,80,80,100,100
qaoa\_n6\_p4,330,138,138,138,131,131,154,154
qaoa\_n8\_p4,440,184,184,184,180,180,208,208
qaoa\_n10\_p4,550,230,230,230,230,230,266,266
qaoa\_n14\_p4,770,322,322,322,318,318,374,374
qaoa\_n20\_p4,1100,460,460,460,911,911,532,532
qaoa\_n30\_p4,1650,690,690,690,1605,1605,803,805
\end{filecontents*}
\csvreader[
        tabular = l|r r r r r r r r, 
        table foot = \hline, 
        table head = \bfseries Circuit & \bfseries Original & \bfseries \voqc & \bfseries Qiskit & \bfseries \tket & \bfseries Quartz & \bfseries Quartz-NoPP & \bfseries \ours & \bfseries \ourspp \\\hline
    ]{plot_generation/all_results_ibm_24hr.csv}{}
    {\csvcoli & \csvcolii & \csvcoliii & \csvcoliv & \csvcolv & \csvcolvi & \csvcolvii & \csvcolviii & \csvcolix}
\end{table}

\begin{table}[h]
    \caption{\textbf{Total} gate count results for Rigetti gate set with \textbf{24 hour timeout}}
    \begin{filecontents*}{plot_generation/all_results_rigetti_24hr.csv}
Circuit,Original,Quilc,Quartz,Quartz-NoPP,QUESO,QUESO-PP
adder\_8,4412,3212,2553,4240,2132,1449
barenco\_tof\_3,268,193,148,242,134,86
barenco\_tof\_4,528,418,272,477,262,160
barenco\_tof\_5,788,553,386,716,390,226
barenco\_tof\_10,2088,1540,960,1896,1030,569
csla\_mux\_3,870,608,654,810,418,370
csum\_mux\_9,1848,1421,1100,1678,932,660
gf2\string^4\_mult,1059,748,796,963,512,454
gf2\string^5\_mult,1633,1197,1231,1479,794,696
gf2\string^6\_mult,2329,1741,1751,2113,1130,990
gf2\string^7\_mult,3147,2371,2371,2853,1529,1338
gf2\string^8\_mult,4213,3057,3185,4104,2066,1739
gf2\string^9\_mult,5149,3796,4130,5041,2558,2184
gf2\string^10\_mult,6333,4724,5120,6258,3172,2685
mod5\_4,305,225,197,279,143,116
mod\_mult\_55,545,408,361,501,263,212
mod\_red\_21,1208,845,738,1081,592,423
qcla\_adder\_10,2535,1769,1616,2331,1266,929
qcla\_com\_7,2048,1462,1095,1874,1020,636
qcla\_mod\_7,4186,3001,2527,4053,2020,1409
rc\_adder\_6,1010,750,606,920,481,353
tof\_3,207,135,135,188,104,76
tof\_4,345,218,199,313,172,114
tof\_5,483,325,271,441,240,155
tof\_10,1173,797,631,1071,580,360
vbe\_adder\_3,740,452,366,680,368,216
qaoa\_n4\_p4,712,332,472,576,268,257
qaoa\_n6\_p4,1068,501,712,864,380,385
qaoa\_n8\_p4,1424,664,952,1152,564,524
qaoa\_n10\_p4,1780,805,1192,1440,669,649
qaoa\_n14\_p4,2492,1171,1704,2016,935,914
qaoa\_n20\_p4,3560,1646,2757,3327,1338,1297
qaoa\_n30\_p4,5340,2375,4480,5236,1999,1966
\end{filecontents*}
\csvreader[
        tabular = l|r r r r r r, 
        table foot = \hline, 
        table head = \bfseries Circuit & \bfseries Original & \bfseries Quilc & \bfseries Quartz & \bfseries Quartz-NoPP & \bfseries \ours & \bfseries \ourspp \\\hline
    ]{plot_generation/all_results_rigetti_24hr.csv}{}
    {\csvcoli & \csvcolii & \csvcoliii & \csvcoliv & \csvcolv & \csvcolvi & \csvcolvii }
\end{table}

\begin{table}[h]
    \caption{\textbf{Total} gate count results for Nam gate set with \textbf{24 hour timeout}}
    \begin{filecontents*}{plot_generation/all_results_nam_24hr.csv}
Circuit,Original,VOQC,Qiskit,Quartz,Quartz-NoPP,QUESO,QUESO-PP
adder\_8,900,682,869,724,900,729,596
barenco\_tof\_3,58,50,56,38,50,48,38
barenco\_tof\_4,114,95,109,68,114,89,68
barenco\_tof\_5,170,140,162,98,170,130,98
barenco\_tof\_10,450,365,427,280,450,335,248
csla\_mux\_3,170,158,168,154,162,148,148
csum\_mux\_9,420,308,420,262,420,406,252
gf2\string^4\_mult,225,192,213,177,219,198,176
gf2\string^5\_mult,347,291,327,279,339,299,273
gf2\string^6\_mult,495,410,465,391,485,441,381
gf2\string^7\_mult,669,549,627,531,657,594,516
gf2\string^8\_mult,883,705,819,703,883,803,680
gf2\string^9\_mult,1095,885,1023,875,1091,943,835
gf2\string^10\_mult,1347,1084,1257,1058,1345,1213,1008
mod5\_4,63,56,62,26,48,45,33
mod\_mult\_55,119,90,117,93,108,98,92
mod\_red\_21,278,214,261,202,267,229,196
qcla\_adder\_10,521,438,512,428,519,466,394
qcla\_com\_7,443,314,428,290,441,335,272
qcla\_mod\_7,884,723,853,719,884,784,644
rc\_adder\_6,200,157,195,154,188,179,152
tof\_3,45,40,44,35,40,40,35
tof\_4,75,65,73,55,69,65,55
tof\_5,105,90,102,75,101,90,75
tof\_10,255,215,247,175,255,215,175
vbe\_adder\_3,150,101,146,85,146,93,76
qaoa\_n4\_p4,220,124,124,102,102,98,126
qaoa\_n6\_p4,330,186,186,186,186,178,202
qaoa\_n8\_p4,440,248,248,415,415,240,272
qaoa\_n10\_p4,550,310,310,533,533,306,346
qaoa\_n14\_p4,770,434,434,767,767,426,482
qaoa\_n20\_p4,1100,620,620,1099,1099,612,692
qaoa\_n30\_p4,1650,930,930,1650,1650,922,1050
\end{filecontents*}
\csvreader[
        tabular = l|r r r r r r r, 
        table foot = \hline, 
        table head = \bfseries Circuit & \bfseries Original & \bfseries \voqc & \bfseries Qiskit & \bfseries Quartz & \bfseries Quartz-NoPP & \bfseries \ours & \bfseries \ourspp \\\hline
    ]{plot_generation/all_results_nam_24hr.csv}{}
    {\csvcoli & \csvcolii & \csvcoliii & \csvcoliv & \csvcolv & \csvcolvi & \csvcolvii & \csvcolviii }
\end{table}

\FloatBarrier
\subsection{Two-qubit gate count 24 hour}

\begin{table}[h]
    \caption{\textbf{Two-qubit} gate count results for \ibm gate set with \textbf{24 hour timeout}}
    \begin{filecontents*}{plot_generation/all_results_ibm_24hr_2q.csv}
Circuit,Original,VOQC,Qiskit,TKET,Quartz,Quartz-NoPP,QUESO,QUESO-PP
adder\_8,409,331,385,383,314,409,339,297
barenco\_tof\_3,24,20,24,24,18,22,21,18
barenco\_tof\_4,48,38,48,48,34,44,41,34
barenco\_tof\_5,72,56,72,72,50,66,61,50
barenco\_tof\_10,192,146,192,192,130,192,161,130
csla\_mux\_3,80,74,71,71,70,70,72,72
csum\_mux\_9,168,168,168,168,168,168,168,168
gf2\string^4\_mult,99,99,99,99,96,99,97,95
gf2\string^5\_mult,154,154,154,154,148,154,151,147
gf2\string^6\_mult,221,221,221,221,215,221,217,211
gf2\string^7\_mult,300,300,300,300,296,300,295,285
gf2\string^8\_mult,405,405,405,402,403,403,397,382
gf2\string^9\_mult,494,494,494,494,492,494,487,466
gf2\string^10\_mult,609,609,609,609,608,609,601,573
mod5\_4,28,28,28,28,27,27,24,21
mod\_mult\_55,48,46,48,48,44,46,46,44
mod\_red\_21,105,99,105,104,91,101,92,81
qcla\_adder\_10,233,207,213,209,193,226,206,181
qcla\_com\_7,186,148,174,174,127,174,146,126
qcla\_mod\_7,382,344,366,366,307,382,353,292
rc\_adder\_6,93,81,81,81,81,89,79,79
tof\_3,18,16,18,18,14,16,16,14
tof\_4,30,26,30,30,22,26,26,22
tof\_5,42,36,42,42,30,36,36,30
tof\_10,102,86,102,102,70,92,86,70
vbe\_adder\_3,70,54,62,62,44,54,48,38
qaoa\_n4\_p4,48,48,48,48,36,36,40,40
qaoa\_n6\_p4,72,72,72,72,65,65,64,64
qaoa\_n8\_p4,96,96,96,96,92,92,88,88
qaoa\_n10\_p4,120,120,120,120,120,120,116,116
qaoa\_n14\_p4,168,168,168,168,164,164,164,164
qaoa\_n20\_p4,240,240,240,240,240,240,232,232
qaoa\_n30\_p4,360,360,360,360,360,360,353,355
\end{filecontents*}
\csvreader[
        tabular = l|r r r r r r r r, 
        table foot = \hline, 
        table head = \bfseries Circuit & \bfseries Original & \bfseries \voqc & \bfseries Qiskit & \bfseries \tket & \bfseries Quartz & \bfseries Quartz-NoPP & \bfseries \ours & \bfseries \ourspp \\\hline
    ]{plot_generation/all_results_ibm_24hr_2q.csv}{}
    {\csvcoli & \csvcolii & \csvcoliii & \csvcoliv & \csvcolv & \csvcolvi & \csvcolvii & \csvcolviii & \csvcolix}
\end{table}

\begin{table}[h]
    \caption{\textbf{Two-qubit} gate count results for Rigetti gate set with \textbf{24 hour timeout}}
    \begin{filecontents*}{plot_generation/all_results_rigetti_24hr_2q.csv}
Circuit,Original,Quilc,Quartz,Quartz-NoPP,QUESO,QUESO-PP
adder\_8,409,385,321,409,409,321
barenco\_tof\_3,24,24,20,24,24,20
barenco\_tof\_4,48,48,40,48,48,40
barenco\_tof\_5,72,72,58,72,72,58
barenco\_tof\_10,192,192,160,192,192,160
csla\_mux\_3,80,71,76,80,80,76
csum\_mux\_9,168,168,168,168,168,168
gf2\string^4\_mult,99,99,99,99,99,99
gf2\string^5\_mult,154,154,154,154,154,154
gf2\string^6\_mult,221,221,221,221,221,221
gf2\string^7\_mult,300,300,300,300,300,300
gf2\string^8\_mult,405,405,405,405,405,405
gf2\string^9\_mult,494,494,494,494,494,494
gf2\string^10\_mult,609,609,609,609,609,609
mod5\_4,28,28,28,28,28,28
mod\_mult\_55,48,48,40,48,48,40
mod\_red\_21,105,105,81,105,105,81
qcla\_adder\_10,233,213,199,233,233,199
qcla\_com\_7,186,174,132,186,186,132
qcla\_mod\_7,382,366,314,382,382,314
rc\_adder\_6,93,81,73,93,93,73
tof\_3,18,18,14,18,18,14
tof\_4,30,30,22,30,30,22
tof\_5,42,42,30,42,42,30
tof\_10,102,102,70,102,102,70
vbe\_adder\_3,70,62,50,70,70,50
qaoa\_n4\_p4,48,48,48,48,48,48
qaoa\_n6\_p4,72,72,72,72,72,72
qaoa\_n8\_p4,96,96,96,96,96,96
qaoa\_n10\_p4,120,120,120,120,120,120
qaoa\_n14\_p4,168,168,168,168,168,168
qaoa\_n20\_p4,240,240,240,240,240,240
qaoa\_n30\_p4,360,360,360,360,360,360
\end{filecontents*}
\csvreader[
        tabular = l|r r r r r r, 
        table foot = \hline, 
        table head = \bfseries Circuit & \bfseries Original & \bfseries Quilc & \bfseries Quartz & \bfseries Quartz-NoPP & \bfseries \ours & \bfseries \ourspp \\\hline
    ]{plot_generation/all_results_rigetti_24hr_2q.csv}{}
    {\csvcoli & \csvcolii & \csvcoliii & \csvcoliv & \csvcolv & \csvcolvi & \csvcolvii }
\end{table}

\begin{table}[h]
    \caption{\textbf{Two-qubit} gate count results for Nam gate set with \textbf{24 hour timeout}}
    \begin{filecontents*}{plot_generation/all_results_nam_24hr_2q.csv}
Circuit,Original,VOQC,Qiskit,Quartz,Quartz-NoPP,QUESO,QUESO-PP
adder\_8,409,337,409,401,409,340,289
barenco\_tof\_3,24,22,24,16,20,20,16
barenco\_tof\_4,48,44,48,30,48,38,30
barenco\_tof\_5,72,66,72,44,72,56,44
barenco\_tof\_10,192,176,192,146,192,146,114
csla\_mux\_3,80,74,80,76,78,70,70
csum\_mux\_9,168,168,168,122,168,168,112
gf2\string^4\_mult,99,99,99,96,99,98,95
gf2\string^5\_mult,154,154,154,154,154,152,148
gf2\string^6\_mult,221,221,221,221,221,221,211
gf2\string^7\_mult,300,300,300,300,300,300,285
gf2\string^8\_mult,405,405,405,405,405,403,382
gf2\string^9\_mult,494,494,494,494,494,491,466
gf2\string^10\_mult,609,609,609,609,609,609,573
mod5\_4,28,28,28,15,22,21,18
mod\_mult\_55,48,40,48,39,43,39,39
mod\_red\_21,105,93,105,81,105,92,79
qcla\_adder\_10,233,207,233,211,231,211,179
qcla\_com\_7,186,148,186,129,186,143,121
qcla\_mod\_7,382,338,382,374,382,352,299
rc\_adder\_6,93,73,93,73,81,79,71
tof\_3,18,16,18,14,16,16,14
tof\_4,30,26,30,22,28,26,22
tof\_5,42,36,42,30,40,36,30
tof\_10,102,86,102,70,102,86,70
vbe\_adder\_3,70,54,70,40,66,44,37
qaoa\_n4\_p4,48,48,48,30,30,30,36
qaoa\_n6\_p4,72,72,72,68,68,64,64
qaoa\_n8\_p4,96,96,96,96,96,88,88
qaoa\_n10\_p4,120,120,120,120,120,116,116
qaoa\_n14\_p4,168,168,168,165,165,160,160
qaoa\_n20\_p4,240,240,240,239,239,232,232
qaoa\_n30\_p4,360,360,360,360,360,352,360
\end{filecontents*}
\csvreader[
        tabular = l|r r r r r r r, 
        table foot = \hline, 
        table head = \bfseries Circuit & \bfseries Original & \bfseries \voqc & \bfseries Qiskit & \bfseries Quartz & \bfseries Quartz-NoPP & \bfseries \ours & \bfseries \ourspp \\\hline
    ]{plot_generation/all_results_nam_24hr_2q.csv}{}
    {\csvcoli & \csvcolii & \csvcoliii & \csvcoliv & \csvcolv & \csvcolvi & \csvcolvii & \csvcolviii }
\end{table}

\FloatBarrier
\subsection{T gate count}

\begin{table}[h]
    \caption{\textbf{T} gate count results for Nam gate set with \textbf{1 hour timeout}}
    \begin{filecontents*}{plot_generation/all_results_nam_tcount.csv}
Circuit,Original,VOQC,Qiskit,PyZX,Quartz,Quartz-NoPP,QUESO,QUESO-PP
adder\_8,399,215,361,170,231,399,265,219
barenco\_tof\_3,28,16,24,16,16,28,16,16
barenco\_tof\_4,56,28,48,28,28,56,28,28
barenco\_tof\_5,84,40,72,40,40,84,60,40
barenco\_tof\_10,224,100,192,100,100,224,160,100
csla\_mux\_3,70,64,68,46,64,70,64,64
csum\_mux\_9,196,84,196,78,84,196,168,84
gf2\string^4\_mult,112,68,96,60,68,112,82,68
gf2\string^5\_mult,175,115,155,94,115,175,135,115
gf2\string^6\_mult,252,150,216,140,150,252,198,150
gf2\string^7\_mult,343,217,301,217,217,343,273,214
gf2\string^8\_mult,448,264,384,264,264,448,360,264
gf2\string^9\_mult,567,351,495,351,351,567,459,351
gf2\string^10\_mult,700,410,600,410,410,700,568,410
mod5\_4,28,16,26,8,14,28,10,10
mod\_mult\_55,49,35,45,28,35,49,39,35
mod\_red\_21,119,73,107,72,77,119,75,73
qcla\_adder\_10,238,164,222,154,164,238,198,164
qcla\_com\_7,203,95,177,92,107,203,135,95
qcla\_mod\_7,413,249,361,227,245,413,321,245
rc\_adder\_6,77,47,67,47,47,77,65,47
tof\_3,21,15,19,15,15,19,15,15
tof\_4,35,23,31,23,23,35,23,23
tof\_5,49,31,43,31,31,49,31,31
tof\_10,119,71,103,71,71,119,71,71
vbe\_adder\_3,70,24,62,24,32,70,28,24
\end{filecontents*}
\csvreader[
        tabular = l|r r r r r r r r, 
        table foot = \hline, 
        table head = \bfseries Circuit & \bfseries Original & \bfseries \voqc & \bfseries Qiskit & \bfseries PyZX & \bfseries Quartz & \bfseries Quartz-NoPP & \bfseries \ours & \bfseries \ourspp \\\hline
    ]{plot_generation/all_results_nam_tcount.csv}{}
    {\csvcoli & \csvcolii & \csvcoliii & \csvcoliv & \csvcolv & \csvcolvi & \csvcolvii & \csvcolviii & \csvcolix}
\end{table}

\end{document}